\documentstyle[aas2pp4,psfig]{article}

\def\dash{\hbox{--}}
\def\degree{{$^\circ$}}
\def\ts{\thinspace}           

\def\puncspace{\ifmmode\,\else{\ifcat.\C{\if.\C\else%
\if,\C\else\if?\C\else\if:\C\else\if;\C\else\if-\C\else%
\if)\C\else\if/\C\else\if]\C\else\if'\C%
\else\space\fi\fi\fi\fi\fi\fi\fi\fi\fi\fi}%
\else\if\empty\C\else\if\space\C\else\space\fi\fi\fi}\fi}%
\def\SP{\let\\=\empty\futurelet\C\puncspace}

\def\ee#1{\ifmmode {} \times 10^{#1} \else ${} \times 10^{#1}$\fi}
\def\sub#1{\ifmmode _{#1} \else $_{#1}$\fi}
\def\sup#1{\ifmmode ^{#1} \else $^{#1}$\fi}

\def\dash{\hbox{--}}
\def\about{\ifmmode \sim \else {$\sim\,$}\fi}
\def\lta{\ifmmode {\,\mathbin{\lower 3pt\hbox   
    {$\,\rlap{\raise 5pt\hbox{$\char'074$}}\mathchar"7218\,$}}}
    \else {${\mathbin{\lower 3pt\hbox
    {$\rlap{\raise 5pt\hbox{$\char'074$}}\mathchar"7218\,$}}}
    $}\fi}
\def\gta{\ifmmode {\mathbin{\lower 3pt\hbox   
    {$\,\rlap{\raise 5pt\hbox{$\char'076$}}\mathchar"7218\,$}}}
    \else {${\mathbin{\lower 3pt\hbox
    {$\rlap{\raise 5pt\hbox{$\char'076$}}\mathchar"7218\,$}}}
    $}\fi}

\def\ast{\mathchar"2203} \mathcode`*="002A   

\def\pd#1{{\partial\displaystyle #1}}
\def\comma{{\;\;,}}

\def\degree{{\ifmmode ^\circ \else $^\circ$\fi}}
\def\eps{{\hbox{ergs\ts s\sup{-1}}}\SP}

\def\gpsqcm{{\hbox{g\ts cm\sup{-2}}}\SP}

\def\Gcmc{{\hbox{G\ts cm\sup3}}\SP}
\def\Hz{{\hbox{Hz}}\SP}
\def\keV{{\hbox{keV}}\SP}
\def\kHz{{\hbox{kHz}}\SP}
\def\km{{\hbox{km}}\SP}

\def\aql#1{\leavevmode\hbox{Aql~X-#1}\SP}

\def\cir#1{\leavevmode\hbox{Cir~X-#1}\SP}
\def\cyg#1{\leavevmode\hbox{Cyg~X-#1}\SP}
\def\fu#1{\leavevmode\hbox{4U~#1}\SP}
\def\gx#1{\leavevmode\hbox{GX~#1}\SP}

\def\ks#1{\leavevmode\hbox{KS~#1}\SP}

\def\sco#1{\leavevmode\hbox{Sco~X-#1}\SP}

\def\exosat{\leavevmode{\it EXOSAT}\SP}
\def\ginga{\leavevmode{\it Ginga\/}\SP}
\def\gro{\leavevmode{\it Compton Gamma-Ray Observatory\/}\SP}

\def\rxte{\leavevmode{\it RXTE\/}\SP}

\def\mdot{{\ifmmode \dot M \else {$\dot M$}\fi}}
\def\mdote{{\ifmmode \dot M_E \else {$\dot M_E$}\fi}}
\def\mdoti{{\ifmmode \dot M_i \else {$\dot M_i$}\fi}}
\def\msun{{\ifmmode M_\odot \else {$M_{\odot}$}\fi}}
\def\nonrot{{\rm 0}}

\begin{document}

\lefthead{Miller, Lamb, \& Psaltis}
\righthead{Sonic-Point Model of Kilohertz QPOs}

\title{Sonic-Point Model of Kilohertz
Quasi-Periodic Brightness\\Oscillations in
Low-Mass X-ray Binaries}

\author{M.\ Coleman Miller\altaffilmark{1}}
\affil{Department of Astronomy and Astrophysics, University of Chicago\\
       5640 South Ellis Avenue, Chicago, IL 60637, USA\\
       miller@bayes.uchicago.edu}
\authoremail{miller@gamma.uchicago.edu}

\and

\author{Frederick K.\ Lamb and 
  Dimitrios Psaltis\altaffilmark{2}}
\affil{Department of Physics and Department of Astronomy\\
       University of Illinois at Urbana-Champaign\\
       1110 W. Green St., Urbana, IL 61801-3080, USA\\
       f-lamb@uiuc.edu}
\authoremail{f-lamb@uiuc.edu}
\altaffiltext{1}{\gro Fellow.}
\altaffiltext{2}{Present address: Harvard-Smithsonian Center
for Astrophysics, 60 Garden St., Cambridge, MA 02138;
dpsaltis@cfa.harvard.edu}


\begin{abstract}

Quasi-periodic brightness oscillations (QPOs)
with frequencies ranging from $\sim\,$300~Hz to
$\sim\,$1200~Hz have been discovered in the
X-ray emission from fourteen neutron stars in
low-mass binary systems and from another neutron
star in the direction of the Galactic center.
These kilohertz QPOs are very strong, with rms
relative amplitudes ranging up to $\sim\,$15\%
of the total X-ray countrate, and are remarkably
coherent, with frequency to FWHM ratios as large
as $\sim\,$200. Two simultaneous kilohertz QPOs
differing in frequency by $\sim\,$250--350~Hz
have been detected in twelve of the fifteen
sources.

Here we propose a model for these QPOs. In this
model the X-ray source is a neutron star with a
surface magnetic field $\sim\!10^7 \dash
10^{10}$~G and a spin frequency of a few hundred
Hertz, accreting gas via a Keplerian disk. Some
of the accreting gas is channeled by the stellar
magnetic field but some remains in a Keplerian
disk flow that penetrates to within a few
kilometers of the stellar surface. The frequency
of the higher-frequency QPO in a kilohertz QPO
pair is the Keplerian frequency at a radius near
the sonic point at the inner edge of the
Keplerian flow whereas the frequency of the
lower-frequency QPO is approximately the difference between
the Keplerian frequency at a radius near the
sonic point and the stellar spin frequency. The
difference between the frequencies of the pair
of QPOs is therefore close to (but not necessarily
exactly equal to) the stellar spin frequency.
The amplitudes of the QPOs at the sonic-point
Keplerian frequency and at the beat frequency
depend on the strength of the neutron star's
magnetic field and the accretion rate and hence
one or both of these QPOs may sometimes be
undetectable. Oscillations at the stellar spin
frequency and its overtones are expected to be
weak but may sometimes be detectable.

This model is consistent with the magnetic field
strengths, accretion rates, and scattering
optical depths inferred from previous modeling
of the X-ray spectra and rapid X-ray variability
of the atoll and Z sources. It explains
naturally the frequencies of the kilohertz QPOs
and the similarity of these frequencies in
sources with different accretion rates and
magnetic fields. The model also explains the
high coherence and large amplitudes of the
kilohertz QPOs, and the steep increase of QPO
amplitude with photon energy. The increase in
QPO frequency with inferred accretion rate seen
in many sources is also understandable in this
model.

We show that if the frequency of the
higher-frequency QPO in a pair is an orbital
frequency, as in the sonic-point model, the
frequencies of these QPOs place interesting upper
bounds on the masses and radii of the neutron
stars in the kilohertz QPO sources and provide
new constraints on the equation of state of
matter at high densities. Further observations
of these QPOs may provide compelling evidence
for the existence of a marginally stable orbit,
confirming a key prediction of general
relativity in the strong-field regime.

\end{abstract}


\section{INTRODUCTION}

Observations of accreting neutron stars in
low-mass X-ray binaries (LMXBs) with the {\em
Rossi X-ray Timing Explorer\/} have revealed that
the persistent X-ray emission of at least fifteen
show remarkably coherent quasi-periodic
brightness oscillations (QPOs), with frequencies
$\nu_{\rm QPO}$ ranging from $\sim\,$300~Hz to
$\sim\,$1200~Hz. These kilohertz QPOs are the
highest-frequency oscillations ever seen in any
astrophysical object.

Eight of the fourteen identified sources in which
kilohertz QPOs have been detected are ``atoll''
sources (\fu{0614$+$091}, \fu{1608$-$52},
\fu{1636$-$536}, \fu{1728$-$34},
\ks{1731$-$260}, \fu{1735$-$444},
\fu{1820$-$30}, and \aql{1}). Kilohertz QPOs
have also been detected in all six of the
originally identified ``Z'' sources (\sco1,
\gx{5$-$1}, \gx{17$+$2}, \gx{340$+$0},
\gx{349$+$2}, and \cyg{2}). (For the definitions
of atoll and Z sources, see Hasinger \& van der
Klis 1989.) Highly coherent brightness
oscillations with frequencies ranging from
$\sim\,$360~Hz to $\sim\,$580~Hz have been
detected during type~I (thermonuclear) X-ray
bursts from four kilohertz QPO sources. Another
neutron star in the direction of the Galactic
center shows burst oscillations with a frequency
of 589~Hz, but this source has not yet been
positively identified. The frequency ranges and
rms amplitudes of the currently known kilohertz
QPOs and burst oscillations are listed in
Table~1, with references.

Two simultaneous kilohertz QPOs have so far been
seen in six of the eight atoll sources in which
kilohertz QPOs have been detected (all except
\fu{1735$-$444} and \aql{1}) and in all six of
the Z sources. The differences $\Delta\nu$
between the frequencies of the two QPOs seen in
these sources all fall in the range $\sim
250$--350~Hz. In the atoll sources
\fu{0614$+$091} (Ford et al.\ 1996, 1997a),
\fu{1636$-$536} (Wijnands et al.\ 1997b), and
\fu{1728$-$34} (Strohmayer et al.\ 1996a, 1996b,
1996c) and in the Z sources \gx{5$-$1} (van der
Klis et al.\ 1996e), \gx{17$+$2} (Wijnands et
al.\ 1997c), and \cyg{2} (Wijnands et al.\ 1998)
$\Delta\nu$ is constant in time, within the
errors. In the Z source \sco1, $\Delta\nu$
changed from $\sim\,$250~Hz when the accretion
rate was probably near and sometimes even
slightly greater than the Eddington critical
rate to $\sim\,$300~Hz when the accretion  rate
was significantly lower (van der Klis et al.\
1997b).

In \fu{0614$+$091}, a brightness oscillation with
a frequency consistent with the difference
$\Delta\nu = 328$~Hz between the frequencies of
the two kilohertz QPOs seen simultaneously in
this source was detected with marginal
significance during one 30-min interval (Ford et
al.\ 1996, 1997a). In \fu{1728$-$34}, a strong,
relatively coherent brightness oscillation was
detected in six of the twelve X-ray bursts so
far observed from this source, with a frequency
that has remained constant, within the errors,
for more than a year and is consistent with the
difference $\Delta\nu = 363$~Hz between the
frequencies of the two kilohertz QPOs seen
simultaneously in this source (Strohmayer et
al.\ 1996b, 1996c; Strohmayer, Zhang, \& Swank
1997; Strohmayer 1997, personal communication).
Burst oscillations have also been seen in
\fu{1636$-$536}, with a frequency $\sim$580~Hz
(Zhang et al.\ 1996, 1997), and in
\ks{1731$-$260}, with a frequency $\sim$520~Hz
(Morgan \& Smith 1996; Smith, Morgan, \& Bradt
1997). In both cases the frequency of the burst
oscillation is approximately {\em twice} the
difference between the frequencies of the two
kilohertz QPOs observed simultaneously in these
sources. 

The peaks in power density spectra associated
with these kilohertz QPOs are relatively narrow,
with FWHM $\delta\nu_{\rm QPO}$ as small as
$0.005\,\nu_{\rm QPO}$. When integrated over all
photon energies, the rms relative amplitudes of
the kilohertz QPOs seen in the atoll sources
range up to $\sim 15$\% of the total 2--60~keV
X-ray countrate and are typically much greater
than the $\lta1$\% amplitudes of the
kilohertz QPOs observed in the Z sources.
The amplitudes of the kilohertz QPOs increase
steeply with photon energy, up to at least
15~keV, in all the atoll and Z sources where the
photon-energy dependence of the kilohertz QPOs
has been studied (\fu{1608$-$52}: Berger et al.\
1996; \fu{1636$-$536}: Zhang et al.\ 1996;
\fu{1728$-$34}: Strohmayer et al.\ 1996c;
\ks{1731$-$260}: Wijnands \& van der Klis 1997;
\gx{5$-$1}: van der Klis et al.\ 1996e;
\gx{17$+$2}: Wijnands et al.\ 1997c; \cyg{2}:
Wijnands et al.\ 1998).

The frequencies of the kilohertz QPOs have been
seen to vary by as much as a factor $\sim 2$. The
frequencies of the two kilohertz QPOs observed in
the persistent X-ray emission between type~I
X-ray bursts of the atoll sources
\fu{0614$+$091} (Ford et al.\ 1996, 1997a),
\fu{1728$-$34} (Strohmayer et al.\ 1996a, 1996b,
1996c), \ks{1731$-$260} (Wijnands \& van der
Klis 1997), and \fu{1820$-$30} (Smale et al.\
1996, 1997) increase steeply with increasing
countrate ($d\log\nu/d\log C\!R\;\gta\;1$). The
frequencies of the pair of kilohertz QPOs seen
during two different observations of
\fu{0614$+$091} (Ford et al.\ 1997a) correlated
tightly and positively with the energy of the
{\it peak\/} in the keV X-ray spectrum (Ford et
al.\ 1997b), even though the tracks in the
frequency-{\it countrate\/} plane made by the
source during the two observations differed
significantly. The frequencies of the pair of
high-frequency QPOs observed in the Z sources
\sco1, \gx{5$-$1}, \gx{17$+$2}, and \cyg{2}
increase steeply with the inferred accretion
rate in these sources (van der Klis et al.\
1996a, 1996b, 1996d, 1996e, 1997a; Wijnands et
al.\ 1997c, 1998; as discussed by van der  Klis
[1989] and Lamb [1989, 1991], there is no simple
relation between countrate and accretion rate in
the Z sources).

Two simultaneous kilohertz QPOs are so common
that attention has focused on models that might
be able to explain the occurrence of such
kilohertz QPO pairs. Strohmayer et al.\ (1996c)
and Ford et al.\ (1997a) have suggested adapting
the magnetospheric beat-frequency model to
explain these QPOs. This model was originally
proposed by Alpar \& Shaham (1985) and Lamb et
al.\ (1985) and developed further by Shibazaki
\& Lamb (1987) to explain the {\em single\/}
15--60~Hz horizontal-branch oscillation (HBO)
discovered earlier in the Z sources (see van der
Klis 1989). As we discuss in \S~2, it is very
difficult to understand several key features of
the kilohertz QPOs, including the simultaneous
occurrence of two kilohertz QPOs, in terms of
the magnetospheric beat-frequency model.

Here we propose a model for the kilohertz QPOs
that is fundamentally different from the
magnetospheric beat-frequency model. In the model
presented here, the X-ray source is a neutron
star with a surface magnetic field $\sim 10^7
\dash 10^{10}$~G and a spin frequency of a few
hundred Hertz, accreting gas via a Keplerian
disk. In this model the key characteristics of
the kilohertz QPOs are explained as follows
(details are presented in \S~3):

{\em Accretion flow}.---Magnetoturbulence,
differential\break cooling, and radiation forces
create density inhomogeneities (``clumps'') in
the gas in the nearly Keplerian disk flow. These
clumps are dissipated by the shear in the
azimuthal velocity combined with gas pressure
forces and turbulence. If the neutron star has a
magnetic field $\gta 10^7$~G, some of the
accreting gas is channeled out of the disk flow
by the stellar magnetic field. Channeling occurs
at several stellar radii, if the stellar magnetic
field is $\sim 10^{9}$--$10^{10}$~G, but only
very close to the stellar surface, if the field
is $\sim10^7$--$10^8$~G. Even if the field is as
strong as $\sim 10^{9}$--$10^{10}$~G, some of the
accreting gas continues to drift inward in a
nearly Keplerian disk flow until either radiation
drag forces become important, the gas reaches the
innermost stable circular orbit, or the gas
begins to interact viscously with the stellar
surface. The fraction of the accreting gas that
remains in the Keplerian flow close to the star
depends primarily on the strength of the stellar
magnetic field and the accretion rate.

{\em Role of radiation drag and general
relativity}.---For the accretion rates typical of
the Z and atoll sources, the nearly Keplerian
flow ends and the inward radial velocity of the
gas in the disk increases sharply as the gas
nears the star, either because of the azimuthal
drag force exerted on the gas in the disk by the
radiation coming from near the stellar surface
or---if the drag force is weak and the radius of
the neutron star is less than the radius of the
innermost stable circular orbit---because of the
general relativistic corrections to Newtonian
gravity that produce an innermost stable orbit.
In either case, the inward radial velocity of
the gas becomes supersonic within a very short
radial distance from the radius at which it
begins to increase.

{\em Effect of inhomogeneities in the
flow}.---Roughly speaking, clumps that form
outside the radius where the radial inflow
becomes supersonic are dissipated before gas
from them reaches the stellar surface, whereas
gas from clumps that form near the radius where
the flow becomes supersonic falls inward and
collides with the stellar surface before the
clumps are destroyed. Where the streams of
denser gas from the clumps orbiting near the
sonic point collide with the stellar surface,
the streams create arc-shaped areas of brighter
X-ray emission that move around the star's
equator with a frequency equal to the orbital
frequency of the clumps, which is approximately
the Keplerian frequency at the sonic point.

{\em Higher-frequency kilohertz QPO}.---The
frequency $\nu_{\rm QPO2}$ of the higher-frequency of the two
kilohertz QPOs is approximately equal to the
Keplerian frequency at the sonic point. A
detectable quasi-periodic oscillation of the
X-ray flux and energy spectrum may be produced at
this frequency by the periodically changing
aspect (and, for most viewing directions, by the
neutron star's periodic eclipse) of the
arc-shaped areas of brighter X-ray emission, as
they move around the star with a frequency equal
to the orbital frequency of clumps in the
Keplerian disk flow near the sonic point. The
oscillation appears quasi-periodic because the
clumps have finite lifetimes and slightly
different orbital frequencies. We find that
$\nu/\delta\nu$ can be as large as \about 100
for these oscillations, consistent with the
observed coherence of the kilohertz QPOs. The
frequency of the sonic-point Keplerian QPO is
almost independent of the spin rate of the
neutron star (frame-dragging causes the
Keplerian frequency at the sonic point, and
hence the QPO frequency, to depend weakly on the
star's spin rate).

{\em Lower-frequency kilohertz QPO}.---The
frequency $\nu_{\rm QPO1}$ of the lower-frequency of the two
kilohertz QPOs is equal to one (or possibly two)
times the difference between the Keplerian
frequency at a radius near the sonic point and
the stellar
spin frequency. A QPO may be produced at this
beat frequency if the magnetic field of the
neutron star is weak enough that a Keplerian disk
flow penetrates close to the star but strong
enough to channel some of the accreting gas
before it collides with the stellar surface. For
the accretion rates of the Z and atoll sources,
this requires surface magnetic fields $B_s$ in
the range $\sim 10^{7} \dash 10^{10}$~G.

Channeling of gas out of the disk generates a QPO
at the beat frequency because it creates slightly
brighter regions that {\it rotate with the
star\/} (these slightly brighter regions are
distinct from the brighter spots that move around
the star at the sonic-point Keplerian frequency).
The slightly brighter regions that rotate with
the star produce a weakly-beamed radiation
pattern that rotates at the stellar spin
frequency. The radiation in this pattern produces
a slight increase of the azimuthal drag force on
the gas in the clumps orbiting near the sonic
point, once or twice each beat period (depending
on the symmetry of the slightly brighter region).

This oscillation of the drag force in turn
creates an oscillation of the inward flux of gas
from a given clump at the sonic point once or
twice each beat period. This modulation of the
mass flux in the gas streams flowing inward from
all clumps orbiting near the sonic point
generates a quasi-periodic oscillation of the
luminosity and spectrum of the X-ray emission
from the stellar surface, with a frequency equal
to one (or possibly two) times the difference 
between the Keplerian frequency
at a radius near the sonic point and
the stellar spin
frequency. The difference between the frequencies
of the two QPOs in a pair is therefore close to
(but not necessarily {\em exactly\/} equal to)
the stellar spin frequency.

{\em Oscillation amplitudes}.---The amplitudes of
the\break QPOs at the sonic-point Keplerian
frequency and at the beat frequency depend on the
strength of the neutron star's magnetic field and
the mass accretion rate, and hence one or both of
these QPOs may sometimes be undetectable.
Oscillations at the stellar spin frequency and
its overtones are expected to be weak but may
sometimes be detectable. 

{\em Similarity of kilohertz QPO frequency behavior in
different sources.}---The similar frequency
ranges and the similarity of the
frequency-accretion rate correlations of the
kilohertz QPOs observed in the Z and atoll
sources---which have accretion rates and magnetic
fields that differ by factors of ten or more---is
explained by the crucial role of radiation drag,
which is only effective within one or two stellar
radii, the approximate proportionality of the
angular momentum carried by the accreting matter
and the radiation drag force, the fact that the
stellar magnetic fields as well as the accretion
rates are larger in the Z sources, and the
tendency for the vertical thickness of the disk
to increase with increasing mass flux.

The radius where the radiation drag or general
relativistic corrections to Newtonian gravity
cause the radial velocity to increase abruptly
plays a key role in the model of the kilohertz
QPOs proposed here. Once the radial velocity
begins to increase, the flow becomes supersonic
within a very short radial distance (the inward
radial velocity increases so abruptly that the
location of the sonic point is insensitive to the
precise definition used), so the sonic point is a
useful indicator of the point in the disk flow
where the radial velocity begins to increase
steeply. Hence, for convenience we shall call the
point where the radial velocity increases sharply
the ``sonic point'' and refer to this model for
the kilohertz QPOs as the ``sonic-point model''.

We show that if the frequency of the
higher-frequency QPO in a pair of kilohertz QPOs
is the orbital frequency of gas in a stable
Keplerian orbit around the neutron star, as in
the sonic-point model proposed here, the
frequencies of these QPOs provide interesting
new upper bounds on the masses and radii of the
neutron stars in the Z and atoll sources and new
constraints on the equation of state of matter
at high densities.

Any model of how the kilohertz QPOs are produced
must be consistent with the known properties of
the atoll and Z sources. Hence, before analyzing
the sonic-point QPO model of the kilohertz QPOs
in more detail, we first summarize in \S~2 the
previously known X-ray spectral and
lower-frequency X-ray variability properties of
the atoll and Z sources and the physical picture
of these sources that has been developed based on
these properties. In \S~3 we analyze the physics
of the sonic-point model and show that it is
consistent with the basic properties of the
kilohertz QPOs. In \S~4 we  demonstrate that the
sonic-point model is also consistent with the
existing physical picture of the Z and atoll
sources and with many of the more detailed
properties of the kilohertz QPOs. In \S~5 we show
how to derive upper bounds on the masses and
radii of the neutron stars in the kilohertz QPO
sources from the frequencies of stable Keplerian
orbits in the kilohertz range and discuss the
constraints on the properties of neutron-star
matter that follow from these bounds; the bounds
we derive include the effects of frame-dragging.
Finally, in \S~6 we discuss several specific
predictions of the sonic-point model.

\section{PROPERTIES OF THE NEUTRON STARS IN LMXBS}

\subsection{Observed Properties of the Atoll and
Z Sources}

The atoll sources are LMXBs that, over time,
trace atoll-shaped patterns in X-ray color-color
diagrams (Hasinger \& van der Klis 1989). They
have luminosities $L \sim
10^{36}\dash10^{37}$~\eps, i.e., \about 1--10\%
of the Eddington critical luminosity $L_E$ of a
neutron star. Power-density spectra of their
brightness variations show broad, band-limited
noise components at frequencies below $\sim
100$~Hz. No QPOs with frequencies $\lta 100$~Hz
have so far been detected in any of the atoll
sources with the exception of \cir1, which has a
QPO  that increases in frequency from 1 to 30~Hz
as the countrate increases, when the source is
very bright (Oosterbroek et al.\ 1995; Shirey et
al.\ 1996; Bradt, Shirey, \& Levine 1998).

Recent comparisons of models of the X-ray
emission of neutron stars in LMXBs with the
X-ray spectra of atoll sources observed with
\exosat (Psaltis \& Lamb 1998a, 1998b, 1998c)
suggest that these sources can be subdivided
into two groups, the ``4U'' atoll sources (such
as \fu{1636$-$53}, \fu{1705$-$44},
\fu{1820$-$30},\break
 \fu{1608$-$52}, and \fu{1728$-$34}) and the
``GX'' atoll sources (such as \gx{9$+$1},
\gx{9$+$9}, \gx{3$+$1}, and \gx{13$+$1}), based
on the strengths of their inferred magnetic
fields; one atoll source, \fu{1735$-$44}, has
intermediate spectral properties and therefore
probably has a magnetic field of intermediate
strength.

The Z sources are LMXBs that produce a
characteristic Z-shaped track in X-ray color-color
diagrams (Hasinger \& van der Klis 1989). They
have luminosities $L \sim 10^{38}$~\eps, i.e.,
comparable to $L_E$. The three branches of the Z
are called the horizontal, normal, and flaring
branches. When a Z source is on the horizontal
branch, a QPO with a frequency in the range
15--60~Hz is observed (van der Klis et al 1985;
van der Klis 1989). This ``horizontal branch
oscillation'' (HBO) is also detectable in some Z
sources when they are on the upper part of the
normal branch. The relative width $\delta\nu/\nu$
of the HBO peak in power density spectra is
typically \about 0.1--0.3 and its centroid
frequency typically increases with increasing
countrate. As a Z source moves down the normal
branch, the HBO becomes weaker and eventually
disappears into the noise continuum.

Near the middle of the normal branch, a
different QPO appears. The peaks in power
density spectra produced by this QPO have
relative widths $\delta\nu/\nu \sim 0.3$ and
centroid frequencies in the range 4--8~Hz
(Middleditch \& Priedhorsky 1986; van der Klis
1989). The properties of this QPO do not vary
appreciably on the lower normal branch. As a Z
source moves from the lower normal branch to the
flaring branch, the frequency of this second QPO
increases abruptly to $\sim 15 \dash 20$~Hz and
the QPO becomes weaker and less coherent,
eventually disappearing into the noise (van der
Klis 1995; Dieters \& van der Klis 1997). This
QPO is the ``normal/flaring branch oscillation''
(N/FBO).

In addition to these two types of quasi-periodic
oscillations, power spectra of the brightness
variations of the Z sources also show three
distinct band-limited noise components at
frequencies below \about 200~Hz (Hasinger \& van
der Klis 1989). The properties of the QPO and
noise components vary systematically with the
position of a source on its Z track (Hasinger \&
van der Klis 1989; see Kuulkers 1995, Dieters \&
van der Klis 1997, and Wijnands et al.\ 1997a
for detailed studies of this behavior and
possible exceptions).

Recent analyses of archival \exosat data
(Kuulkers 1995; Kuulkers et al.\ 1995, 1996)
suggest that the Z sources can be subdivided
into two groups, based on the morphology of
their Z tracks and power spectra: the
``Cyg-like'' Z sources (\cyg2, \gx{5$-$1}, and
\gx{340$+$0}) and the ``Sco-like'' Z sources
(\sco1, \gx{17$+$2}, and \gx{349$+$2}). At a
more detailed level, \gx{17$+$2} shares some of
the characteristics of the ``Cyg-like'' sources
and therefore probably has intermediate
properties (Hasinger \& van der Klis 1989;
Wijnands et al.\ 1997a).

\subsection{Current Physical Picture of the Atoll
and Z Sources}

The atoll and Z sources are neutron stars
accreting gas from a Keplerian disk fed by a
low-mass companion star (see van der Klis 1989).
The magnetic fields and the accretion rates of
these neutron stars are thought to be the most
important parameters that determine their X-ray
spectral and temporal characteristics. According
to the most complete and self-consistent current
model of these sources, the so-called {\it
unified model\/} (Lamb 1989, 1991), the atoll
sources have dipole magnetic fields $\lta
5\ee{9}$~G and luminosities \about 1--10\% of the
Eddington critical luminosity, whereas the Z
sources have dipole fields $\sim 10^{9} \dash
10^{10}$~G and luminosities very close to (and
sometimes slightly above) the Eddington critical
luminosity.

{\it X-ray spectra.}---The X-ray spectra 
of the Z and atoll sources and
the low upper limits on the amplitudes of any
periodic variations of their persistent X-ray
brightness (less than 1\% in some cases; see,
e.g., Vaughan et al.\ 1994) constrain the
properties of these neutron stars. The low upper
limits on brightness variations at their spin
frequencies can be understood if the magnetic
fields of these neutron stars are $\lta
10^{10}$\,G and the magnetospheres and inner disks
are surrounded by a central coronae with
electron scattering optical depths $\sim 3 \dash
5$, even at low accretion rates (Lamb et al.\
1985; Lamb 1989, 1991), because such coronae
strongly suppress the X-ray flux oscillation
produced by any radiation pattern that rotates
with the star (Brainerd \& Lamb 1987). This
effect is discussed further in \S~3.6.

When, as in the Z sources, the total luminosity
of the neutron star, inner disk, and central
corona becomes comparable to $L_E$, the vertical
radiation force drives gas upward, out of the
disk and radiation drag removes its angular
momentum in less than one orbit, creating a
region of approximately radial inflow that
extends out to $\sim 300$~km (Lamb 1989, 1991).
Oscillations in this radial flow are thought to
be responsible for the N/FBO (Lamb 1989; Fortner
et al.\ 1989; Fortner 1992; Miller \& Lamb 1992).

Detailed physical modeling of the X-ray spectra
of the atoll and Z sources indicates that soft
($\sim 0.5 \dash 1$~keV) photons are produced by
optically-thick bremsstrahlung and other
processes at the surface of the star and by
self-absorbed, high-harmonic cyclotron emission
in the inner magnetosphere (Psaltis, Lamb, \&
Miller 1995; Psaltis \& Lamb 1998a, 1998b,
1998c). As we discuss below, the accretion rate
and the strength of the stellar magnetic field
determine whether emission from the stellar
surface or cyclotron emission in the
magnetosphere is the dominant source of soft
photons. These soft photons are upscattered by
electrons in the magnetosphere and the central
corona that surrounds the neutron star to
produce the X-ray spectrum that emerges from the
corona. In the Z sources, scattering by cooler
electrons in the region of approximately radial
inflow that surrounds the inner disk and corona
further deforms the X-ray spectrum.

Cyclotron emission is the dominant source of soft
photons if upscattered cyclotron photons are able
to supply the full accretion luminosity (see
Psaltis et al.\ 1995). The range of stellar
magnetic fields for which cyclotron emission is
dominant can be estimated as follows. Within the
magnetosphere, the spectrum of self-absorbed
cyclotron photons can be approximated by a
blackbody spectrum truncated at the energy at
which cyclotron emission becomes optically thin.
Comptonization within the magnetosphere and
central corona increases the energy of a typical
photon by a factor $\sim e^y$, where $y \equiv
(4k_B T_e/m_e c^2) \max\{\tau,\tau^2\}$ is the
Compton $y$ parameter, $T_e$ and $m_e$ are the
electron temperature and rest mass, $\tau$ is the
electron scattering optical depth, and $k_B$ is
the Boltzmann constant (see Rybicki \& Lightman
1979, pp 195--223). Hence cyclotron emission is the
dominant source of soft photons if $e^yL_{\rm cyc}
\approx GM\mdot/R$ or, equivalently,
 \begin{eqnarray}
  \hskip+0.05 truein \mu_{\rm cyc,27} &
  \hskip-0.05 truein \gtrsim &
  \hskip-0.05 truein 16\,e^{-y/3}
   \left(\frac{\dot{M}_{\rm ns}}{\dot{M}_E}\right)^{\!\!1/3}
   \!\!\left(\frac{T_e}{5~\mbox{keV}}\right)^{\!\!-1/3}\!\!\times\nonumber\\
   &&\hskip-0.7 truein 
   \times\left(\frac{R_{\rm cyc}}{10^6~\mbox{cm}}\right)^{\!\!-2/3}
   \!\!\left(\frac{n}{15}\right)^{\!\!-1}
   \!\!\left(\frac{M}{1.4~\msun}\right)^{\!\!1/3}
   \!\!\left(\frac{R}{10^6~\mbox{cm}}\right)^{\!\!-1/3}\!\!,\nonumber\\
   \label{eq:muSpectrum}
 \end{eqnarray}
 where $\dot{M}_{\rm ns}$ is the mass accretion
rate onto the neutron star surface, $R_{\rm cyc}$
is the effective radius of the cyclotron
photosphere, $n$ is the harmonic number at which
the transition from optically thick to optically
thin emission occurs, and $R$ is the radius of
the neutron star. Figure~\ref{fig:ParameterSpace}
shows how $\mu_{\rm cyc,27}$ depends on \mdot\
for the range of electron temperatures expected
in the magnetospheres and central coronae of the
atoll and Z sources.

As Figure~\ref{fig:ParameterSpace} shows, if the
star's magnetic field is \hbox{$\lta 5\ee{8}$~G}
and the accretion rate is \hbox{$\sim 0.01 \dash
0.03~\mdote$}, the dominant source of photons is
the thermal emission from the surface of the
neutron star. These photons are then upscattered
by the electrons in the magnetosphere and the
central corona. Numerical calculations of the
X-ray spectra produced by stars with these field
strengths and accretion rates agree well with
\exosat observations of the spectra of the ``4U''
atoll sources when they are in the so-called
``banana'' spectral state (Psaltis \& Lamb
1998c).

If instead the star's magnetic field is $\sim
5\ee{8} \dash 5\ee{9}$~G and the accretion rate
is $\lta 0.1 \mdote$, cyclotron emission is the
dominant source of photons. These photons are
then upscattered by the electrons in the
magnetosphere and the central corona. Numerical
calculations of the X-ray spectra produced by
stars with these field strengths and accretion
rates agree well with \exosat observations of
the spectra of the ``GX'' atoll sources (Psaltis
\& Lamb 1998a, 1998b, 1998c).

If the stellar magnetic field is \hbox{$\sim
10^{9} \dash 10^{10}$~G} and the mass accretion
rate is \hbox{$> 0.5~\mdote$}, electron cyclotron
emission in the magnetosphere is very efficient
in producing soft photons. At these accretion rates
soft photons are Comptonized not only by the hot
electrons in the magnetosphere and central corona
but also by the cool electrons in the
approximately radial inflow. Numerical
computations of the X-ray spectra and color
tracks predicted by this physical model agree
well with $\exosat$ and $\ginga$ measurements of
the X-ray spectra and color tracks of the Z
sources (Psaltis et al.\ 1995; Psaltis \& Lamb
1998a, 1998b, 1998c).

{\it Accretion flows}.---If the magnetic field of the neutron star is
$\lta 10^{6}$~G and the accretion rate is $\gta
0.01\,L_E$, the Keplerian disk flow may extend
inward all the way to the surface of the neutron
star without being channeled by the stellar
magnetic field. (Here and below we refer to
accretion flows confined near the orbital plane
as {\it disk flows}, whether or not the azimuthal
velocity field is nearly Keplerian, and as {\it
Keplerian disk flows} if the azimuthal velocity
field is nearly Keplerian, that is, if the flow
is nearly circular.) If instead the stellar
magnetic field is $\gta 10^{11}$~G, most of the
gas in the disk will couple to the stellar field
and be funneled out of the disk plane toward the
magnetic poles of the star at a characteristic
cylindrical radius $\varpi_0$ (Ghosh \& Lamb
1979a, 1979b, 1992) that is much larger than the
radius of the star. Hence, the disk flow around
such a star ends far above the stellar surface.

For stellar magnetic fields of intermediate
strength, some of the accreting gas is likely to
couple to the stellar magnetic field beginning
at $\varpi_0$, which will then channel it toward
the star's magnetic poles, but some is also
likely to continue as a disk flow inside
$\varpi_0$, as a result of Rayleigh-Taylor
instability and incomplete coupling of the flow
in the inner disk to the stellar magnetic field
(see, e.g., Scharlemann 1978; Ghosh \& Lamb
1979a, 1991; Lamb 1984; Spruit \& Taam 1990).
In this case we  expect some of the gas in the
disk flow to be channeled out of the disk by the
magnetic field over a range of radii inside
$\varpi_0$, with the rest of the gas remaining
in a geometrically thin Keplerian flow that
penetrates close to the stellar surface, as
shown in Figure~\ref{fig:SideViews}. This last
case is the one that we expect to be relevant to
the Z sources and most of the atoll sources.

Ghosh \& Lamb (1992) derived expressions for the
characteristic cylindrical radius $\varpi_0$ at
which gas in the disk couples strongly to the
stellar magnetic field and begins to be
channeled out of the disk flow, as a function of
the stellar dipole magnetic moment $\mu$, the
mass accretion rate \mdoti\ through the inner
disk, and the neutron star gravitational mass
$M$, both for gas-pressure-dominated (GPD) and
for radiation-pressure-dominated (RPD) inner
disks, assuming $\varpi_0$ is much larger than
the radius $R$ of the star. The GPD model is
expected to be valid for the atoll sources. The
disks around the Z sources are expected to be
RPD. The scaling of $\varpi_0$ with $\mu$,
\mdoti, and $M$ given by the analytical RPD
model is expected to be valid for the Z sources,
but the vertical thickness of the boundary layer
given by the model is not expected to be
reliable.  We have therefore reduced the vertical thickness 
of the boundary layer in the model so that the Keplerian
frequency at the radius where the gas couples strongly
to the magnetic field agrees quantitatively with the
Keplerian frequency inferred from spin frequency of
\cyg2, based on the frequencies of its kilohertz QPOs
(Wijnands et al.\ 1998; Psaltis et al.\ 1998) and the
magnetospheric beat-frequency interpretation of its HBO
(Psaltis et al.\ 1998), for a magnetic moment equal to
the moment inferred from its X-ray spectrum (Psaltis \&
Lamb 1998c). Specifically, the Keplerian frequency at
the coupling radius was assumed to be 380~Hz, as
indicated by the sum of the spin and HBO frequencies
when \cyg2 is on the normal branch, for a magnetic
moment of $3.5\ee{27}\,\Gcmc$ and a mass flux through
the inner disk equal to the mass accretion rate \mdote\
onto the neutron star that produces the Eddington
critical luminosity.

The resulting two expressions for $\varpi_0(\mu,\mdoti,M)$
can be solved for the dipole moment $\mu_0$ that
gives a Keplerian frequency $\nu_{\rm K0}$ at
the coupling radius; this characteristic dipole
moment is
 \begin{equation}
 \mu_{0,27}
  = \left\{
   \begin{array}{ll}
 \!\! 1.5\,
 \left(\nu_{\rm K0}/1100~\mbox{Hz}\right)^{-1.1}
   \left(\mdoti/\dot{M}_E\right)^{0.43}\times\\
    \quad\times\left(M/1.4~\msun\right)^{0.94},
    \quad\mbox{for GPD disks}\;; \\ \\
 \!\! 0.88\,
 \left(\nu_{\rm K0}/1100~\mbox{Hz}\right)^{-1.3}
   \left(\mdoti/\dot{M}_E\right)^{0.29}\times\\
    \quad\times\left(M/1.4~\msun\right)^{0.91}
   \quad\mbox{for RPD disks}\;.
   \end{array}
 \right.
 \label{eq:muKepler}
 \end{equation}
 Here $\mu_{0,27}\equiv \mu_0/10^{27}\,\Gcmc$.
If the star's dipole moment exceeds $\mu_0$,
most of the gas in the disk flow will couple to
the magnetic field and be channeled out of the
disk at a radius where the Keplerian frequency
is less than $\nu_{\rm K0}$.
Figure~\ref{fig:ParameterSpace} shows the dipole
moment that gives Keplerian frequencies of 500
and 1100~Hz at the main coupling radius, as a
function of the mass accretion rate.  These two 
curves show that for the neutron star magnetic
fields inferred from X-ray spectral modeling, the
Keplerian frequency at the gas coupling radius 
cannot be made simultaneously consistent with the
$\sim$500--1100~Hz ranges of the kilohertz QPOs
in the 4U and Z sources.

Accreting gas that is not channeled out of the
disk plane at $\varpi_0$ is expected to continue
to drift slowly inward in a Keplerian disk flow
until (a)~it is channeled out of the disk at
smaller radii, (b)~it begins to flow rapidly
inward, either because it loses angular momentum
to the radiation drag force or because it crosses
the radius of the marginally stable orbit, or
(c)~it collides with the stellar surface. Gas
that is flowing rapidly inward remains in a
(non-Keplerian) disk flow unless it is channeled
out of the disk by the stellar magnetic field. We
emphasize that {\em the accretion disk ends above
the stellar surface only if all the gas in the
disk is channeled out of the disk flow before it
reaches the star}.

In the magnetospheric beat-frequency model of the
15--60~Hz HBO observed in the Z sources (Alpar
\& Shaham 1985; Lamb et al.\ 1985; Shibazaki \&
Lamb 1987), a substantial fraction of the gas in
the Keplerian disk flow is channeled out of the
disk at $\sim 2 \dash 3$ stellar radii and flows
toward the star's magnetic poles. This model
therefore requires that the Z sources have
magnetic fields $\sim 10^{9} \dash 10^{10}$~G.
The atoll sources have not previously shown any
direct indications that they have dynamically
important magnetic fields (e.g., no HBO-like QPOs
have so far been observed in an atoll source),
and hence their magnetic fields are thought to be
\lta 5\ee{9}~G. However, the existence of
kilohertz QPOs at the difference between an
orbital frequency and the spin frequency
of the neutron star indicates that most if
not all of the atoll sources have magnetic fields
$\gta10^{7}$~G. This is consistent with the detailed
modeling of their X-ray spectra described above,
which indicates that some atoll sources have
magnetic fields as strong as 5\ee{9}~G. These
sources may produce weak HBOs, which may be
detected in the future.

\subsection{Magnetospheric Beat-Frequency Model
and the Kilohertz QPOs}

As noted in \S~1, Strohmayer et al.\ (1996c) and
Ford et al.\ (1997a) have suggested adapting the
magnetospheric beat-frequency QPO model to explain
the two simultaneous kilohertz QPOs. As described
in \S~2.2, the magnetospheric beat-frequency
model was originally proposed by Alpar \& Shaham
(1985) and Lamb et al.\ (1985) and developed
further by Shibazaki \& Lamb (1987) to explain
the {\em single}, variable frequency, 15--60~Hz
HBO seen in the Z sources. Strohmayer et al.\ and
Ford et al.\ have suggested that the frequency of
the higher-frequency QPO in a kilohertz QPO pair
might be the Keplerian orbital frequency at the
radius where gas in the accretion disk begins to
couple to the stellar magnetic field and that the
frequency of the lower-frequency QPO is the beat
between this frequency and the stellar spin
frequency. There are several difficulties with
this application of the magnetospheric
beat-frequency model.

{\em Widespread occurrence of two
simultaneous, highly coherent kilohertz QPOs}. The
occurrence of two simultaneous, strong (amplitudes
up to $\sim 15$\% rms), and highly coherent
($\nu/\delta\nu \sim 50$--100) kilohertz QPOs
appears inconsistent with the magnetospheric beat
frequency model, because a strong, highly coherent
oscillation with a frequency equal to the orbital
frequency where gas first couples to the stellar
magnetic field is not expected, nor is it observed
in other types of sources where the magnetospheric
beat frequency mechanism is thought to be
operating.

The magnetospheric beat-frequency mechanism is
expected to produce a single, strong, moderately
coherent oscillation with a frequency equal to
the beat frequency (power is of course also
expected at overtones of this frequency), but it
is not expected to produce a strong, highly
coherent oscillation  with a frequency equal to
the orbital frequency at the initial gas
coupling radius, for two reasons. First, no
effect is known that would select a sufficiently
narrow range of radii near this radius. Second,
even if a narrow range of radii were selected,
no mechanism is known that would make the
orbital frequencies at these radii visible.

Gas is expected to leave the disk along field
lines that thread a very narrow annulus where
the gas couples strongly to the magnetic field
of the star (Ghosh \& Lamb 1979a). The range of
orbital frequencies in this annulus is likely to
be very small, consistent with the
$\sim\,$5--10~Hz observed width of the HBO (Lamb
et al.\ 1985). The magnetic field is expected to
couple more weakly to the gas in the disk over a
wider annulus (Ghosh \& Lamb 1979a). However, in
both annuli the gas-magnetic field interaction
repeats each beat period, not each orbital
period (Lamb et al.\ 1985), so there is no
obvious way that this interaction would make the
orbital period visible in X-rays. No other
mechanism is known that would single out a
narrow range of radii in the inner disk.

Even if a narrow range of radii were singled out,
little luminosity is expected to be generated in
such a narrow annulus, as was explained in the
context of the HBO by Lamb (1988). Generation of
a QPO by beaming of radiation via periodic
occultation of the emission from the neutron
star by clumps orbiting at the initial coupling
radius would require us to be viewing {\it
all\/} the kilohertz QPO sources from within a
small range of special inclinations and would
also require the height of the disk at its inner
edge to vary with accretion rate in just the
right way (again see Lamb 1988). Moreover, such
beaming would be strongly suppressed by
scattering in the ionized gas that surrounds the
neutron star (see Lamb et al.\ 1985; Brainerd \&
Lamb 1987; Kylafis \& Phinney 1989; Lamb 1988,
1989; Psaltis et al.\ 1995).

A QPO that appears to be the magnetospheric beat
frequency QPO has been observed in several
accretion-powered pulsars (see, e.g., Angelini,
Stella, \& Parmar 1989; Lamb 1989; Shibazaki
1989; Shinoda et al.\ 1990; Lamb 1991; Finger,
Wilson, \& Harmon 1996; Ghosh 1996). In these
pulsars only a single, fairly coherent
oscillation has been seen, with properties which
indicate that it is the magnetospheric
beat-frequency QPO; no QPO is observed at what
would be the Keplerian frequency at the coupling
radius.

To summarize this point, the theory of the
magnetospheric beat-frequency QPO mechanism
predicts, and observations of accretion-powered
pulsars confirm, that this mechanism produces
only a {\em single\/} strong, moderately
coherent oscillation, with a frequency equal to
the beat frequency. In order to explain the two
simultaneous kilohertz QPOs, the magnetospheric
beat-frequency mechanism would have to generate
two strong, {\em highly coherent\/} oscillations,
at the beat frequency {\em and\/} at the
Keplerian frequency at the radius where gas in
the disk first strongly couples to the magnetic field.
This is the most important reason why the
magnetospheric beat-frequency mechanism appears
unpromising as an explanation of the two
kilohertz QPOs.

{\em Simultaneous occurrence of two
kilohertz QPOs and the HBO in the Z sources}.
Identification of the 15--60~Hz HBO in the Z
sources as the magnetospheric beat-frequency
oscillation is supported by the fact that
(1)~the magnetic field strengths required are in
agreement with those inferred from the spin-down
rates of the millisecond recycled
rotation-powered pulsars and detailed modeling
of the X-ray spectra of the Z sources (see Alpar
\& Shaham 1985; Lamb et al.\ 1985; Ghosh \& Lamb
1992; and \S~2.2), (2)~the $\sim\,$200--350~Hz
neutron star spin frequencies predicted by the
magnetospheric beat-frequency model (Lamb et
al.\ 1985; Ghosh \& Lamb 1992) are consistent
with the spin rates of the recycled
rotation-powered pulsars (Ghosh \& Lamb 1992)
and the $\sim\,$250--350~Hz spin frequencies
inferred from the difference between the
frequencies of the two kilohertz QPOs seen in
the Z sources (see \S~1), and (3)~only the
fundamental and (sometimes) the first overtone
of the HBO have ever been detected in the Z
sources when they are on the horizontal branch,
as one would expect if the HBO is the
magnetospheric beat-frequency QPO (see van der
Klis 1989). If the HBO {\em is\/} the
magnetospheric beat-frequency oscillation in the
Z sources, then {\em neither\/} of the two
kilohertz QPOs can be the magnetospheric
beat-frequency oscillation, because the HBO was
always present when the kilohertz QPOs were
detected (van der Klis et al.\ 1996e).

{\em Steep increase of kilohertz QPO
frequencies with increasing mass accretion rate}.
In at least some sources, the frequencies of both
kilohertz QPOs increase more steeply with
inferred mass flux through the inner disk than is
expected in the magnetospheric beat-frequency
model, which predicts $\nu_{\rm K0} \propto 
\mdoti^{0.4}$ for a GPD disk and $\nu_{\rm K0}
\propto \mdoti^{0.2}$ for an RPD disk (Ghosh \&
Lamb 1992; see Figure~2). It is therefore difficult to explain
why the frequencies of the kilohertz QPOs
increase so steeply with \mdoti\ using this
model.

The magnetospheric beat-frequency model {\it
can\/} explain the steep increase of the HBO
frequency with \mdoti\ observed in the Z
sources, because these neutron stars are likely
to have been spun up to frequencies comparable
to the Keplerian frequency at the gas-magnetic
field coupling radius, so the HBO frequency in a
Z source is expected to be small compared to
either the star's spin frequency or the
Keplerian frequency at the gas coupling radius
(see Lamb et al.\ 1985; Shibazaki \& Lamb 1987).
This expectation is consistent with the
$\sim\,$250--350~Hz spin frequencies inferred
from the differences between the frequencies of
the QPOs in the kilohertz QPO pairs.

In contrast to the comparatively low frequency of
the HBO, the frequency of the lower-frequency QPO
in the kilohertz QPO pairs (which is the beat
frequency in the magnetospheric beat-frequency
interpretation) is up to $\sim\,$0.7 times the
frequency of the higher-frequency QPO (which in
this interpretation is
the Keplerian frequency where the gas in the disk couples to
the magnetic field). It follows that in this
interpretation the neutron star spin frequency is
in some cases only $\sim0.3$ times the Keplerian
frequency at the coupling radius. Hence the
frequencies of both of the two kilohertz QPOs
should increase as $\mdoti$ increases in the same
way that $\nu_{K0}$ increases, which is much more
slowly than observed, as Figure~2 shows.

To summarize, whereas the inferred dependence of
the HBO on the mass flux through the inner disk
is explained naturally by the magnetospheric
beat-frequency model of the HBOs, the inferred
dependence of the kilohertz QPOs on this mass
flux is difficult to explain using this model.

{\em Observed range of kilohertz QPO
frequencies}. The frequencies of the QPOs at what
would be the Keplerian orbital frequency in the
magnetospheric beat-fre\-quency interpretation
all range up to $\sim\,$1000--1200~\Hz. This is
understandable if these QPOs are generated
within a few kilometers of the surface of the
neutron star, as in the sonic-point model,
whereas in the magnetospheric beat-frequency
model the frequencies of these QPOs would be
expected {\em a priori\/} to range from
$\sim\,$50~Hz up $\sim\,$1500~Hz for the wide
range of neutron star magnetic field strengths
and accretion rates inferred in the atoll and Z
sources.

The expected range of orbital frequencies could
be reduced if the mass accretion rate and the
strength of the star's dipole magnetic field are
tightly correlated in just the right way.
However, the correlation that is indicated by
spectral modeling (Psaltis et al.\ 1995; Psaltis
\& Lamb 1998a, 1998b, 1998c) gives magnetospheric beat
frequencies in the range $\sim\,$20--200~Hz and
orbital frequencies in the range
$\sim\,$300--500~Hz, much smaller than the
frequencies of the kilohertz QPOs.

{\em Anticorrelation between kilohertz QPO
amplitudes and inferred magnetic field
strengths}. In the magnetospheric beat-frequency
interpretation, the\break
 kilohertz QPOs should have higher amplitudes in
systems with stronger neutron star magnetic
fields. However, as discussed further in \S~4.4,
the rms amplitudes of the kilohertz QPOs in the
atoll sources, which are thought to have the
weakest neutron star magnetic fields, are
considerably stronger than the rms amplitudes of
the kilohertz QPOs in the Z sources, which are 
thought to have the strongest neutron star
magnetic fields.

In the next section we describe a model that is
consistent with both the physical picture of the
atoll and Z sources developed over the past
decade and with the properties of the kilohertz
QPOs.

\section{THE SONIC-POINT MODEL}

We now analyze the physics of the sonic-point
model of the kilohertz QPOs. In \S~3.1 we
describe the key elements of the sonic-point
model and summarize the results of the general
relativistic gas dynamical and radiation
transport calculations that are described in
more detail later in this section. In \S~3.2 we
investigate the motion of the gas in the inner
disks of the Z and atoll sources and show that
at a characteristic angular momentum loss radius
$R_{\rm aml}$ that is typically several
kilometers larger than the radius of the neutron
star, radiation drag extracts angular momentum
from the gas so quickly that the gas accelerates
sharply inward. As a result, density
fluctuations in the gas near the sonic point
create brighter footprints that rotate around
the stellar equator at the sonic point Keplerian
frequency $\nu_{Ks}$, producing a QPO at this
frequency. We show further that a decrease in
$R_{\rm aml}$ with accretion rate is to be
expected.

In \S~3.3 we show that the sonic-point Keplerian
frequency is expected to be between
$\sim\,$300~Hz and $\sim\,$1200~Hz in all
sources and should increase with increasing mass
flux through the inner disk. The model therefore
explains one of the most important features of
the kilohertz QPOs. We also show that the
sonic-point mechanism naturally produces a
second QPO at the beat frequency between
$\nu_{Ks}$ and the stellar spin frequency
$\nu_{\rm spin}$. In \S~3.4 we show that this
mechanism can produce QPOs with $\nu/\delta\nu
\sim 100$, comparable to the largest Q values
observed. In \S~3.5 we consider radiation
transport in the footprints and surrounding hot
central corona and show that the amplitude of
the QPO at the sonic-point Keplerian frequency
is expected to increase steeply with increasing
photon energy. Also, the Keplerian-frequency
oscillations at higher photon energies are
expected to be delayed by $\sim10\,\mu$sec
relative to the oscillations at lower energies,
to the extent that the effects of Comptonization
dominate.
In \S~3.6 we consider the effect of gas
surrounding the neutron star and magnetosphere
on the amplitudes of the kilohertz QPOs and
demonstrate that QPO amplitudes as large as
$\sim 20$\% are understandable in the
sonic-point model.

The light travel time across a neutron star is a
fraction $\lta 10^{-15}$ of the time required for
accretion to change the star's mass and spin
rate, so the exterior spacetime is stationary to
extremely high accuracy. The spacetime outside a
steadily and uniformly rotating axisymmetric
star with gravitational mass $M$ and angular
momentum $J$ is unique to first order in the
dimensionless angular momentum parameter $j
\equiv J/M^2$ (Hartle \& Thorne 1968) and is the
same as the Kerr spacetime to this order (the
spacetime outside a rotating star differs from
the Kerr spacetime in second and higher orders;
see, e.g., Cook, Shapiro, \& Teukolsky 1994).
All, or almost all, of the atoll and Z sources
appear to have spin frequencies \lta 350~Hz (see
\S~2 and \S~4), in which case $j$ is \hbox{$\lta
0.3$} for these stars (see \S~5.2). Hence the
Kerr spacetime is a reasonably accurate
approximation to the exterior spacetime.
Therefore, in this work we use the familiar
Boyer-Lindquist coordinates.

{\em All analytical expressions given in this
paper are accurate only to first order in $j$}.
These expressions would therefore be the same if
written in terms of the circumferential radius.
Calculation of physical quantities to higher
order is straightforward, given a stellar model
and interior and exterior metrics valid to the
higher order, but no analytical expressions are
known, so such calculations must be carried out
numerically. For conciseness, in this section
(only) we use units in which $c=G=k_B=1$, where
$c$ is the speed of light, $G$ is the
gravitational constant, and $k_B$ is Boltzmann's
constant.

\subsection{Physical Picture and Summary of
Calculations}

Before presenting our calculations of the most
important elements of the sonic-point model, we
first describe the model in more detail,
summarizing the results of our calculations and
indicating where these results are presented in
the subsections that follow. We shall assume that
the spin axes of the neutron stars in the
kilohertz QPO sources are closely aligned with
the rotation axes of their accretion disks. This
is expected to be the case in these LMXBs,
because mass transfer is expected to produce an
accretion disk in which the gas circulates in
the same sense as the orbital motion of the
system. The torque on the neutron star created
by such a disk will align the spin axis of the
star with the axis of the accretion disk in a
time short compared to the duration of the mass
transfer phase in such systems (see Daumerie et
al.\ 1996).

{\em Accretion flow near the star}.---As
discussed in \S~2.2, if the stellar magnetic
field is of intermediate strength we expect a
Keplerian disk flow to penetrate inside the magnetosphere.
What happens to the flow there depends on the
magnitude of the radiation drag force (Miller \&
Lamb 1993, 1996) and on whether the radius $R$
of the neutron star is larger or smaller than
the radius $R_{\rm ms}$ of the innermost stable
circular orbit. There are four possibilities (see
Table~\ref{table:FlowCases} for a summary).

Suppose first that $R < R_{\rm ms}$. This will be
the case if the equation of state of neutron-star
matter is relatively soft and the star in
question has an intermediate-to-high mass and is
not spinning near its maximum rate. There are
then two possibilities (\S~3.2):

 \noindent (1a)~If the drag force exerted by
radiation coming from near the neutron star's
surface is strong enough to remove $\gta 1$\% of
the angular momentum of the gas in the Keplerian
disk flow just outside $R_{\rm ms}$, gas in the
inner disk drifts slowly inward in nearly
circular orbits until it reaches the critical
angular momentum loss radius $R_{\rm aml}$,
where it transfers so much angular momentum to
the radiation so rapidly that centrifugal
support fails and the radial velocity of the gas
increases sharply. Inside $R_{\rm aml}$ the gas
spirals inward in a disk flow in which the
radial velocity is supersonic until the gas is
channeled by the star's magnetic field or
collides with the star's surface.

 \noindent (1b)~If instead the radiation drag is
so weak that it removes $\ll 1$\% of the angular
momentum of the gas in the Keplerian disk flow by
the time the gas reaches $R_{\rm ms}$, gas in the
inner disk drifts slowly inward in nearly
circular orbits until it approaches $R_{\rm
ms}$, where general relativistic corrections to
Newtonian gravity cause the radial velocity of
the flow to increase steeply (see, e.g.,
Muchotrzeb 1983; Muchotrzeb-Czerny 1986). Inside
$R_{\rm ms}$, the gas spirals inward in a disk
flow in which the radial velocity is supersonic
until the gas is channeled by the star's
magnetic field or collides with the star's
surface.

 \noindent Whether case~1a or 1b applies depends
primarily on the strength of the star's magnetic
field and on the mass accretion rate.

Suppose now that $R > R_{\rm ms}$. This will be
the case if the equation of state of neutron-star
matter is relatively stiff and the star in
question has an intermediate-to-low mass or is
spinning very rapidly. There are again two
possibilities (\S~3.2): 

 \noindent (2a)~If the radiation drag force is at
least moderately strong, gas in the inner disk
drifts slowly inward until it reaches the
critical radius $R_{\rm aml}$, where it
accelerates rapidly inward. Inside $R_{\rm aml}$
the gas spirals inward
 in a disk flow in which the radial velocity is
supersonic until the gas is channeled by the
star's magnetic field or collides with the star's
surface. 

 \noindent (2b)~If instead the radiation drag is
weak, gas in the inner disk drifts slowly inward
in a disk flow in which the radial velocity is
supersonic until the gas is channeled by the
star's magnetic field or interacts directly with
the star's surface.

 \noindent Again, whether case~2a or 2b applies
depends primarily on the strength of the star's
magnetic field and on the mass accretion rate.

Regardless of whether the slow inward drift of
the gas in the disk is terminated by loss of
angular momentum to the radiation field or
because the gas crosses the radius of the
innermost stable circular orbit, the inward
radial velocity of the gas increases abruptly at
$R_{\rm aml}$ or $R_{\rm ms}$ and becomes
supersonic within a very small radial distance.
As mentioned in \S~1 and earlier in this section
and described in detail below, the radius where
either radiation drag or general relativistic
corrections to Newtonian gravity cause the
radial velocity to increase sharply and become
supersonic plays a key role in the model of the
kilohertz QPOs proposed here. The radius of the
sonic point is a useful indicator of the
location of this transition. 

Radiation drag can remove at most only a fraction
of the specific angular momentum of the gas and
therefore can create a radially supersonic
inflow in the inner disk only within a few
stellar radii (\S~3.2). The reason is that the
specific angular momentum of gas in circular
Keplerian orbits at radii much larger than the
radius of the neutron star is much greater than
in orbits just outside the stellar surface, so
removal of a small fraction of the angular
momentum of the gas at large radii cannot cause
a large fraction of it to fall to the stellar
surface. In contrast, the specific angular
momentum of gas orbiting near the star is not
much greater than the angular momentum of gas
orbiting at the stellar surface, so radiation
drag can cause gas in orbit near the star to
plunge to its surface.

Radiation drag can cause gas to plunge inward
from a radius larger than one would estimate
using the Newtonian approximation, because
(1)~special and general relativistic effects on
the gas dynamics and radiation transport
significantly increase the fraction of the
angular momentum of the accreting gas that can
be removed by the radiation and (2)~in general
relativity the specific angular momentum of gas
in circular Keplerian orbits with radii \hbox{$r
\lta 3\,R_{\rm ms}$} varies only slowly with
$r$. For this reason, the sonic-point model
predicts that if radiation drag produces a
transition to rapid radial inflow, the
transition will occur within \hbox{$\sim
36~\km$} (for a $1.4\,\msun$ neutron star) and
hence that the Keplerian frequency at the
transition radius will be \hbox{$\gta 300\,\Hz$}
(\S~3.3).

Inside the sonic point, the vertical optical
depth of the disk flow falls steeply with
decreasing radius, usually to a value small
compared to unity. The optical depth of the disk
flow measured radially from the stellar surface
to the sonic radius typically also becomes less
than unity within a very small radial distance,
unless the accretion rate is very high or the
geometrical thickness of the disk in the
vertical direction is very small. When the
change from ``optically thick'' to ``optically
thin'' disk flow is caused by the radiation drag
force, the transition is somewhat analogous to
the ionization front at the boundary of an
H{\ts\scriptsize II} region. In the disk flow
transition, the photon mean free path becomes
longer because the radiation is removing angular
momentum and the flow is accelerating inward,
causing the density to fall sharply, whereas in
an ionization front the radiation is removing
bound electrons from atoms and molecules,
causing the opacity to fall sharply.

{\em Generation of the QPO at the sonic-point
Keplerian frequency}.---Suppose first that the
magnetic field of the neutron star is dynamically
negligible. An element of gas in the disk flow
outside the sonic radius diffuses inward
subsonically. When it reaches the sonic radius,
it falls supersonically to the stellar surface
along a spiral trajectory like that shown in
Figure~\ref{fig:Spirals}a. This trajectory was
computed in full general relativity for a
nonrotating, isotropically emitting star, using
the numerical algorithm described in Miller \&
Lamb (1996). (The shape of the trajectory
depends on the luminosity, spin, and other
properties of the source; the example shown in
Fig.~\ref{fig:Spirals}a is only illustrative.)
The surface density of the disk flow is much
smaller inside the sonic radius than outside
because of the sharp increase in the radial
velocity at the sonic point. Gas falls inward
from the sonic radius and impacts the star all
around its equator, producing a bright
equatorial ring of X-ray emission.

The flow in the inner part of the accretion disk
is expected to have density fluctuations
(``clumps'') produced by a variety of
mechanisms, such as thermal instability,
Kelvin-Helmholtz instability, and
magnetoturbulence (see Lamb et al.\ 1985;
Shibazaki \& Lamb 1987). There may be as many as
several hundred such clumps at a given radius in
the disk outside the sonic point. The
velocity-independent radially outward component
of the radiation force and the
velocity-dependent radiation drag force both
tend to cause gas in the shadow of a clump to
overtake the clump in azimuth and radius, and
hence the radiation force tends to increase
clumping.

As discussed in \S~3.4, a clump that forms
outside the sonic radius is dissipated by gas
pressure forces, turbulence, and the shear in
the velocity field before the gas from it can
reach the stellar surface, so such clumps do not
produce significant inhomogeneities in the
inflow from the sonic radius. In contrast, a
clump that forms near the sonic radius can
persist for up to $\sim 100$ times the infall
time to the stellar surface, so such a clump
generates a stream of denser infalling gas from
the sonic point to the stellar surface that
lasts as long as the clump survives.

The shape of the pattern formed by the denser gas
falling inward from a clump depends on how the
angular velocity of the infalling gas varies with
radius. For example, if the angular velocity of
the gas were independent of radius, the pattern
of higher gas density would be a straight,
radial line from the clump to the stellar
surface. In general, the pattern formed by the
denser gas from a clump is a fairly open curve.
The shape of the pattern, like the trajectories
that produce it, depends on the luminosity,
spin, and other properties of the source.
Figure~\ref{fig:Spirals}b shows the pattern
formed by inflow of gas along spiral
trajectories with the shape shown in
Figure~\ref{fig:Spirals}a. In this example the
pattern of denser gas is also a spiral. It is
more open than the spiral trajectories of the
individual elements of denser gas that produce
it, because the source of the denser gas (the
clump) orbits the star only slightly slower than
the infalling gas.

We expect the time-averaged radiation field
around the star to be nearly axisymmetric.
Moreover, the gas from a given clump typically
orbits the star $\sim 5 \dash 10$ times before
colliding with the stellar surface, so the
effects of any azimuthal variations in the
radiation drag force are averaged out.
Therefore, the spiral pattern of denser gas
produced by the inflow from a given clump
rotates nearly uniformly around the star with a
rotation frequency equal to the orbital
frequency of the clump that is producing it,
which is the Keplerian frequency at a radius
near the sonic point.

The time evolution that generates an X-ray
oscillation at the sonic-point Keplerian
frequency is illustrated by the four panels of
Figure~\ref{fig:KeplerQPO}. In this figure a
single clump is shown advancing in its orbit by
90\degree\ from one panel to the next. The
pattern of higher density gas rotates uniformly
around the star with a frequency equal to the
orbital frequency of the clump, so it also
advances by 90\degree\ from one panel to the
next. Where the denser gas from a clump collides
with the stellar surface, it produces an
arc-shaped area of brighter X-ray emission. This
arc-shaped brighter ``footprint'' moves around
the star's equator with a frequency equal to the
rotation frequency of the pattern, which is the
orbital frequency of the clump and is therefore
approximately equal to the Keplerian frequency
$\nu_{\rm Ks}$ at the sonic point. In reality,
many clumps are crossing the sonic radius at any
given time, and hence there are many bright
footprints moving around the star's equator. The
radiation from these footprints carries the
kinetic energy of infall that is released when
the gas that has fallen inward from the clumps
at the sonic radius collides with the stellar
surface.

As seen by a distant observer whose line of sight
is inclined with respect to the orbital axis of
the disk, the aspect presented by a given bright
footprint varies periodically and the footprint
is eclipsed with a frequency equal to the
rotation frequency of the pattern, which is
approximately the Keplerian orbital frequency at
the sonic point. Footprints come and go as
clumps form near the sonic point and then
dissipate. Also, the orbital frequencies of the
clumps that are producing footprints at any
given time differ slightly. As a result, a
distant observer sees a strong, {\em
quasi-periodic\/} oscillation of the X-ray flux
and spectrum at the pattern rotation frequency,
which is close to $\nu_{\rm Ks}$. {\em This is
the the sonic-point Keplerian-frequency QPO}.

The spin of the neutron star is not involved in
generating the sonic-point Keplerian frequency
QPO. The precise frequency of the sonic-point
Keplerian QPO does depend weakly on the spin
frequency of the neutron star, because the
frame-dragging caused by the star's spin has a
small effect on the frequencies of Keplerian
orbits near the neutron star. The sonic-point
Keplerian frequency QPO mechanism described
here, which generates a QPO with a frequency
equal to the orbital frequency at the sonic
point, differs fundamentally from the suggestion
of Paczynski (1987), who speculated that when the
sonic point is located at the radius of the
marginally stable orbit, a QPO might be produced
by unsteady flow through it with a low ($\sim 20
\dash 50\,\Hz$) frequency that would have
nothing directly to do with the orbital
frequency of the marginally stable orbit.

{\em Generation of the QPO at the sonic-point
beat frequency}.---Suppose now that the magnetic
field of the neutron star is weak enough that a
prograde Keplerian disk flow penetrates near the
surface of the star but strong enough that close
to the star some of the gas is channeled by the
field. The channeled gas produces slightly
brighter spots where it collides with the stellar
surface. If these slightly brighter spots are
offset from the star's spin axis, the enhanced
radiation from them generates a beamed pattern of
radiation that rotates with the star. This
rotating radiation pattern creates a periodic
oscillation in the radiation force that is acting
on the clumps of gas crossing the sonic radius.
Hence, if conditions near the star are such that
the nearly Keplerian motion of the gas in the
clumps orbiting near the sonic radius is
terminated by loss of angular momentum to the
radiation,
the inward flux of mass from the clumps near the
sonic radius will increase once or twice each
beat period, depending on the symmetry of the
radiation pattern that rotates with the star.

As a specific example, suppose that the radiation
drag force experienced by the gas in a given
clump orbiting near the sonic radius peaks each
time the beam of radiation that rotates with the
star sweeps across the clump. This will occur
with a frequency equal to one or two times the difference between
the orbital frequency of the clump, which is the
Keplerian orbital frequency near the sonic point,
and the spin
frequency $\nu_{\rm spin}$ of the neutron star.
{\em We call $\nu_{\rm Ks}-\nu_{\rm spin}$
the sonic-point beat frequency and denote it
$\nu_{\rm Bs}$}. If the transition to supersonic
inflow is caused by radiation drag, the
oscillation of the radiation drag force will
cause the supersonic inflow of gas from the
clumps orbiting near the sonic point to oscillate
quasi-periodically with a frequency equal to one
or two times the sonic-point beat frequency.

The time evolution that produces this oscillation
is shown schematically in the six snapshots of
Figure~\ref{fig:BeatQPO}. For clarity, only a
single clump is shown orbiting at the sonic
radius and the six snapshots show the sequence
of events as seen in a frame corotating with the
star.
  In panel~(a) the clump is in the beam of
radiation that rotates with the star and hence
the inward flux of denser gas from the clump is
greater than average.
  In panel~(b) the clump has moved out of the
beam and hence the inward flux of gas from the
clump is smaller than average; the pulse of gas
that was dragged off the clump by the stronger
radiation flux in frame~(a) is now falling
inward from the clump.
  In panel~(c) the clump has moved ahead of the
beam by 180\degree; the inward flux of gas from
the clump remains smaller than average and the
pulse of denser gas that began to fall inward in
frame~(a) is further away from the clump.
  In panel~(d) the clump is now ahead of the beam
by 270\degree; the inward flux of gas from the
clump remains smaller than average and the pulse
of denser gas that began to fall inward in
frame~(a) is approaching the stellar surface.
  In panel~(e) the clump has now moved back into
the beam of radiation and a new pulse of denser
gas is leaving the clump; the pulse of denser gas
that began to fall inward in frame~(a) is now
very close to the stellar surface.
  In panel~(f) the clump has again moved ahead of
the beam by 90\degree\ and the inward flux of gas
from the clump is again smaller than average; the
pulse of denser gas that began to fall inward in
frame~(a) is now colliding with the stellar
surface, producing a brighter beam of radiation
from the footprint of the stream.

In reality, the radiation drag force, and hence
the inward mass flux from a clump, may be
greatest not when the clump is fully illuminated
by the beam that rotates with the star, as
assumed in this illustration, but at some other
relative phase, because the radiation drag force
depends in a complicated way on the various
components of the radiation stress-energy tensor
(see Miller \& Lamb 1996). In any case, the
inward mass flux from the orbiting clumps
oscillates quasi-periodically with a frequency
approximately equal to the sonic-point beat frequency $\nu_{\rm
Bs}$, producing a quasi-periodic oscillation in
the luminosity and spectrum of the X-ray emission
from the stellar surface. {\em This is the
sonic-point beat-frequency QPO}.

{\em Properties of the sonic-point QPOs}.---In
\S~3.3 we show that the sonic-point mechanism
naturally generates Keplerian-frequency QPOs with
frequencies $\sim 300 \dash 1200$~Hz for a wide
range of stellar magnetic fields and accretion
rates and that it is natural for the frequencies
of the sonic-point Keplerian and beat-frequency
QPOs to increase steeply with increasing mass
flux through the inner disk. The angular
distribution of the emission from each emitting
point within a footprint at the stellar surface
is broad, so the power at $\nu_{\rm Ks}$ is
likely to be much greater than the power at
overtones of $\nu_{\rm Ks}$. The modulation of
the inflow from clumps orbiting at the sonic
radius is unlikely to be perfectly sinusoidal,
so overtones of the sonic-point beat frequency
$\nu_{\rm Bs}$ as well as $\nu_{\rm Bs}$ itself
may be detectable.

In \S~3.4 we argue that the sonic-point mechanism
can produce QPOs with $\nu/\delta\nu$ ratios as
large as $\sim 100$, i.e., similar to those
observed in the atoll and Z sources. The
oscillation at the sonic-point beat frequency is
generated by the beating of the stellar spin
frequency, which is periodic, against the
quasi-periodic sonic-point Keplerian frequency
oscillation. It is therefore natural to expect
the $\nu/\delta\nu$ values of the oscillations
in a pair to be roughly similar, in the absence
of other disturbing effects. In \S~3.5 we show
that the steep increase in QPO amplitude with
photon energy observed in many kilohertz QPOs
is understandable in the sonic-point model.

The visibilities of the various QPOs generated by
the sonic-point mechanism depend on the mass flux
in the Keplerian disk flow that penetrates to the
sonic point, the number and distribution of
clumps in the disk at the sonic point, the
brightness of the footprints produced by the
streams from the clumps relative to the
brightness of the rest of the stellar surface,
the geometry and optical depth of the scattering
material around the star, and the inclination of
the system.

If radiation forces are required to produce large
clumps with substantial density contrasts, the
amplitudes of the QPOs at the sonic-point
Keplerian and beat frequencies are likely to
become much smaller and may become undetectable
once the angular momentum loss radius $R_{\rm
aml}$ has retreated inward to the radius $R_{\rm
ms}$ of the innermost stable circular orbit,
because once this has happened, the gas in the
disk near the sonic point is increasingly
shielded from radiation coming from the stellar
surface by gas in the disk closer to the star.
However, if large clumps with substantial density
contrasts are formed by other mechanisms, the
sonic-point Keplerian frequency QPO may still be
detectable if the sonic point is at $R_{\rm ms}$,
even if the sonic-point beat frequency
QPO---which can only be generated if radiation
forces have a significant dynamical effect at the
sonic radius---has disappeared.

In \S~3.6 we show that scattering by the
electrons in the central corona that surrounds
the neutron stars in the Z and atoll sources
strongly attenuates the already intrinsically
weak beaming oscillation at the stellar spin
frequency as well as the weak beaming oscillations at
other frequencies, making all but the strongest
beaming oscillation---the one at the sonic-point
Keplerian frequency---difficult to detect with
current instruments. In contrast, scattering
only weakly attenuates luminosity oscillations,
so the luminosity oscillations at the
sonic-point beat frequency and its overtones are
likely to be detectable. Taking into account
both the generation of these various
oscillations and their attenuation in the
circumstellar environment, Keplerian-frequency
and beat-frequency QPOs with rms amplitudes as
large as 15\% appear possible.

As discussed in \S~2, the Z sources appear to
have significantly stronger magnetic fields than
the atoll sources. The magnetic fields of the Z
sources are therefore likely to channel a larger
fraction of the accreting gas out of the disk
plane before it penetrates close to the stellar
surface, so we expect the amplitudes of the kilohertz QPOs
to be smaller in the Z sources than in the atoll
sources. The magnetic field of the neutron star
in some atoll sources may be so weak that the
QPO at the sonic-point beat frequency is
undetectable, even though a QPO at the
sonic-point Keplerian frequency is visible. The
sonic-point beat frequency may also be
undetectable at some times if the scattering
optical depth between the stellar surface and
the sonic point is too large. In other sources
or at other times, the luminosity oscillation at
the sonic-point beat frequency may be detectable
even though the beaming oscillation at the
sonic-point Keplerian frequency has been
suppressed by scattering in the corona
surrounding the star.

\subsection{Transition to Rapid Radial Inflow}

In both the atoll and the Z sources, we expect
radiation drag to cause gas in the Keplerian disk
flow to make an abrupt transition from slow
inward drift to rapid radial inflow several
kilometers above the surface of the neutron
star. This transition occurs at the radius where
the drag exerted by radiation from the star
removes enough angular momentum from the gas
quickly enough that it falls to the surface of
the star unimpeded by a significant centrifugal
barrier. We first describe the consequences of
the transfer of angular momentum from the gas in
the inner disk to the radiation and then present
a fully general relativistic calculation of the
gas dynamics and radiation transport in the
inner disk that shows the nature of this
transition in the atoll sources, which have
luminosities much less than the Eddington
critical luminosity. We next give approximate
analytical expressions for the location and
width of this transition, then discuss why the
nature and location of the transition may be
similar in the Z sources even though their
accretion rates and luminosities are much higher
than the accretion rates and luminosities of the
atoll sources, and finally summarize the
implications for the sonic-point QPOs.

{\em Transfer of angular momentum to the
radiation field}.---The azimuthal drag force
exerted by radiation from the star can create a
radially supersonic inflow in the inner disk.
However, this is possible only within a few
stellar radii if the luminosity of the star is
produced by accretion and most of the accreting
gas flows through the inner disk. The reason is
that radiation drag can remove at most a
fraction \hbox{$\eta_{\rm rad}({\rm max})
\approx$20\%} of the specific angular momentum
of the accreting gas (see Fortner et al.\ 1989;
for more detailed discussions, see Miller \&
Lamb 1993, 1996). The specific angular momentum
of gas in circular Keplerian orbits at radii
large compared to the radius $R$ of the neutron
star is much larger than at $R$, so radiation
drag cannot cause more than $\sim$20--30\% of the
gas at such large radii to fall to the stellar
surface. In contrast, the specific angular
momentum of gas in circular Keplerian orbits
very near the star is not much larger than the
angular momentum of orbits at the stellar
surface, so radiation drag can cause gas in
Keplerian orbits near the star to plunge to the
surface.

Radiation drag can create a radially supersonic
inflow at a larger radius than one would estimate
using the Newtonian approximation, because
special and general relativistic effects
significantly increase the fraction of the
angular momentum of the accreting gas that is
transferred to the radiation (Miller \& Lamb
1993). Also, in general relativity the specific
angular momentum of gas in circular Keplerian
orbits with radii \hbox{$r \lta 3\,R_{\rm ms}$}
varies only slowly with radius. The latter point
is illustrated by
Figure~\ref{fig:AngMomVsRadius}, which shows the
specific angular momentum of gas in circular
Keplerian orbit around a nonrotating and a
slowly rotating $1.4\,\msun$ neutron star, as a
function of the radius of the orbit. The
specific angular momentum varies fairly steeply
at \hbox{$r \gg R_{\rm ms}$}, but only slowly
with radius between \hbox{$r=R_{\rm ms}$} and
\hbox{$r \sim 3\,R_{\rm ms}$}. For example, the
specific angular momentum of circular orbits
changes by less than $\sim10$\% from
\hbox{$r=12M$} (25~km) to \hbox{$r=6M$} (12~km).
These curves are of course physically meaningful
only at radii greater than the radius $R$ of the
neutron star. The relatively flat shape of the
angular momentum curve for circular orbits near
a neutron star means that if gas orbiting there
loses even $\sim 20$\% of its angular momentum,
it will plunge to the stellar surface.

We can estimate the outer boundary of the region
where radiation coming from near the stellar
surface can remove sufficient angular momentum
from the gas orbiting in the disk so that it
falls supersonically to the stellar surface by
comparing $\eta_{\rm rad}({\rm max})$, the
largest fraction of angular momentum that can be
removed by radiation coming from the surface of
the neutron star, with $\eta_{\rm flow}$, the
fraction of the specific angular momentum of gas
in a circular Keplerian orbit at radius $r$ that
must be removed in order for the gas to fall
from $r$ to $R_{\rm ms}$ (if the radius of the
star is less than $R_{\rm ms}$, gas that has
reached $R_{\rm ms}$ can fall to the stellar
surface without losing any more angular
momentum). These two quantities are compared in
Figure~\ref{fig:AngMomFractionVsRadius}. Here
$\eta_{\rm rad}({\rm max})$ was computed by
assuming that the gravitational binding energy at
the stellar surface is converted into radiation
there, taking into account the relevant angular,
special relativistic, and general relativistic
factors (see Miller \& Lamb~1996), and assuming
that all of the radiation emitted in the
direction of the disk interacts with the disk
flow ($f=1$ in the notation of Miller \&
Lamb~[1993]).
Figure~\ref{fig:AngMomFractionVsRadius} shows
that $\eta_{\rm rad}({\rm max})$ exceeds
$\eta_{\rm flow}$ only inside \hbox{$\sim 17 M$}
(\hbox{$\sim 36\,\km$}). This characteristic
radius depends on the radius of the neutron
star, because the luminosity is proportional to
the gravitational binding energy at the stellar
surface, but this dependence is relatively weak
(the characteristic radius decreases from $21M$
to $13M$ as the stellar radius increases from
$4M$ to $7M$). Hence, {\em the sonic-point model
predicts that if radiation drag produces a
transition to rapid radial inflow, the
transition will occur within \hbox{$\sim 15\,M$}
and hence that the Keplerian frequency at the
transition radius will be \hbox{$\gta
300\,\Hz$}} (see \S~3.3). Determining the actual
radius of the transition requires gas dynamical
and radiation transport calculations.

{\em Model calculations}.---The nature of the
transition to supersonic radial inflow is
illustrated by the following fully general
relativistic calculations of the gas dynamics and
radiation transport in the innermost part of the
accretion disk flow. In these calculations the
azimuthal velocity of the gas in the disk is
assumed to be nearly Keplerian far from the star.
Internal shear stress is assumed to create a
constant inward radial velocity $v^{\hat r}$ in
the disk, as measured in the local static frame,
of $10^{-5}$ (here and below we take $v^{\hat r}$
to be positive for radially inward flow). The
half-height $h(r)$ of the disk flow at radius $r$
is assumed to be $\epsilon r$ at all radii, where
$\epsilon$ is a constant and $r$ is the radius,
and the kinetic energy of the gas that collides
with the surface of the star is assumed to be
converted to radiation that emerges from a band
around the star's equator with a half-height
equal to $\epsilon R$. For simplicity, and to
show the effects of radiation forces more
clearly, any effect of the stellar magnetic
field on the dynamics of the disk flow near the
sonic transition is neglected in this model
calculation.

The surface density \hbox{$\Sigma_{\rm co} \equiv
2h\rho_{\rm co}$} of the disk flow as measured
by a local observer comoving with the gas is
related to the surface density $\Sigma_{\rm
stat}$ of the disk flow as measured by a local
static observer by a Lorentz boost, so
 \begin{equation}
  \Sigma_{\rm co} = \gamma^{-1}\Sigma_{\rm stat}
  \;,
 \label{SigmaCo}
 \end{equation}
 where $\gamma \equiv [1-(v^{\hat r})^2 -
(v^{\hat\phi})^2]^{-1/2}$ and we have neglected
$v^{\hat z}$. For stationary disk accretion,
$\Sigma_{\rm stat}$ at radius $r$ is given by
(see Novikov \& Thorne 1973, eq.~5.6.3)
 \begin{equation}
  \Sigma_{\rm stat} =
  (\mdot_{\infty}/2\pi r v^{\hat r})
  (1-2M/r)^{-1/2}
  \;,
 \label{SigmaStatic}
 \end{equation}
 where $\mdot_{\infty}$ is the mass accretion
rate through the disk, as measured at infinity. 

Close to the star, the photon mean-free-path is
limited primarily by Thomson scattering and is
therefore $1/n_e\sigma_T$, where $n_e$ is the
electron number density in the disk flow as
measured in the frame comoving with the flow and
$\sigma_T=6.65\ee{-25}$~cm$^2$ is the Thomson
scattering cross section. Hence, for a steady
flow, the radial optical depth from the stellar
surface at radius $R$ through the disk to radius
$r$ is
 \begin{eqnarray}
  \tau_{\rm radial}(r) & \equiv &
  \sigma_T \int_R^r n_e(r^p)\,dr^p =\nonumber\\
   & & \hskip -0.8 truein
  \left(\frac{\sigma_T\mdot_{i,\infty}}
             {4\pi m_p}\right)
  \int_R^r
  \frac{[1-v^{\hat r}(r')]^{-1}}
       {\gamma_r(r')\gamma(r')}\,
  \frac{(1-2M/r')^{-1}}{r'h(r')v^{\hat r}(r')}
  \,dr',\nonumber\\
  \label{eq:RadialOpticalDepth}
 \end{eqnarray}
 where $\mdot_{i,\infty}$ is the mass accretion
rate through the inner disk, as measured at
infinity. The radial coordinate $r^p$ in the
first expression on the right is the proper
radial distance in the frame comoving with the
gas. In the second expression on the right we
have estimated $n_e$ in the disk flow between
$r$ and the stellar surface by assuming for
simplicity that the gas in the disk is fully
ionized hydrogen and using the continuity
equation. Then $n_e = \Sigma_{\rm co}/2h m_p =
\gamma^{-1} (\mdot_{i,\infty}/4\pi m_p v^{\hat
r} rh)(1-2M/r)^{-1/2}$, where $m_p$ is the
proton mass and $h$ is the half-thickness of the
disk flow. The differential proper radial
distance $dr^p$ in the frame comoving with the
gas is related to the differential radial
distance in the local static frame by a Lorentz
boost, so $dr^p = (1-v^{\hat
r})^{-1}{\gamma_r}^{-1} dr_{\rm stat}$, where
$v^{\hat r}$ is the inward radial velocity and
$\gamma_r \equiv [1-(v^{\hat r})^2 ]^{-1/2}$.
The differential radial distance in the local
static frame is in turn related to the
differential radial distance $dr'$ in the global
(Boyer-Lindquist) coordinate system by $dr_{\rm
stat} = (1-2M/r')^{-1/2} dr'$, so $dr^p =
(1-v^{\hat r})^{-1}\, {\gamma_r}^{-1}\,
(1-2M/r')^{-1/2}\, dr'$.

Once the drag force exerted by the radiation from
the stellar surface begins to remove angular
momentum from the gas in the Keplerian disk,
centrifugal support is lost and the gas falls
inward, accelerating rapidly. Radiation that
comes from near the star and is scattered by the
gas in the disk is generally scattered out of the
disk plane and hence does not interact further
with the gas in the disk. Moreover, second and
successive scatterings do not contribute
proportionately to the azimuthal radiation drag
force on the gas because the radiation field is
aberrated by the first scattering and afterward
carries angular momentum (see Miller \& Lamb
1993, 1996).  We therefore treat the interaction
of the radiation with the gas in the disk by
assuming that the intensity of the radiation
coming from the star is attenuated as it passes
through the gas in the disk, diminishing as
$\exp(-\tau_{\rm radial})$, where $\tau_{\rm
radial}(r)$ is given by
equation~(\ref{eq:RadialOpticalDepth}), and that
scattered radiation does not contribute to
removal of angular momentum from the gas. In
calculating the radiation drag force, we assume
for simplicity that the differential scattering
cross section is isotropic in the frame comoving
with the accreting gas (this gives results very
close to those using a Thomson differential cross
section; see Lamb \& Miller
1995). The radiation field and the motion of the
gas are computed in full general relativity.

The optical depth of the disk flow near the star
is generally much less than one would estimate by
calculating it without taking into account the
radiation forces. The reason is that the loss of
centrifugal support caused by transfer of angular
momentum to the radiation causes the inward
radial velocity of the gas in the disk at radii
less than the radius $R_{\rm aml}$ of the
radiation-induced transition to supersonic
inflow to be orders of magnitude higher, and the
density of the gas to be orders of magnitude
lower, than if radiation forces were neglected
(see Miller \& Lamb 1996). Hence the mean free
path in the disk flow is much larger, and the
optical depth from the stellar surface to a
given radius in the flow is much smaller, than
they would be in the absence of radiation drag.
In this way the radiation increases the
transparency of the disk flow, so its effects
are felt much farther out in the flow than one
would estimate from the properties of the
undisturbed flow. 

Figure~\ref{fig:SimpleExamples} shows results
obtained by solving self-consistently for the gas
dynamics and the radiation field in this simple
model, for a disk flow of semi-thickness
$\epsilon \equiv h/r = 0.1$, a nonrotating
neutron star with gravitational mass
\hbox{$M=1.4\,\msun$} and radius \hbox{$R=5M$},
and accretion rates through the inner disk of
0.02, 0.03, 0.04, and $0.05\,\mdote$, where
$\mdote$ is the mass accretion rate that
produces the Eddington critical luminosity at
infinity (see Lamb \& Miller 1995). These
accretion rates are typical of the
lower-luminosity atoll sources. The calculations
were carried out using the numerical algorithm
described in Miller \& Lamb (1996).

The results shown in
Figure~\ref{fig:SimpleExamples} demonstrate that
radiation forces and general relativistic effects
create a sharp transition to supersonic inflow
several kilometers above the stellar surface and
that the transition radius decreases with
increasing mass flux when the transition is
caused by radiation drag. These results are not
intended to represent the inflow in any
particular source. For \hbox{$\mdot_i =
0.02\,\mdote$} and $0.03\mdote$, the transition
to rapid radial inflow occurs at $7.3\,M$ and
$6.3\,M$, respectively, and is caused by
transfer of angular momentum from the gas to the
radiation field. For \hbox{$\mdot_i =
0.04\,\mdote$} and $0.05\,\mdote$, most of the
radiation from the star does not penetrate as
far out as the radius $R_{\rm ms}=6M$ of the
innermost stable circular orbit, so the
transition to rapid radial inflow occurs at
$R_{\rm ms}$ and is caused by the absence at
\hbox{$r < R_{\rm ms}$} of circular orbits with
the specific angular momentum of the orbit at
\hbox{$r = R_{\rm ms}$}.

The radial velocity profiles plotted in
Figure~\ref{fig:SimpleExamples}a show that at the
sonic point the radial velocity typically
increases by about two orders of magnitude in a
very small radial distance \hbox{($\delta r \sim
0.001 \dash 0.01\,r$)}. The angular velocity
profiles plotted in
Figure~\ref{fig:SimpleExamples}b show that the
effect of radiation forces on the angular
velocity is smaller, but is still very
significant. For \hbox{$\mdot_i = 0.04\,\mdote$}
and $0.05\,\mdote$, the angular velocities
outside $6M$ are nearly Keplerian, because the
gas there is shielded from the radiation by the
gas further in. However, for \hbox{$\mdot_i =
0.02\,\mdote$} and $0.03\mdote$, the angular
velocity drops below the Keplerian value well
outside $6M$. It is this departure from
Keplerian orbital motion, and the associated
loss of centrifugal support, that causes the
flow to accelerate radially inward. All the
angular velocity profiles turn downward near the
stellar surface because of the strong azimuthal
radiation drag force there. The decrease in the
kinetic energy and angular momentum of the
inflowing gas is exactly compensated by the
increase in the energy and angular momentum
carried outward by the escaping radiation (see
Miller \& Lamb 1993, 1996).

Figure~\ref{fig:SimpleExamples}c shows that the
vertical column density of all four disk flows
drops abruptly from $\sim10^4~\gpsqcm$ outside
the sonic radius to $\lta\,$10~\gpsqcm inside.
The radial optical depth measured outward from
the stellar surface is shown in
Figure~\ref{fig:SimpleExamples}d; it decreases
by about two orders of magnitude at the sonic
radius, but not as sharply as the vertical
optical depth. Even when the transition to rapid
radial inflow occurs at $R_{\rm ms}$, the flow
between $R_{\rm ms}$ and the stellar surface is
generally strongly affected by radiation forces,
as shown by the differences between the velocity
and optical depth profiles for \hbox{$\mdot_i =
0.04\,\mdote$} and $0.05\,\mdote$. We caution
that these results are only illustrative. For
example, X-rays may be emitted from a larger
fraction of the stellar surface than we have
assumed, in which case $R_{\rm aml}$ will be
larger for a given accretion rate.

{\em Radius of the transition}.---In the model of
gas dynamics and radiation transport just
described, under conditions such that loss of
angular momentum to radiation drag is important
outside $R_{\rm ms}$, loss of centrifugal support
and rapid radial inflow begins about five photon
mean-free-paths from the stellar surface, i.e.,
the transition occurs where \hbox{$\tau_{\rm
radial}(r) \approx 5$}. In this model the inward
radial velocity of the gas in the disk inside
$R_{\rm aml}$ typically rises very sharply to
$\about 0.1\,c$ and then changes more slowly
until the gas collides with the stellar surface
(see also Miller \& Lamb 1993, 1996). Hence we
can estimate $R_{\rm aml}$ by setting
\hbox{$\tau_{\rm radial} = 5$} in
equation~(\ref{eq:RadialOpticalDepth}), scaling
$v^{\hat r}$ and $h/r$ in units of $10^{-2}\,c$ and
$0.1$, respectively, and solving for the
radius. The result is
 \begin{eqnarray}
  R_{\rm aml} &\approx& R  + 5\, 
  \left(\frac{\mdoti}{0.01\mdote}\right)^{\!-1} 
  \left({R\over{10~\km}}\right)\times\nonumber\\
  &&\qquad\qquad\times
  \left({h/R\over{0.1}}\right)
  \left({v^{\hat r}\over{10^{-2} c}}\right)
     \;{\rm km}\;.
  \label{eq:EstimateRaml}
 \end{eqnarray}
 This estimate is in rough agreement with the
transition radius found in our numerical
calculations.

{\em Width of the transition}.---Under conditions
such that the transition to rapid radial inflow
is caused by radiation drag, the width $\delta$
of the radial velocity transition is determined
by $\lambda^{\rm K}_\gamma$, the photon
mean-free-path in the Keplerian disk flow at
$R_{\rm aml}$, because this is the characteristic
distance over which the gas first becomes exposed
to radiation from the star, and by the radial
distance $\delta_{\rm aml}$ over which the gas in
the disk loses its angular momentum once it is
exposed to radiation from the star. From
expressions similar to those used above to
compute the radial optical depth in the
Keplerian disk flow,  we find
 \begin{equation}
  \frac{\lambda^{\rm K}_\gamma}{r} \sim 10^{-4}
  \Bigg(\frac{\mdoti}{0.01\mdote}\Bigg)^{-1}
  \Bigg(\frac{h/R}{0.1}\Bigg)
  \Bigg({v^{\hat r}\over{3\ee{-5}\,c}}\Bigg)\;,
  \label{eq:EstimateDeltaTau1}
 \end{equation}
 where we have scaled the quantities that enter
this expression in terms of their approximate
values in the transition region. This estimate is
consistent with the results of our numerical
calculations. From the expression for the
radiation drag rate given by Miller \& Lamb
(1993, 1996), we find 
 \begin{eqnarray}
 {\delta_{\rm aml}\over {r}} &\sim& 
 3\,c_s^{3/2}
 \left(\frac{\mdot}{\mdote}\right)^{-1/2}
 \nonumber\\
 &\sim&
 10^{-3}
 \left(\frac{c_s^{}}{10^{\!-3}_{}}\right)^{3/2}
\left(\frac{\mdot}{0.01\,\mdote}\right)^{\!-1/2},
 \label{eq:EstimateDeltaTau2}
 \end{eqnarray}
 where we have scaled $c_s$ in terms of the
typical value for a gas-pressure-dominated inner
disk. These estimates show that the radial
velocity is expected to increase over a very
small radial distance $\delta \sim 10^{-3}\,r$--
$10^{-4}\,r$. This expectation is in
accord with the results of our numerical
calculations presented above. As noted in
\S~3.1, this transition is somewhat analogous to
the transition at the boundary of a Str\"omgren
sphere, except that here the radiation removes
angular momentum from the gas rather than
stripping electrons from atoms or molecules and
causes a sharp increase in the inward radial
velocity rather than in the degree of ionization.

If the neutron star radius is smaller than
$R_{\rm ms}$ and conditions are such that the
transition to rapid radial inflow is caused by
general relativistic corrections to Newtonian
gravity, the radial velocity increases rapidly
near $R_{\rm ms}$ because gas inside $R_{\rm
ms}$ can fall to the stellar surface without
losing any angular momentum. The precise
location of the sonic transition is slightly
affected by pressure forces and, if the shear
stress is large, by outward angular momentum
transport within the disk flow (see Muchotrzeb
1983; Muchotrzeb-Czerny 1986). In the absence of
shear stresses and pressure forces, the radial
velocity $v^{\hat r}$ of an element of gas at
radius $r=R_{\rm ms}(1-\xi)$, where \hbox{$\xi
\ll 1$}, that has fallen inward from a circular
orbit at $R_{\rm ms}$ is approximately
$(\xi/2)^{3/2}$ (Miller \& Lamb 1996). Hence,
for $c_s \sim 10^{-3}$, the inflow becomes
supersonic over a radial distance $\delta \sim
\xi r \sim (c_s)^{2/3}r \sim 10^{-2}\,r$.

{\em Summary and implications for kilohertz
QPOs}.---For sources in which the radius $R$ of
the neutron star is larger than the radius
$R_{\rm ms}$ of the innermost stable circular
orbit and accretion is under conditions such
that angular momentum loss to the radiation field
terminates the Keplerian flow at a radius
larger than $R$, our results show that
the Keplerian disk flow near the star is
terminated by angular momentum loss to radiation
and that the radius of the transition to rapid
radial inflow decreases steeply with increasing
mass flux through the inner disk. Hence the
orbital frequency at the sonic radius, and
therefore the frequency of the associated QPO,
increases steeply with increasing mass flux. If
instead conditions are such that radiation forces
do not terminate the Keplerian flow,
the Keplerian flow is probably
terminated by interaction with the stellar
surface, in which case generation of a coherent
QPO with a frequency equal to the orbital
frequency at the stellar surface appears very
unlikely.

For sources in which $R$ is smaller than $R_{\rm
ms}$ and accretion occurs under conditions such
that radiation forces terminate the flow outside
$R_{\rm ms}$, our
results show that the Keplerian flow is
terminated by angular momentum loss to radiation
and that the radius of the transition to rapid
radial inflow decreases steeply with increasing
mass flux. Hence the frequency of the QPO
associated with the orbital frequency at the
sonic radius again increases steeply with
increasing mass flux. If instead conditions are
such that radiation forces do not terminate
the Keplerian flow outside $R_{\rm ms}$, the
Keplerian flow is terminated by general
relativistic corrections to Newtonian gravity
and the radius of the transition to rapid radial
inflow is approximately independent of the mass
flux. Hence, under these conditions the
frequency of any QPO with the orbital frequency
at the sonic point will be approximately
independent of the accretion rate.

Although the numerical results presented here are
for accretion rates typical of the atoll sources,
which have very weak magnetic fields and are
accreting at rates much less than the Eddington
critical rate, we expect the structure of the
accretion flow and the behavior of the
sonic-point Keplerian QPO frequency to be
similar in the Z sources, even though they are
thought to have stronger magnetic fields and are
accreting at much higher rates. There are three
reasons.

(1)~The radius of the transition to rapid radial
inflow that we are considering is necessarily
close to the neutron star whatever the
luminosity, because the fraction of the angular
momentum in the Keplerian disk flow that can be
removed by radiation drag is small and general
relativistic corrections to Newtonian gravity
are important only near the neutron star.
(Although the luminosities of the Z sources are
much higher, the angular momentum flux that must
be removed to create a radial inflow is
correspondingly higher. In sources with
luminosities very close to the Eddington
critical luminosity, radiation drag can
significantly affect the motion of up to
\hbox{$\sim 20$\%} of the accreting gas further
away, but can only affect the motion of most of
the accreting gas close to the star [see Fortner
et al.\ 1989; Lamb 1989].) 

(2)~The inferred magnetic fields of the neutron
stars in the Z sources are stronger than those
of the neutron stars in the atoll sources, and
hence a larger fraction of the disk flow is
likely to be channeled out of the disk by the
stellar magnetic field, causing the mass flux
through the inner disks of the Z sources to be
comparable to the mass fluxes through the inner
disks of the atoll sources. 

(3)~The vertical thickness of the inner disk
flow is likely to be larger in the Z sources
than in the atoll sources.

All of these effects act in the direction of
making the radius of angular momentum loss in
the Z sources similar to its value in the atoll
sources. We do expect significant departures
from the behavior found in the present
calculations when the luminosity equals or
exceeds the Eddington critical luminosity. These
points are discussed in more detail in \S~4.

\subsection{Frequencies of the Sonic-Point QPOs}

We consider now the frequencies that are
generated by the sonic-point mechanism. There
are two fundamental QPO frequencies: (1)~the
Keplerian frequency at or near the sonic radius
and (2)~the sonic-point beat frequency, which is
generated by interaction of radiation from the
neutron star with the accretion flow near the
sonic radius. QPOs may also be detectable at
overtones of the Keplerian and beat frequencies.
QPOs at sideband frequencies and oscillations at
the spin frequency of the neutron star and its
overtones are likely to be very weak.

In analyzing the frequencies generated by the
sonic-point mechanism, we shall for definiteness
use Boyer-Lindquist global coordinates, which are
familiar because they are commonly used to
describe the spacetime of rotating black holes.
However, all of the expressions we give are
accurate only to first order in $j$ and hence
would be unchanged if we used circumferential
rather than Boyer-Lindquist coordinates.

In Boyer-Lindquist coordinates, the circular
frequency of an element of matter on the surface
of the rotating star or of an element of gas in
orbit around it is
 \begin{equation}
 \nu \equiv (1/2\pi)(d\phi/dt) \comma
 \label{eq:CircFreq}
 \end{equation}
 where $\phi$ and $t$ are the azimuthal and time
coordinates of the element of matter or gas.
When measured in Boyer-Lindquist coordinates,
any rotational frequency appears to be the same
at every point in space and is equal to the
frequency measured by a distant observer because
of the symmetries of the spacetime outside a
steadily rotating star\footnote{The spacetime
outside a steadily rotating star has both
timelike and spacelike Killing vector fields
${\xi}_{(t)}$ and ${\xi}_{(\phi)}$, which
reflect the stationarity and axial symmetry of
the spacetime (see, e.g., Misner, Thorne, \&
Wheeler 1973, pp. 892--895). In any coordinate system
in which the time and azimuthal coordinates are
based on ${\xi}_{(t)}$ and ${\xi}_{(\phi)}$,
such as the Boyer-Lindquist coordinate system,
the time interval required for one rotation of
an element of the star or one orbit of an
element of gas orbiting the star is the same
everywhere, when measured in the global time
coordinate. Stated more concretely, if an
element of gas emits a pulse of X-rays each time
its azimuthal coordinate $\phi$ increases by
$2\pi$, the time interval $\Delta t$ between the
arrival of successive pulses will be the same
everywhere, when measured in the global time
coordinate $t$.} and the fact that the
Boyer-Lindquist time coordinate is the proper
time of an observer at radial infinity. We
emphasize that the frequency~(\ref{eq:CircFreq})
is {\em not\/} the frequency that would be
measured by a local observer (i.e., in a local
orthonormal tetrad), unless the observer is at
infinity.

{\em Sonic-point Keplerian frequency}.---As
explained earlier in this section, the radiation
drag force acting on the gas in a clump near the
sonic radius generates a supersonic stream of
denser gas that spirals inward toward the stellar
surface. To the extent that the radiation field
is axisymmetric, the inspiral trajectory of the
gas from each such clump is identical and hence
the azimuthal separation $\Delta\phi_{\rm str}$
between any two such streams is the same at
every radius inside the sonic point and equal to
the azimuthal separation $\Delta\phi_{\rm cl}$
of the two clumps at the sonic radius $r_{\rm
s}$ that are producing the two streams.
Therefore a collection of clumps distributed
around the star at the sonic radius will
generate a pattern of inspiral streams that
rotates around the star at the orbital frequency
$\nu_{\rm Ks} \equiv \nu_{\rm K}(r_{\rm s})$ of
the gas at the sonic radius.

We can estimate $\nu_{\rm Ks}$ by noting that,
to first order in $j$, the the spacetime outside
a slowly and uniformly rotating star is the same
as the Kerr spacetime (Hartle \& Thorne 1968;
the two spacetimes are not the same to higher
orders in $j$). Hence, to first order in $j$ the
orbital frequency of an element of gas in a
prograde Keplerian circular orbit at
Boyer-Lindquist radius $r_{\rm s}$ in the
rotation equator of a slowly rotating star is
(see Bardeen, Press, \& Teukolsky 1972)
 \begin{eqnarray}
 \nu_{\rm Ks}\hskip-0.1 truein &=&
 \hskip-0.1 truein 
 \frac{1}{2\pi}\frac{d\phi_{\rm Ks}(r)}{dt} =
 \frac{1}{2\pi}
 \left[
   1-j\left(\frac{M}{r_{\rm s}}\right)^{\!3/2}
 \right]
 \left(
   \frac{M}{r_{\rm s}^3}
 \right)^{\!1/2} \nonumber\\
 &\approx&
 1181\,\left(M\over{1.4\,\msun}\right)^{\!1/2}
 \left(r_{\rm s}\over\mbox{15\ km}\right)^{\!\!-3/2}
 \,{\rm Hz}.
 \label{eq:SonicKeplerFreq}
 \end{eqnarray}
 The {\em pattern\/} of the gas streaming inward
from the clumps at $r_{\rm s}$ rotates uniformly
around the star with frequency $\nu_{\rm Ks}$.
The pattern frequency is different from the {\em
orbital\/} frequency $\nu_{\rm orb}(r)$ of the
gas inside the sonic radius, which varies with
radius. In particular, the pattern frequency is
different from the orbital frequency of the gas
just before it collides with the stellar surface.

The brightness pattern made by the collision of
the inspiraling streams of denser gas with the
stellar surface rotates around the star at
frequency $\nu_{\rm Ks}$, as measured by an
observer at infinity. This is therefore the
frequency at which the radiation pattern
produced by the bright footprints of the streams
rotates around the star and hence also the
centroid frequency of the resulting
quasi-periodic oscillation in the X-ray flux and
spectrum seen by a distant observer whose line
of sight is inclined to the rotation axis of the
disk. Therefore the radiation pattern produced
by the bright footprints of the streams
generally does {\em not\/} corotate with the
star, but instead moves around the stellar
surface.

{\em Stellar spin frequency}.---As explained in
\S~3.1, if the star's magnetic field is too weak
to affect the accretion flow at the sonic point
but is strong enough to partially channel the
flow close to the stellar surface, it will
create a weakly beamed pattern of radiation that
rotates {\em with the star}. The surface
magnetic fields of the neutron stars in the
atoll and Z sources, which are thought to be
several hundred million years old, may be nearly
dipolar. If the magnetic field is dipolar
but offset from the center of the star, the
radiation pattern it produces will have two
unequal maxima
around the star, whereas if the field is
a centered but tilted dipole field, the
radiation pattern will have two nearly
equal maxima around the star. We therefore
expect that there will
generally be a brightness oscillation with a
frequency equal to one or two times the spin
frequency $\nu_{\rm spin}$ of the star. However,
as we discuss in detail in \S~4.3, we expect the
relatively weak magnetic fields of the atoll and
Z sources to produce radiation patterns that are
only weakly beamed, even near the star. When the
substantial attenuation of this inherently weak
beaming that is caused by scattering in the
central corona is taken into account, the X-ray
oscillation at $\nu_{\rm spin}$ or $2\nu_{\rm
spin}$ seen by a distant observer may be very
weak or even undetectable with current
instruments.

{\em Sonic-point beat frequency}.---The weakly
beamed pattern of radiation rotating with the
star causes the radiation drag force acting on
the gas in a given clump orbiting near the sonic
radius to peak with a frequency equal to one or
two times the
difference (beat) frequency $\nu_{\rm Bs}$ between the
Keplerian frequency at the sonic point and the
spin frequency of the star,
as explained in \S~3.1. The peak in
the drag force will cause a temporary increase
in the flux of denser gas streaming
supersonically inward from the clump with a
frequency
 \begin{equation}
 \nu_{\rm peak}=k\nu_{\rm Bs} \equiv
 k(\nu_{\rm Ks} - \nu_{\rm spin})
 \comma
 \label{eq:BeatFreq}
 \end{equation}
 where $k$=1 or 2, depending on the symmetry
(see Lamb et al.\ 1985). The inflow time from the
sonic point is typically \about 5--10 beat
periods, so there are typically this many
density enhancements in the stream of gas from a
given clump to the stellar surface. These
density enhancements do not lie along a single
gas streamline, but instead along a sequence of
streamlines. As they collide with the stellar
surface they produce a sequence of brighter
impact arcs around the surface in the plane of
the disk. As the quasi-periodic increase in the
mass flux from a given clump arrives at the
stellar surface, it produces a quasi-periodic
modulation in the luminosity of the star with
frequency $\nu_{\rm peak}$. Because this
frequency is the difference between an orbital
and a spin frequency, the frequency of the
resulting quasi-periodic modulation of the X-ray
flux and spectrum seen by an observer at
infinity is $\nu_{\rm peak}$.

{\em Expected range of sonic-point QPO
frequencies.}---In \S~3.2 we showed that either
radiation forces or general relativistic
corrections to Newtonian gravity cause an abrupt
transition to rapid radial inflow and that the
radius of this transition is bounded below by the
radius of the marginally stable orbit, which is
\hbox{$12~\km$} for a nonrotating $1.4\,\msun$
star, and above by the \hbox{$\sim 20$\%} upper
bound on the fraction of the angular momentum of
the accreting gas that can be transferred to the
radiation field, which constrains the transition
radius to be \hbox{$\lta 30~\km$} for a
$1.4\,\msun$ star. As discussed in \S~4.2, there
are several effects that are likely to reduce
further the allowed range of the transition
radius, perhaps to \hbox{$\sim 15 \dash
25~\km$}. Inserting these radii in
equation~(\ref{eq:SonicKeplerFreq}) gives an
expected range for the sonic-point Keplerian
frequency of \hbox{$\sim 500 \dash 1200~\Hz$}.
For stars with spin frequencies \hbox{$\sim 200
\dash 300~\Hz$}, the expected range of beat
frequencies is then \hbox{$\sim 200 \dash
1000~\Hz$}. These ranges are similar to the
frequencies of the kilohertz QPOs detected so
far.

{\em Overtones and sidebands of the sonic-point
QPO frequencies.}---In addition to the beaming
oscillations at $\nu_{\rm Ks}$ and $\nu_{\rm
spin}$ and the luminosity oscillation at
$\nu_{\rm Bs}$, we expect accretion onto weakly
magnetic neutron stars to generate brightness
oscillations at several other frequencies. For
example, the X-ray brightness modulation
produced by the motion of the brighter impact
footprints around the star with frequency
$\nu_{\rm Ks}$ is unlikely to be {\em exactly\/}
sinusoidal, so there may be some power at
overtones of $\nu_{\rm Ks}$. However, the
angular distribution of the X-ray emission from
the impact footprints is expected to be very
broad (see \S~3.1), so the power at overtones of
$\nu_{\rm Ks}$ is likely to be much smaller than
the power at $\nu_{\rm Ks}$. The radiation
pattern that rotates with the star is also
expected to be very broad by the time the
radiation escapes from the central corona, so
the power at overtones of $\nu_{\rm spin}$ is
likely to be much smaller than the power at
$\nu_{\rm spin}$, which is itself expected to be
small for the reasons discussed above. However,
the radiation pattern that rotates with the star
is likely to be narrower at the sonic radius
than outside the central corona, causing the
modulation of the mass flux from the sonic point
to be somewhat non-sinusoidal and giving rise to
weak overtones of the sonic-point beat
frequency. Luminosity oscillations are not as
strongly attenuated by scattering in the central
corona (see \S~4.1), so it may be easier to
detect overtones of $\nu_{\rm Bs}$.

The sonic-point mechanism causes a modulation of
the inward mass flux and hence the stellar
luminosity at the sonic-point beat frequency;
from the point of view of a distant observer,
this luminosity is then modulated at the
sonic-point Keplerian frequency by the motion of
the brighter footprints around the star. As a
result, QPOs are generated at some sideband
frequencies. However, by far the strongest QPOs
are generated at the sonic-point Keplerian and
beat frequencies. To see this, consider as an
example the X-ray flux waveform
 \begin{eqnarray}
 \hskip -0.2 truein 
  f(t) & = &
  [1 + K_1\cos (2\pi\nu_{\rm K}t + \phi_{\rm K1})\nonumber\\
     &&\quad\mbox{} + K_2\cos (4\pi\nu_{\rm K}t + \phi_{\rm K2})]
  \times\nonumber\\&&\quad\times\;
  [1 + B_1\cos (2\pi\nu_{\rm B}t + \phi_{\rm B1})\nonumber\\
     &&\quad\qquad \mbox{} + B_2\cos (4\pi\nu_{\rm B}t + \phi_{\rm B2})]
 \;,
 \label{FluxWave}
 \end{eqnarray}
 which describes these modulations. In
equation~(\ref{FluxWave}) we have included for
the sake of illustration only the fundamental and
first overtones of the Keplerian and beat
frequencies; higher harmonics may of course also
be present. The frequencies that are generated
are listed in Table~\ref{table:FreqGen}. The only
first-order QPOs are those at the sonic-point
beat and Keplerian frequencies. There are also
QPOs at the second harmonics of the beat and
Keplerian frequencies. The lowest-order QPOs
generated by the modulation of the
beat-frequency ``carrier'' by the motion of the
footprints around the star are of second order
and are at the stellar spin frequency $\nu_{\rm
spin}$ and at the very high frequency $2\nu_{\rm
Ks} - \nu_{\rm spin}$. The only QPO of
intermediate frequency generated by this
modulation is a QPO at $\nu_{\rm Ks} - 2\nu_{\rm
spin}$ that is third-order and hence likely to
be very weak. All the other QPOs generated are
also third-order and in addition have very high
frequencies. They are therefore likely to be
intrinsically weak, strongly attenuated by
scattering in the central corona, and difficult
to detect.

As we discuss in \S~4.3, only brightness
oscillations that are fairly large near the
neutron star and are not strongly attenuated by
the gas surrounding the star will be detectable
by a distant observer. In the sonic-point model,
only the sonic-point beat frequency QPO and its
overtones are luminosity oscillations; the other
QPOs are beaming oscillations and are therefore
more attenuated by scattering. For this reason,
we expect strong QPOs only at  $\nu_{\rm Bs}$ and
$\nu_{\rm Ks}$. Nonetheless, QPOs at other
frequencies, such as $\nu_{\rm spin}$, $2\nu_{\rm
Bs}$, and $2\nu_{\rm Ks}$ may be detectable.
Detection of oscillations at any of these
frequencies would corroborate the sonic-point
model.

\subsection{Coherence of Sonic-Point QPOs}

A key question any theory of the kilohertz QPO
sources that relates a QPO frequency to orbital
motion must address is why narrow peaks are seen
in power spectra of the brightness variations of
these sources, rather than a broad continuum
corresponding to the range of orbital frequencies
in the inner disk. Indeed, the remarkably high
coherence ($\nu_{\rm QPO}/\delta\nu_{\rm QPO}
\gta 100$) of some of the kilohertz QPOs places
very strong constraints on any model of these
QPOs, because there are many physical effects
that tend to decrease the coherence of
oscillations at kilohertz frequencies. We discuss
some of the most important effects, derive the
resulting constraints on the accretion flow, and
show that the sonic-point mechanism can produce
narrow peaks consistent with the observed
coherence of the kilohertz QPOs.

Gas in the inner disk orbits the neutron star
with a wide range of frequencies, so at first
glance one might think that fluctuations in the
gas density throughout the inner disk would
generate a broad spectrum of brightness
variations. For the reasons discussed in \S~2.3,
it is very unlikely that such fluctuations can
produce large amplitude brightness variations
via direct emission or occultation. We have
therefore focused our attention on the
brightness variations that such density
fluctuations produce indirectly, as gas from
them falls inward and impacts the stellar
surface. Here we show that the accretion flow in
the vicinity of the sonic radius acts as a
filter that {\em selects\/} brightness
oscillations at the sonic-point orbital
frequency while {\em suppressing\/} brightness
oscillations at higher and lower frequencies. In
order to see how this filtering occurs, we
analyze the X-ray emission at the stellar
surface produced by clumps that form at
different radii in the inner accretion disk.

{\em Effects of azimuthal shear}.---Let us
suppose for the sake of argument that the
thickness of the inner disk is infinitesimal and
that the flow there is laminar (we discuss the
effects of turbulence shortly). Suppose
further that a very small, roughly spherical
density fluctuation (clump) has formed well
outside the sonic radius and consider what will
happen to the gas in this clump and how it will
affect X-ray emission from the star. When a
small clump forms, the orbital phases of the
elements of gas that comprise it are necessarily
very similar, because of the small azimuthal
extent of the clump. The elements of gas that
comprise the clump also have very similar
orbital frequencies, because the radial extent
of the clump is small. As the clump drifts
inward, elements of gas at different radii are
sheared relative to one another in the azimuthal
direction.

The frequency of the X-ray brightness
oscillation that is generated by the inspiraling
gas from a clump is approximately equal to the
Keplerian frequency at the radius where the
clump originally formed (see the discussion in
\S~3.3).\footnote{At first
glance, one might think that if the inward
radial drift time from a given radius $r$ in the
inner disk to the stellar surface is large,
azimuthal shearing of a clump formed at $r$
would by itself reduce the amplitude of the
brightness variation produced by gas from the
clump, but this is not so. The reason is that to
first order, the azimuthal velocity shear in a
Keplerian flow conserves the volume of the
clump, because the velocity field is
divergence-free to this order. Hence, the
density of the gas in a clump relative to the
density of the background gas is affected only
weakly by the differential rotation. In fact, if
the flow were incompressible as well as laminar,
the angular width of the brighter impact arc on
the stellar surface would also be unaffected,
even if the {\em total\/} angular extent of the
clump at all radii becomes large, because the
angular extent of that part of the clump that is
crossing the sonic radius at any given time is
unchanged by azimuthal shear.} A clump that
forms very near the sonic radius will therefore
produce an X-ray brightness oscillation at the
sonic-point Keplerian frequency. If clumps can
form inside the sonic radius, such clumps would
produce X-ray brightness oscillations at the
orbital frequencies where they formed.

This analysis shows that if the clumps were small
and the flow laminar, differential rotation of
the gas would not by itself pick out any
particular orbital frequency. Consequently, if
clumps were to form at a wide range of radii and
nothing besides inward drift and azimuthal shear
were to happen, gas inspiraling from the clumps
and colliding with the stellar surface would
produce brightness fluctuations with a wide
range of frequencies up to the \hbox{$\sim 1500
\dash 2000\,\Hz$} orbital frequency at the
stellar surface (see, e.g., Kluzniak, Michelson,
\& Wagoner 1990).

{\em Effects of clump destruction}.---In reality,
clumps that form {\em outside\/} the sonic radius
are destroyed before they drift inward to the
sonic point and gas in them can reach the stellar
surface. Clumps (if any) that form {\em inside\/}
the sonic radius orbit the star at a rapidly
changing frequency and collide with the stellar
surface before gas from them can produce a long
wavetrain. Hence, only clumps that form very near
the sonic radius can produce strong, relatively
coherent brightness oscillations. In this way the
disk flow filters out brightness oscillations at
orbital frequencies other than the sonic-point
Keplerian frequency.

In order to illustrate the effects of clump
destruction in the Keplerian flow, we consider a
simple model in which clumps form and are then
destroyed by small-scale turbulence in the disk,
such as that produced by the magnetoturbulence
that is thought to be responsible for angular
momentum transport within the disk flow (Balbus
\& Hawley 1991, 1992, 1997; Brandenburg et al.\
1995, 1996; Hawley, Gammie, \& Balbus 1995,
1996). In this simple model we assume that the
forces that formed the clump in the first place
no longer act to hold it together. If the clump
has size $\xi$, turbulent motions on scales
$\ell\gta\xi$ will tend to advect the clump
without destroying it; only turbulent motions on
scales $\ell\ll\xi$ will disrupt the clump. We
assume that  the effect of these small-scale
turbulent motions on the clump can be described
by a diffusion coefficient \hbox{$D \sim \beta
\xi c_s$}, where $\beta$ is a dimensionless
parameter that describes the strength of the
small-scale turbulence and $c_s$ is the thermal
sound speed in the disk; for turbulent motions
with scales $\ell \ll \xi$, $\beta$ is small
compared to unity. As a result of outward
angular momentum transport by the
magnetoturbulence, a clump of initial size $\xi$
that forms at radius $r$ will drift slowly
inward with radial velocity $v^{\hat r} \sim
\alpha (h/r) c_s$, where $\alpha$ is the usual
viscosity parameter and $h$ is the
half-thickness of the disk. Such clumps will be
dissipated by turbulent diffusion in a time
$t_{\rm diss} \sim (\xi/\beta c_s) \sim
(\xi/\beta h\Omega_{\rm K})$.

In this model, a clump that forms a radial
distance $\delta r$ {\em outside\/} the sonic
radius $r_{\rm s}$ can reach the sonic radius and
generate a supersonically inspiraling stream of
gas before being destroyed only if $\delta r <
v^{\hat r} t_{\rm diss}$, i.e., only if the clump
forms within a radial distance
 \begin{equation}
 \delta r \lta (\alpha/\beta) (h/r_{\rm s})
 (\xi/r_{\rm s}) r_{\rm s} \lta
 10^{-2}\,r_{\rm s}
 \end{equation}
 of the sonic radius. Turbulence on small scales
may be weak, and the effects that form the
clumps in the first place---such as thermal
instability, magnetic stresses, and radiation
forces---may tend to continue to hold them
together as they drift inward toward the sonic
point, so that clumps can survive somewhat
longer than $t_{\rm diss}$. Even so, it is clear
that clumps that form even a small distance
outside the sonic radius will be disrupted by
turbulence before they reach the sonic point.
This process suppresses brightness oscillations
with frequencies less than the sonic-point
Keplerian frequency.

It appears improbable that clumps will form in
the hypersonic radial inflow between the sonic
radius and the stellar surface, but even if
clumps do form in this region, they will not
produce appreciable brightness oscillations, for
two reasons. 
First, a clump that forms inside the sonic
radius will collide with the stellar surface
after completing $\sim$1--10 orbits. A
clump that lives only for a time $t_{\rm
lifetime}$ will generate power over a range of
frequencies $\delta \nu_{\rm lifetime} \sim (\pi
t_{\rm lifetime})^{-1}$. Hence the power
produced by a clump that forms inside the sonic
radius will be spread over a frequency range
$\delta \nu_{\rm lifetime} \gta 0.03\,\nu $. 
Second, a clump that forms inside the
sonic radius orbits the star at a frequency that
rapidly increases from, for example, the $\sim
500\,\Hz$ orbital frequency at the sonic radius
to the $\sim 1500\,\Hz$ orbital frequency at the
stellar surface, as the clump spirals inward.  For
these reasons, even if clumps do form inside the
sonic radius, any power they generate will be
spread over a wide range of frequencies rather
than concentrated in a narrow peak.

{\em Coherence of the sonic-point Keplerian
frequency QPO}.---Clumps that form outside the
sonic radius but close enough to it to reach it
before being destroyed have initial orbital
frequencies in the relatively narrow range
 \begin{eqnarray}
 \delta\nu \sim
 \left(\frac{\delta r}{r_{\rm s}}\right)
 \,\nu_{\rm Ks} \sim
 \left(\frac{\alpha}{\beta}\right) 
 \left(\frac{h}{r_{\rm s}}\right)
 \left(\frac{\xi}{r_{\rm s}}\right)\,
 \nu_{\rm Ks}
 \nonumber\\
 \nonumber\\
 \leq 10^{-2}\, \nu_{\rm Ks},
 \end{eqnarray}
 which is consistent with the
observed narrowness of the kilohertz QPO peaks.
Clumps orbiting at different distances above or
below the midplane of the disk will generally
also have slightly different orbital
frequencies. To estimate the resulting spread in
frequencies, we assume that at each distance $z$
above and below the disk plane the inward radial
velocity increases sharply from subsonic to
supersonic at some cylindrical radius $R_{\rm
s}$. We call the axisymmetric two-dimensional
surface defined by $R_{\rm s}(z)$ the sonic
surface. If the sonic surface is cylindrical,
the oscillation frequency generated by the
clumps orbiting in the midplane of the disk is
slightly greater than the oscillation frequency
generated by the clumps orbiting above and below
the midplane. The spread in oscillation
frequencies depends on the precise shape of the
sonic surface, but to lowest order in $(h/r_{\rm
s})$ the spread in frequencies caused by this
effect is\break
 $\delta\nu \sim (h/r_{\rm s})^2\,
\nu_{\rm Ks}$, which is consistent with the
observed narrowness of the kilohertz QPO peaks
if $h_{\rm s}/r_{\rm s} \lta 0.1$.

The persistence of a clump orbiting at the sonic
radius is limited by its decay as gas inspirals
from it to the stellar surface and by its
disruption by small-scale turbulence, pressure
forces, and other effects. In the simple model
considered here, the clump decay timescale is
$t_{\rm decay} \sim \xi/v^{\hat r} \sim
(1/\alpha) (\xi/h_{\rm s}) (r_{\rm s}/h_{\rm s})
(1/\Omega_{\rm K})$, which broadens the QPO peak
by an amount $\delta\nu_{\rm decay} \sim \alpha
(h_{\rm s}/\xi) (h_{\rm s}/r_{\rm s})\, \nu_{\rm
Ks} \lta 10^{-2}\, \nu_{\rm Ks}$, which is
consistent with the observed narrowness of the
kilohertz QPO peaks. The dissipation of clumps by
turbulent diffusion contributes a relative width
$\delta\nu_{\rm diss} \sim \beta (h_{\rm
s}/\xi)\,\nu_{\rm Ks}$, which is consistent with
the observed narrowness of the Keplerian
frequency QPO peaks if $\beta \ll 1$ or the
effects that form the clumps in the first place,
such as thermal instability, magnetic stresses,
and radiation forces, continue to hold them
together as they orbit at the sonic point.

This simple model shows that if a clump is too
small, it will persist for such a short time that
the power it generates will be spread over a
broad frequency range, whereas if a clump is too
large it will persist for such a long time that
its orbital frequency will change appreciably
during its lifetime and the power it generates
will again be spread over a broad frequency
range. It is the clumps of intermediate size
that generate relatively coherent oscillations.

In addition to frequency variation and lifetime
broadening, other effects may increase the width
of the Keplerian frequency QPO peak. For
example, the radiation field inside the sonic
radius is not perfectly axisymmetric, and hence
the radiation forces acting on the inspiraling
gas will vary slightly with azimuth. As a
result, the radius of the sonic point and the
density patterns produced by the inspiraling gas
will be slightly different at different
azimuths. A quantitative treatment of these
effects is well beyond what is possible with
current accretion disk models.

The simple model we have discussed shows the
crucial role played by the sharp transition to
rapid radial inflow at the sonic radius that was
found in \S~3.2. In the absence of such a
transition, density and magnetic field
fluctuations in the inner disk would produce weak
brightness variations over a wide range of
frequencies rather than in the narrow range of
frequencies needed to create a QPO peak. In the
presence of such a transition, on the other hand,
clumps orbiting in a narrow range of radii near
the sonic radius produce strong X-ray brightness
variations with a correspondingly narrow range of
frequencies. Simple estimates of the broadening
produced by the different orbital frequencies of
these clumps and by their finite lifetimes appear
consistent with the high observed coherence of
the Keplerian frequency QPOs.

{\em Coherence of related QPOs}.---If there is a
QPO at the beat frequency $\nu_{\rm Bs}$, we
expect that its FWHM will in general be comparable
to the FWHM of the QPO at $\nu_{\rm Ks}$. This is
because the peak at $\nu_{\rm Bs}$ is at the beat
frequency of the sonic-point Keplerian frequency
with the stellar spin frequency,
which is nearly coherent and therefore adds
relatively little to the width of the QPO peak.
However, we do {\em not\/} expect the widths of the
QPO peaks at $\nu_{\rm Bs}$ and $\nu_{\rm Ks}$ to
be identical, because scattering by the moving gas
in the central corona affects the oscillations at
$\nu_{\rm Bs}$ and $\nu_{\rm Ks}$ differently.

\subsection{Photon-Energy Dependence of the
Sonic-Point QPOs and Time Lags}

The X-ray spectra of systems powered by
accreting, weakly magnetic neutron stars, such
as the kilohertz QPO sources, are formed
primarily by Comptonization, as described in
\S~2.2. Cyclotron and bremsstrahlung photons
with energies \lta 1~keV are produced near the
neutron star surface and are then Comptonized by
hot electrons in the region where the accretion
flow interacts with the stellar surface and in
the hot central corona that surrounds the star
and its magnetosphere, yielding the observed
X-ray spectra.

During a sonic-point Keplerian or beat-frequency
oscillation, the rotation of the patterns of
denser gas spiraling inward from clumps orbiting
at the sonic radius causes the optical depth
along the line of sight from the neutron star
surface to the observer to vary
quasi-periodically. The rate of production of
soft photons and the physical properties of the
Comptonizing gas may also oscillate, especially
during the sonic-point beat-frequency
oscillation, which is mainly a luminosity
(accretion rate) oscillation. These oscillations
cause the X-ray spectrum produced by the system
to oscillate quasi-periodically with various
frequencies. A small fractional variation in the
optical depth causes a much larger fractional
variation in the number of photons at high
photon energies, because of the characteristic
way in which the spectrum produced by
Comptonization changes as the optical depth
oscillates (see Miller \& Lamb 1992). To the
extent that the effects of Comptonization 
dominate, the
oscillations at $\gta 10~\keV$ will lag the
oscillations at $\sim 5~\keV$, because the
photons above 10~keV have scattered more times
in escaping from the source.

An accurate, quantitative treatment of the X-ray
spectral oscillations produced by the
inhomogeneous and time-dependent accretion flow
in the kilohertz QPO sources would require a
three-dimensional, time-dependent radiation
hydrodynamic calculation in full general
relativity. This is beyond present computational
abilities. However, one can obtain a qualitative
understanding of the dependence of QPO
amplitudes on photon energy and the time lags to
be expected using the following simplified model
of the Comptonization process.

Assume that soft photons with a characteristic
energy $E_{\rm in}$ that is much smaller than
the electron temperature $T_e$ are injected at
the center of a static, uniform, spherical
Comptonizing region of radius $R_{\rm C}$,
electron density $n_e$, and scattering optical
depth $\tau\equiv n_e \sigma_T R_{\rm C}$, where
$\sigma_T$ is the Thomson scattering cross
section. Assume further that $y \equiv 4 T_e
\tau^2/m_e\lesssim 1$, where $m_e$ is the
electron rest mass, and that the electrons have
a negligible bulk velocity. These are good
approximations for modeling formation of the
time-averaged X-ray spectra of the atoll sources
(see Psaltis \& Lamb 1998b). The region of
approximately radial inflow that develops when
the mass accretion rate becomes comparable to
$\dot{M}_{\rm E}$ (see Lamb 1989, Lamb 1991, and
\S~2) introduces additional effects on the
spectra of the Z sources that cannot be treated
in this way; we do not consider these effects
here (but see Psaltis \& Lamb 1998b). We treat
the effects of oscillations in the rate of
soft-photon production and in the properties of
the Comptonizing region near the star by varying
these quantities in the spherical model. In
particular, we mimic the effects of the
quasi-periodic variation of the optical depth
along the line of sight by varying the optical
depth of the model. We assume that the
oscillation period is longer than the mean time
for photons to escape from the Comptonizing
region, an approximation that appears to be
excellent for the kilohertz QPO sources.

{\em Photon-energy dependence of the sonic-point
Keplerian and beat-frequency oscillation
amplitudes}.---The X-ray number spectrum that
emerges from the Comptonizing region at photon
energies $E \gg E_{\rm in}$ can be approximated
by (see, e.g., Rybicki \& Lightman 1979, pp. 221-222)
 \begin{equation}
  f(E) \propto \left\{
  \begin{array}{ll}
  \!\!\left(\frac{E}{E_a}\right)^{2+\alpha},
      \;\; E\ll E_b\\ \\
  \!\!\left(\frac{E_b}{E_a}\right)^{\alpha}
  \!\!\left(\frac{E}{E_a}\right)^2
    \!\exp[-(E-E_b)/T_e],
      \;\; E\gg E_b
  \end{array}
  \right.
  \label{eq:Spectrum}
 \end{equation} 
 where $T_e$ is the electron temperature, $E_a$ is
a normalization constant that is typically a few
keV, $E_b \sim (1 \dash 3)T_e$ is the cutoff
energy, and
 \begin{equation}
   \alpha =
    -\frac{3}{2}-\sqrt{\frac{9}{4}+\frac{4}{y}}
  \;.
 \end{equation}

For small-amplitude oscillations in which the
variation of the electron temperature in the
Comptonizing region is negligible (see below),
the relative amplitude $\gamma(E)$ of the
oscillation is approximately (Miller \& Lamb
1992)
 \begin{equation}
   \gamma(E)=\frac{\Delta 
  \dot{N}_{\rm s}}{\dot{N}_{\rm s}}
   +\frac{d}{d\tau}\ln[f(E)] \,\Delta \tau\;.
   \label{eq:AmpGeneral}
 \end{equation}
 Here $\Delta \dot{N}_{\rm s}/\dot{N}_{\rm s}$ is
the relative change in the rate at which soft
photons are injected into the Comptonizing region
during an oscillation and $\Delta\tau$ is the
change in the optical depth of the region.
The relative amplitude $\gamma(E)$ is computed 
by considering a full oscillation period.

In general, the soft photon injection rate and
the optical depth both vary during the
oscillation. If so, equations
(\ref{eq:Spectrum})--(\ref{eq:AmpGeneral}) give
 \begin{equation}
   \vert\gamma(E)\vert\simeq\left\vert
    \frac{\Delta \dot{N}_{\rm s}}{\dot{N}_{\rm s}}
      +\gamma_{\rm \tau}(E) \right\vert\;,
   \label{eq:AmpAbsGeneral}
 \end{equation} where
  \begin{eqnarray}
   \label{eq:AmpKepler}
   \gamma_{\rm \tau}(E)
   \hskip-0.1 truein
   &\simeq&
   \hskip-0.1 truein
   -0.6 \left(\frac{T_e}{10~\hbox{keV}}\right)^{-1/2}
      \left(\frac{3}{\tau}\right)
     \left(\frac{\Delta \tau/\tau}{0.05}\right)
      \times\nonumber\\&&
   \hskip-0.5 truein\times
      \left[\frac{\ln(E_b/E_a)}{2}\right]
  \left\{
  \begin{array}{ll}
      \frac{\ln(E/E_a)}{\ln(E_b/E_a)}\;, & E\ll E_b\\
  \\
      \quad\; 1 \;, & E\gg E_b
  \end{array}
  \right.
  \label{eq:GammaTau}
  \end{eqnarray}
 in terms of the expected properties of the
Comptonizing region and we have neglected the weak
dependence of the normalization constant $E_a$ on
$\tau$. Equations~(\ref{eq:AmpAbsGeneral})
and~(\ref{eq:GammaTau}) show that in this case the
relative amplitude of the QPO depends both on the
relative change in the soft-photon injection rate,
$\Delta \dot{N}_s/\dot{N}_s$, and on the relative
change in the optical depth, $\Delta \tau/\tau$.
In writing equation~(\ref{eq:AmpGeneral}) 
we have assumed that
the electron temperature in the Comptonizing cloud
remains constant during the oscillation. If the 
electron temperature were allowed to vary, then
the relative amplitude of the oscillation 
would not be constant at high photon
energies, in disagreement with observations.

The quantity $\gamma_{\rm \tau}$ is negative for
the spectrum~(\ref{eq:Spectrum}) and is small at
low photon energies. If the oscillation in the
injection rate of the soft photons is in phase with
the oscillation in the optical depth, i.e., if
$\Delta \dot{N}_s/\dot{N}_s$ and $\Delta
\tau/\tau$ have the same sign, and if $\gamma_\tau$
is larger in magnitude than $\Delta
\dot{N}_s/\dot{N}_s$ at high photon energies, then
the relative amplitude of the oscillation in the
photon number will not be a monotonic function of
photon energy. For $\Delta \dot{N}_s/\dot{N}_s$
comparable in magnitude to $\Delta \tau/\tau$ and
a Comptonizing region with the properties used in
scaling equation~(\ref{eq:GammaTau}), the relative
amplitude of the oscillation has a minimum in the
energy range $\sim 5 \dash 10~\keV$. The
sonic-point beat frequency QPO is primarily a
luminosity oscillation (see \S~3.1), so the photon
production rate is likely to vary significantly at
the beat frequency. {\em The relative amplitude of
the beat-frequency QPO may therefore have a
minimum in the $5 \dash 10~\keV$ energy range}.

If the relative change in the photon injection
rate during an oscillation is very small, the
first term on the right side of equation 
(\ref{eq:AmpAbsGeneral}) is likely to be
negligible compared to the second term. In this
case the relative amplitude of the oscillation
will increase monotonically with increasing photon
energy, for energies less than $E_b \simeq
10-30$~keV, but will become independent of photon
energy for energies greater than $E_b$. We expect
this to be the case for the oscillation at the
sonic-point Keplerian frequency, which is mainly a
beaming oscillation (see \S~3.1) and hence involves
a relatively small
variation in the photon production rate.
{\em We therefore
expect the relative amplitude of the
Keplerian-frequency QPO to increase steeply in
the $5 \dash 10~\keV$ energy range and then
flatten at high energies}.

When the photon injection rate and spectrum of
injected soft photons do not change appreciably
during a QPO cycle, the amplitude of the
oscillation in luminosity caused by the
oscillation in optical depth is comparable to the
amplitude of the optical depth oscillation and is
given by
 \begin{equation}
 \frac{\Delta L}{L} \sim \Delta y = 0.07
   \left( \frac{T_e}{10~\mbox{keV}} \right)
   \left( \frac{\tau}{3} \right)^2
   \left( \frac{\Delta \tau/\tau}{0.05} \right) \;.
  \label{eq:LumOsc}
 \end{equation}
 Equations~(\ref{eq:AmpKepler}) and
(\ref{eq:LumOsc}) demonstrate that {\em a
relatively weak oscillation in optical depth of
only a few percent that is accompanied by a
luminosity oscillation of moderate amplitude can
produce naturally a much larger oscillation in the
countrate at high photon energies}.

Equation (\ref{eq:AmpKepler}) also shows that the
relative amplitude of the oscillation at a given
photon energy is different for different
electron temperatures and optical depths, unless
the relative amplitude of the oscillation in
optical depth $\Delta \tau/\tau$ varies in such
a way as to compensate for this effect. We
therefore expect that as the mass accretion rate
onto the neutron star changes (on timescales
much longer than a beat-frequency period), the
electron temperature and optical depth in the
Comptonizing region will change, causing the
photon-energy dependence of the beat-frequency
and Keplerian-frequency oscillation amplitudes
to change.

{\em Photon-energy dependence of the
sonic-point Keplerian and beat-frequency
oscillation phases}.---At photon energies \lta
30~keV, the cross section and hence the electron 
scattering mean free path of
the photons in the central corona is nearly
independent of photon energy. Therefore, the
average escape time of photons from the corona is
also independent of their energy. However, the
photons that stay in the corona longer experience
more scatterings on average, and therefore gain 
more energy by scattering off hot electrons and
emerge from the medium with a larger energy. As a
result, the oscillation at high photon energies is
expected to lag the oscillation at low photon
energies. If the injection rate of soft photons
and the properties of the corona are constant in
time, the magnitude of the time lag is determined
mainly by the properties of the corona (see, e.g.,
Wijers, van Paradijs, \& Lewin 1987; Bussard et
al.\ 1988). If instead the source of soft photons
or the properties of the corona are
time-dependent, then the time lag will depend on
the details of this time-dependence.

For the simplified model described earlier in
this section, the average energy after $u$
scatterings  of a photon injected at energy
$E_{\rm in}$ is
 \begin{equation}
  E(u) \sim {\rm min}[ E_{\rm in}
  \exp(4T_e u/m_e), E_b]\;.
 \end{equation} 
 Therefore, for photons with energies smaller
than $\sim E_b$, the ratio of the energies of
two photons that have experienced a different 
number of scatterings is $E_2/E_1\sim \exp(4 T_e
\Delta u/m_e)$, where $\Delta u$ is the
difference in the number of scatterings. The
photon mean-free time is $\sim (n_e \sigma_T
c)^{-1}$ and hence the time lag introduced
between the oscillations at the two photon
energies is
 \begin{eqnarray}
 \hskip-0.3 truein
 \delta t
 &\hskip-0.1 truein \sim &\hskip-0.1 truein
 \frac{R}{c \tau} \frac{m_e}{4 T_e}
   \ln \left(\frac{E_2}{E_1}\right)
   \nonumber\\
 &\hskip-0.1 truein = &\hskip-0.1 truein
  \frac{400}{\tau}
  \left(\frac{R}{10^6~\mbox{cm}}\right)
  \left(\frac{T_e}{10~\mbox{keV}}\right)
  \ln \left(\frac{E_2}{E_1}\right)~\mu\mbox{s},
 \label{eq:timelag}
 \end{eqnarray} 
 when both $E_1$ and $E_2$ are smaller than $\sim
E_b$. Because of the diffusion in energy and the
systematic downscattering of photons with
energies $\gtrsim E_b$,
equation(\ref{eq:timelag}) is not valid at
energies $\gtrsim E_b$.

We emphasize that, for a variety of reasons,
equation~(\ref{eq:timelag}) gives only an upper
bound to the expected time lag at photon energies
$\lesssim E_b$. First, this equation assumes that
all photons are injected at the center of a
spherical medium. However, if most of the photons
are produced near the surface of the neutron star
and escape a few kilometers above it, the time
lag can be significantly smaller. Second,
equation (\ref{eq:timelag}) was derived under the
assumption that photons slowly diffuse outward.
This is not a very good approximation for
electron scattering optical depths $\tau
\lesssim 3$; if a fraction of the photons escape
directly from the corona without interacting
with the electrons, the average escape time
could be significantly smaller. Finally,
equation (\ref{eq:timelag}) was derived under
the assumption of a uniform electron density. If
instead most of the optical depth is
concentrated near the center of the corona, this
will also reduce significantly the average
escape time.

{\em In summary, to the extent that the effects of
Comptonization dominate, we expect the sonic-point
Keplerian- and beat-frequency oscillations at
high photon energies should lag the corresponding
oscillations at lower photon energies and the
time lag between a few keV and $\sim 15$~keV should
be a fraction of a millisecond}, for the
densities and temperatures of the electrons
expected around the neutron stars in the Z and
atoll sources.

\subsection{Attenuation of Kilohertz QPOs}

In the sonic-point model, gas spirals inward from
density inhomogeneities in the accretion flow at
the sonic radius and collides with the stellar
surface, producing a radiation pattern that
rotates around the star with the Keplerian
frequency at or near the sonic point $\nu_{\rm
Ks}$. Because the rate of mass accretion over
the whole neutron star surface remains
approximately constant in time as the radiation
pattern rotates, the total luminosity emerging
from the system also remains approximately
constant in time. For conciseness we call the
oscillations produced by rotation of this
radiation pattern a {\em beaming\/} oscillation,
because it is caused by the angular variation
of the radiation field, even though the radiation
pattern is unlikely to be a narrow beam. In a
pure beaming oscillation, the total luminosity
emerging from the source is independent of time.
As described earlier, the accretion flow is also expected
to produce a much weaker beaming oscillation at
the stellar spin frequency $\nu_{\rm spin}$.

If the magnetic field of the neutron star is
strong enough to channel the accretion flow near
the stellar surface, the sonic-point model
predicts that the resulting radiation pattern,
which rotates with the star, will modulate
inflow from the inner disk at the beat frequency
$\nu_{\rm Ks}-\nu_{\rm spin}$, producing a {\em
luminosity\/} oscillation with this frequency.
In a pure luminosity oscillation, the total
luminosity of the system changes with time, but
the angular pattern of the radiation field
remains unchanged. In the sonic-point model, the
only luminosity oscillation is the beat
frequency oscillation.\footnote{To clarify
further the difference between luminosity and
beaming oscillations, consider an observer whose
line of sight to the neutron star is along the
rotation axes of the disk and the star. X-rays
from beams rotating with the gas in the disk or
with the star will appear time-independent to
such an observer and therefore will not generate
any X-ray flux or color oscillations. However,
the luminosity oscillation at the beat frequency
caused by the oscillation of the mass flux at
this frequency will be visible. Such an observer
will therefore see an oscillation at the beat
frequency and its overtones, but not at any
other frequency.}

When the luminosity oscillation is present, it
is necessarily modulated by the motion of the
emission regions around the stellar surface,
which generates second-order beaming oscillations
at the stellar spin frequency $\nu_{\rm spin}$
and the difference frequency $2\nu_{\rm
Ks}-\nu_{\rm spin}$ (see \S~3.3 and Table~3). In
general, higher-order oscillations are also
generated at other (mostly much higher)
frequencies.

Scattering generally attenuates beaming
oscillations much more than luminosity
oscillations, for the conditions relevant to
LMXBs (Lamb 1988). The reason is that an
anisotropic radiation pattern becomes
approximately isotropic after only a few
scatterings, whereas a luminosity oscillation is
attenuated by scattering only if the time
required for photons to escape from the
scattering region is much larger than the period
of the oscillation, which requires a
large optical depth for the oscillation
frequencies and coronal dimensions of interest
to us.

The relative amplitude $A_\infty$ of a luminosity
oscillation with angular frequency
$\omega=2\pi\nu$ outside a  spherical scattering
cloud of radius $R_{\rm C}$ and  optical depth
$\tau$ is related to the relative  amplitude
$A_0$ at the center of the cloud by the
expression (Kylafis \& Phinney 1989)
 \begin{equation}
 A_{\infty,{\rm lum}}
 \approx \left(2^{3/2}x
 e^{-x}+e^{-\tau}\right)A_{0,{\rm lum}}\;,
 \label{eq:LumAtten}
 \end{equation}
 where $x\equiv[(3/2)\omega R_{\rm
C}\tau/c]^{1/2}$. This expression is accurate to
better than 6\% for $x>1.3$ (Kylafis \& Phinney
1989). The first term in parenthesis on the
right side of equation~(\ref{eq:LumAtten})
describes the amplitude attenuation caused by
the spread of escape times from the cloud
whereas the second term (which has been added in
by hand) describes the amplitude of the
oscillation produced by the photons that escape
from the cloud without scattering.

For a beaming oscillation caused by rotation of a
narrow pencil beam, the amplitude of the
oscillation outside the cloud is (Brainerd \&
Lamb 1987; Kylafis \& Phinney 1989)
 \begin{equation}
 A_{\infty,{\rm beam}} \simeq
 \left[\left(\frac{2}{1+\tau}\right)2^{3/2}x
 e^{-x}+e^{-\tau}\right]A_{0,{\rm beam}}\;.
 \label{eq:BeamAtten}
 \end{equation} The factor multiplying $A_{0,{\rm
beam}}$ in equation (\ref{eq:BeamAtten})
describes the tendency of scattering to
isotropize the photon distribution. The
amplitudes of oscillations produced by the broad
radiation patterns or the radiation patterns
with more than one lobe that we are concerned
with here are reduced even more by scattering
(Brainerd \& Lamb 1987).

Of all the beaming oscillations that may be
generated by the accretion flow, only the one at
the sonic-point Keplerian frequency is likely to
have a large enough amplitude at the neutron
star to produce an oscillation that is strong
enough, after attenuation by passage through the
central corona, to be observed easily with
current instruments. In contrast, the
oscillation at the sonic-point beat frequency,
which is the only luminosity oscillation 
produced by the flow, should be strong enough to
be observed even if its amplitude at the neutron
star surface is moderate. As a result, we expect
that from the collection of possible frequencies
of oscillations, the only oscillations that will
appear strong far from the star will be those at
the sonic-point Keplerian and beat frequencies.

\section{COMPARISON WITH OBSERVATIONS}

In this section we describe the observational
implications of the sonic-point model and
compare them with observations of the kilohertz
QPOs. We first list the important properties of
these QPOs that any model must explain. These
properties are (see \S~1): (1)~high (\about
300--1200~Hz) frequencies, which can vary by
several hundred Hertz in a few hundred seconds
(see Wijnands et al.\ 1998 and van der Klis
1995); (2)~similar frequency ranges in stars
with significantly different magnetic field
strengths and accretion rates; (3)~relatively
high coherence ($Q \equiv \nu/\delta\nu$ up to
$\sim 100$); (4)~2--60~keV rms amplitudes
$\lta1$\% in the Z sources but much higher (up
to \about 15\%) in the atoll sources; (5)~the
common occurrence of two (but never more than
two) simultaneous kilohertz QPOs in a given
source; (6)~a frequency separation $\Delta\nu$
between the two kilohertz QPOs that is
approximately constant in a given source;
(7)~approximate consistency of $\Delta\nu$ with
the stellar spin frequency inferred from burst
oscillations; (8)~the similarity of most atoll
and Z source spin frequencies; (9)~the
frequently similar FWHM of the two QPO peaks
seen simultaneously; (10)~the increase of the
frequencies of the kilohertz QPOs with
increasing inferred accretion rate that is
observed in many sources; and (11)~a steep
increase in the amplitude of the
higher-frequency of the two kilohertz QPOs with
increasing photon energy in the 2--15~keV energy
band in many sources.

In \S~4.1 we show that the stellar magnetic field
strengths and electron scattering optical depths
required for the sonic-point mechanism to operate
are consistent with the stellar magnetic field
strengths and optical depths of the compact
central corona inferred from previous modeling of
the X-ray spectra and $\lta100$~Hz X-ray
variability of the atoll and Z sources. In \S~4.2
we compare the kilohertz QPO and neutron star
spin frequencies expected in the sonic-point
model with the kilohertz QPO and neutron star spin
frequencies observed in the atoll and Z sources.
We also discuss the expected and observed
dependence of the QPO frequencies on accretion
rate. In \S~4.3 describe the amplitude and
coherence of the various QPOs expected in the
sonic-point model, comparing them with the
observed amplitudes and coherence of the
kilohertz QPOs. In particular, we show that the
sonic-point model explains naturally why at most
two kilohertz QPOs have been detected in the
power spectrum of each source. We also discuss
the relative coherence of QPOs at different
frequencies and the expected dependence of QPO
amplitudes on accretion rate and photon energy.
In \S~4.4 we describe the inverse correlation
between QPO amplitude and stellar magnetic field
expected in the sonic-point model and show that
there is already substantial evidence for such a
correlation.

\subsection{Consistency with Previously
 Inferred\break
Properties of the Atoll and Z Sources}

The operation of the sonic-point mechanism for
producing strong, coherent kilohertz QPOs
requires that a number of conditions be
satisfied: 

(1)~In order to produce the higher-frequency
kilohertz QPOs with the large amplitudes
observed in some sources, the neutron star
magnetic fields in these sources must be
relatively weak, so that a substantial fraction
of the accreting gas penetrates close to the
star in a Keplerian disk flow (\S~3.1).

(2)~In order to produce the lower-frequency
kilohertz QPOs, the neutron star magnetic field,
although relatively weak, must be strong enough
to channel some of the accreting gas near the
stellar surface to produce bright spots that
rotate with the star and modulate the accretion
rate at the sonic-point beat frequency (\S~3.1). 

(3)~In order to generate QPOs with the high
coherence observed in most sources, the disk
flow at the sonic point must be geometrically
thin (\S~3.4). 

(4)~In order that only the QPOs at the
sonic-point Keplerian and beat frequencies be
detectable and that X-ray oscillations at the
stellar spin frequency be very weak or
undetectable at present sensitivities, electron
scattering in the central corona must help to
attenuate the weak beaming oscillations at other
frequencies and at the stellar spin frequency.
The central corona must therefore have an
electron scattering optical depth $\gta 3$
(\S~3.6).

As we now explain briefly, all of these
conditions follow naturally from the unified
model of weak-field accreting neutron stars
described in \S~2.2.

The magnetic field strengths in the 4U atoll
sources are thought to be low enough ($\lta\,
10^9$~G) that at their inferred accretion rates
($\sim 0.01$--$0.1\,{\dot M}_E$) the cylindrical
radius $\varpi_0$ at which the Keplerian disk
flow couples strongly to the stellar magnetic
field is $\lta 2\ee6$~cm. The magnetic field
strengths of the neutron stars in the Z sources
are thought to be a few times  larger than in
the 4U atoll sources, but the accretion rates
are much larger as well, so $\varpi_0$ is
comparable to its value in the atoll sources.
Thus, in both the 4U atoll sources and the Z
sources, the gas in the Keplerian disk
penetrates close to the star before any of it
couples strongly to the stellar magnetic field.
We therefore expect that a significant fraction
of the accreting gas will remain in a Keplerian
disk flow down to the sonic point and will
continue in a disk flow very close to the
stellar surface, as required in the sonic-point
model.

Evidence that the magnetic fields of neutron
stars in LMXBs, while typically weak, are
nevertheless strong enough in many sources to
channel the flow near the star and hence to
produce a QPO at the sonic-point beat frequency
comes from power spectra constructed from \rxte
observations of \gx{5$-$1} (van der Klis et al.\
1996e), \sco1 (van der Klis 1996a, 1996b, 1996d,
1997b), \gx{17$+$2}  (Wijnands et al.\ 1997c),
and \cyg{2} (Wijnands et al.\ 1998). In addition
to two simultaneous kilohertz QPOs, these power
spectra show horizontal branch oscillations,
which appear to be magnetospheric beat-frequency
oscillations (see \S~2 and Psaltis et al.\ 1998).

These power spectra indicate that in the Z
sources, some of the gas in the disk couples
strongly to the weak stellar magnetic field at
two or three stellar radii and is funneled into
hot spots that could modulate the mass flux from
the sonic point at the sonic-point beat
frequency by periodically irradiating the clumps
at the sonic radius (see \S~2). It is natural to
expect that the magnetic fields of many of the
atoll sources are only a little weaker than
those in the Z sources and are therefore still
strong enough to produce a QPO at the
sonic-point beat frequency. This expectation is
supported by the spectral modeling described in
\S~2.2.

Accretion via a geometrically-thin Keplerian disk
flow, which is required to explain the high coherence
of the kilohertz QPOs, is expected in the atoll
sources, because at their low accretion rates the
inner part of the accretion disk is likely to be
gas-pressure-dominated. In the Z sources, which are thought to be
accreting at close to the Eddington critical rate,
gas in the Keplerian disk flow that is not channeled
by the stellar magnetic field is likely to be pinched
by the stellar field into a geometrically thin disk. 

The final requirement listed above for the
sonic-point model, namely that the optical depth
of the hot central corona exceed $\sim 3$,
follows directly from the spectral modeling
discussed in \S~2.2.

In summary, the physical picture of the atoll and
Z sources that was developed prior to the
discovery of the kilohertz QPOs, based on their
2--20~keV X-ray spectra and 1--100~Hz X-ray
variability, give the magnetic field strengths,
accretion flows, and electron scattering optical
depths required by the sonic-point model.

\subsection{Kilohertz QPO and Neutron Star Spin
Frequencies}

In the sonic-point model, the frequency of the
higher-frequency kilohertz QPO (when two are
present) is the Keplerian orbital frequency
$\nu_{\rm Ks}$ at the point in the disk flow
where the inward radial velocity increases
rapidly within a small radial distance, whereas
the frequency of the lower-frequency kilohertz
QPO is approximately the difference between
$\nu_{\rm Ks}$ and the stellar spin frequency
$\nu_{\rm spin}$. In this section we describe the
range of sonic-point Keplerian and neutron star
spin frequencies and the variation of the
sonic-point Keplerian and beat frequencies with
accretion rate expected in the model and then
compare these expectations with the observations.
We discuss the expected amplitudes of
oscillations at other frequencies in \S~4.3.

{\em Similarity of the sonic-point Keplerian QPO
frequencies in different sources}.---A key
question is why the higher-frequency kilohertz
QPOs in different sources always have
frequencies between $\sim\,$500~Hz and\break
\hbox{$\sim\,$1200~Hz}, even though the neutron
stars in these sources have time-averaged mass
accretion rates that differ by more than a
factor of 100 and magnetic fields that are
thought to differ by more than a factor of 10.
In the sonic-point model there are three main
reasons why the Keplerian-frequency QPO is
restricted to this frequency range:

(1) As discussed in \S~3.2, there is an upper
bound on the radius $R_{\rm s}$ of the sonic
point, which is set by the maximum fraction of
the angular momentum of the accreting gas that
can be removed by radiation from the stellar
surface and is $\sim 3R_{\rm ms}$, where $R_{\rm
ms}$ is the radius of the marginally stable
orbit. There is also a lower bound on $R_{\rm
s}$, which is $R_{\rm ms}$ if the radius $R$ of
the neutron star is smaller than $R_{\rm ms}$, or
$R$ otherwise. Hence, the Keplerian
frequency at the sonic point is confined to a
similar interval in the atoll and Z sources,
regardless of their accretion rates and magnetic
field strengths. As a result, if the neutron
stars in the kilohertz QPO sources have typically
accreted a few tenths of a solar mass and
therefore have masses $\sim 1.7\,\msun$, these
constraints on the sonic-point Keplerian
frequency mean that the frequencies of their
sonic-point Keplerian QPOs will fall between
$\sim\,$400~Hz and $\sim\,$1300~Hz.

(2) Modeling of the 2-20~keV X-ray spectra and
1--100~Hz power spectra of the atoll and Z sources
indicates that the magnetic fields of these
neutron stars are positively correlated with
their time-averaged accretion rates, i.e.,
sources with higher accretion rates appear to
have stronger magnetic fields (see \S~2.2 and
Psaltis \& Lamb 1998a, 1998b, 1998c). This has an
important implication for the frequency range of
the kilohertz QPOs, because the stronger the
stellar magnetic field, the {\em larger\/} the
fraction of the gas in the disk that is likely to
couple to the magnetic field at two or three
stellar radii and be channeled out of the disk
flow there. As a result, the stronger the
magnetic field, the {\em smaller\/} the mass flux
${\dot M}_{\rm sp}$ through the disk at the sonic
point. Hence, ${\dot M}_{\rm sp}$ differs by a
much smaller factor in the Z and atoll sources than
does the total mass flux ${\dot M}$ onto the
neutron star. The sonic-point Keplerian frequency
$\nu_{\rm Ks}$ is governed more by ${\dot M_{\rm
sp}}$ than by ${\dot M}$, so the positive
correlation between magnetic field and accretion
rate means that the frequency ranges of the
sonic-point QPOs are likely to be more similar in
a collection of neutron-star LMXBs than one would
expect, based only on the different ranges of
${\dot M}$ in different systems.

(3)~In addition, the inner disk is expected to
be thicker if ${\dot M_{\rm sp}}$ is large. As a
result, the radial optical depth through the disk
at a given accretion rate is smaller, which allows
radiation from the neutron star surface to
penetrate further out into the disk than one
would expect based only on the different values
of ${\dot M}$ in the different systems. This
tendency also acts to make the range of
sonic-point radii, and hence the frequency ranges
of the Keplerian-frequency QPOs, more similar in
different systems than consideration of the
accretion rates alone would suggest (see \S~3.2).

{\em Expected and inferred neutron star spin
rates}.---In the sonic-point model, the
separation between the frequencies of the QPOs
in a pair is approximately equal to the stellar
spin frequency. The frequency separations
observed in the kilohertz QPO sources so far (see
Table~\ref{table:SpinFrequencies}) all indicate
spin rates of a few hundred Hertz, consistent
with the spin rates expected in the
magnetospheric beat-frequency model of the HBO
(see \S~2, Lamb et al.\ 1985, and Ghosh \& Lamb
1992).

Oscillations have been observed during
thermonuclear X-ray bursts from five kilohertz
QPO sources so far (again see
Table~\ref{table:SpinFrequencies}). Only a
single oscillation has been observed from each
source during a given burst, the oscillations
in the tails of bursts appear to be highly
coherent (see, e.g., Smith et al.\ 1997), and the
frequencies are always the same in a given
source. Furthermore, comparison of burst
oscillations from \fu{1728$-$34} over about a
year shows that the timescale for any variation
in the oscillation frequency is $\gta3,000$~yr
(see, e.g., Strohmayer 1997). The burst
oscillations are thought to be caused by emission
from one or two regions of brighter X-ray
emission on the stellar surface (see Strohmayer
et al.\ 1997b), producing oscillations at the
stellar spin frequency or its first overtone,
respectively.

The 363~Hz frequency of the burst oscillation
observed in \fu{1728$-$34} is consistent with the
separation frequency of its two simultaneous
kilohertz QPOs. The 524~Hz and 581~Hz frequencies
of the burst oscillations seen in \ks{1731$-$260}
and\break
 \fu{1636$-$536} are probably twice the spin
frequencies of these neutron stars (see Morgan \&
Smith 1996; Smith et al.\ 1997; Wijnands \& van
der Klis 1997; Zhang et al.\ 1996, 1997). If so,
the spin rates of these stars are 262~Hz and
290~Hz, respectively. The frequencies of the
fairly coherent oscillations seen at 549~Hz in
\aql1 (Zhang et al.\ 1998) and at 589~Hz in the
so-far-unidentified source near the Galactic
center (Strohmayer et al.\ 1996d) may also be
twice the spin frequencies of these neutron
stars. These spin rates are all consistent with
those expected in the sonic-point model.

{\it Similarity of the neutron-star spin
frequencies in different sources}.---If the
neutron stars in the kilohertz QPO sources have
spin frequencies comparable to their equilibrium
spin frequencies, then we expect their spin
frequencies to be a few hundred Hertz. This can
be seen from equation~(\ref{eq:muKepler}) in
\S~2.2, which can be solved for the equilibrium
spin frequencies at which continued accretion at
the given rate leaves the spin frequency
unchanged, with the result
 \begin{equation}
 \nu_{\rm eq}
   = \left\{
   \begin{array}{ll}
   \!\!1590\,\omega_{\rm c}
   \left(\mu_{0,27}\right)^{-0.87}\!
   \left({\mdoti}/
        {\dot{M}_E}\right)^{0.39}\times
   \nonumber\\
   \hskip0.05 truein
   \times\left({M}/
        {1.4\msun}\right)^{0.85}\;{\rm Hz},
   \quad\mbox{for GPD disks;}
   \nonumber\\
   \nonumber\\
   \!\!430\,\omega_{\rm c}
   \left(\mu_{0,27}\right)^{-0.77}\!
   \left({\mdoti}/
        {\dot{M}_E}\right)^{0.23}\times
   \nonumber\\
   \hskip0.05 truein
   \times\left({M}/
        {1.4\msun}\right)^{0.70}\;{\rm Hz},
   \quad\mbox{for RPD disks.}
   \end{array}
 \right.
 \label{eq:NuEquil}
 \end{equation}
 Here $\omega_c$ is the critical fastness (Ghosh
\& Lamb 1979b). Hence, for $\omega_{\rm c}
\approx 1$, a $1.7\,\msun$ atoll source with
$\mu_{0,27}=0.5$ accreting from a GPD disk at a
rate $\mdoti=0.001\,{\dot M}_E$ has an equilibrium
spin frequency of 230~Hz, which is similar to the
240~Hz equilibrium spin frequency of a
$1.4\,\msun$ Z source with $\mu_{0,27}=2$
accreting from an RPD disk at a rate
$\mdoti=0.5\,{\dot M}_E$.

A recent analysis (Psaltis et al.\ 1998) of the
properties of the HBOs and kilohertz QPOs in a
collection of five Z sources shows that they are
consistent with many of the predictions of the
magnetospheric beat-frequency model of the HBO
and that the agreement is best if, as expected
in this model, they are all in
spin equilibrium (see also Ghosh \& Lamb 1992).
In particular, this analysis shows that {\em the
narrow range of spin frequencies observed in the Z
sources is to be expected if they are in spin
equilibrium}.

{\it Variation of kilohertz QPO frequencies with
mass accretion rate}.---In the sonic-point
model, the frequencies of the kilohertz QPOs are
expected to rise as the accretion rate
increases, within the bounds set by the orbital
frequencies at the angular momentum loss radius
and at the radius of the marginally stable orbit
(see above and \S~3.2). In order to see why an
increase is expected, suppose that the sonic
radius is at $R_{\rm s1}$ for a given accretion
rate. An increase in the accretion rate will
cause the radial optical depth from the stellar
surface to $R_{\rm s1}$ to increase, if all
other physical quantities remain constant, and
hence the sonic point will move inward  to the
smaller radius $R_{\rm s2}$ at which the optical
depth from the stellar surface is approximately
the same as before, causing an increase in the
orbital frequency at the sonic radius. The X-ray
luminosity between bursts is produced almost
entirely by accretion and hence an increase in
the accretion rate causes an increase in the
luminosity. We therefore expect a strong,
positive correlation between the frequencies of
the sonic-point Keplerian- and beat-frequency
QPOs and the persistent X-ray luminosity.

These basic ideas are illustrated by the general
relativistic accretion flow calculations of
\S~3.2. As shown in Figure~\ref{fig:nuKvsL}, the
sonic-point Keplerian and beat frequencies given
by these calculations increase steeply with
increasing mass accretion rate until the sonic
point reaches $R_{\rm ms}$, at which point their
frequencies stop changing. This flattening of the
two frequency vs.\ accretion luminosity relations
is one possible signature of the existence of a
marginally stable orbit (see also \S~5.4).
Coherent Keplerian- and beat-frequency QPOs are
unlikely to be produced if the sonic point moves
inward, close to the stellar surface, because of
the disruptive effect of the surface magnetic
field and the strong viscous shear layer that is
expected to develop if the Keplerian flow
interacts directly with the neutron star surface.

We emphasize that these calculations are intended
to be illustrative rather than to reproduce the
QPO frequency behavior of any particular source.
If the mass of the neutron star is $1.7\,\msun$,
rather than $1.4\,\msun$ as assumed in these
calculations, the limiting frequency would be
1.3~\kHz rather than 1.6~\kHz (see Miller, Lamb,
\& Psaltis 1998). Again, if the accretion disk is
thicker near the star than is assumed in these
calculations, more of the inner disk will be
illuminated and hence the angular momentum loss
radius $R_{\rm aml}$ will be larger, causing the
sonic-point Keplerian frequency at a given
accretion rate to be lower. If the disk thickness
changes with accretion rate, then the variation
of the sonic-point Keplerian frequency with
accretion rate will be different from the
variation found in the calculations reported
here, which treat the disk thickness as constant.
Despite these uncertainties, the calculations
described here show that an increase in the mass
accretion rate leads naturally to an increase in
the sonic-point Keplerian frequency.

We stress that caution must be used in
comparing the predicted variation of QPO frequencies with
accretion with the
countrate measured by a given X-ray detector,
because countrates are known to be poor
indicators of the mass accretion rate and the
accretion luminosity, at least for some sources
at some times. The most likely reason is that a
particular instrument measures the photon number
flux over only a restricted energy range and
that a change in the mass accretion rate
typically causes a change in the shape of the
X-ray spectrum as well as its normalization.
Hence the detector countrate typically is not
proportional to the accretion rate.

In the Z sources, the mass accretion rate is
known to be different when the \exosat\
countrate is the same but the source is on a
different branch of the Z track (Hasinger \&
van der Klis 1989; Lamb 1989; Psaltis et al.\
1995). Moreover, in the ``Cyg-like'' Z sources,
the normal/flaring branch vertex, which is
thought to correspond to a mass accretion rate
approximately equal to $\mdote$ (Lamb 1989),
occurs at different countrates during different
observations (Kuulkers 1995).

The observations of Ford et al.\ (1997a) show
that the properties of the accretion flow are not
uniquely related to the X-ray countrate in the
atoll sources. In these observations, the
dependence of the kilohertz QPO frequencies on
countrate seen in data on \fu{0614$+$091} taken
with \rxte\ in 1996 August differs, both in
slope and in normalization, from the dependence
of the QPO frequencies on countrate seen in data
taken in 1996 April. In contrast, when the
\fu{0614$+$091} QPO frequencies are plotted
against the energy of the spectral peak, which
may be a good indicator of the accretion rate
(Psaltis \& Lamb 1998c), the slope and
normalization of the curves are the same for
both observations (Ford et al.\ 1997b).

Clearly, bolometric and other corrections are
typically important, so comparison of QPO
frequency versus countrate data with the QPO
frequency versus accretion rate predictions of
theoretical models must be approached with
caution. Despite this caveat, it is worth
emphasizing that there is a strong, positive
correlation between the frequencies of the
kilohertz QPOs seen in \fu{1728$-$34},
\fu{0614$+$091}, \fu{1820$-$30}, and
KS~1731$-$260, and the countrate measured by
\rxte, as expected in the sonic-point model if
the countrate increases with increasing accretion
rate.

It is interesting to speculate about the
dependence of kilohertz QPO frequencies on
luminosity that is to be expected if such
QPOs are detected during a thermonuclear X-ray
burst. In this case the mass accretion rate and
the radial optical depth are no longer tied to
the luminosity, and hence the expected
dependence of $\nu_{\rm Ks}$ on luminosity is
different from what is expected during the
periods between type~I X-ray bursts. When
radiation forces are extremely strong, such as
at the peak of a type~I X-ray burst that causes
photospheric radius expansion, the radiation
controls the accretion flow out to large radii
and we therefore do not expect any QPO at the
sonic-point Keplerian frequency or at the
sonic-point beat frequency. However, QPOs at
these frequencies may reappear in the tail of
the burst. While the burst luminosity is several
times the persistent accretion luminosity, we
expect the sonic point to be farther from the
star than when the luminosity is lower. Thus, we
expect that if sonic-point QPOs are detected in
the tail of a thermonuclear burst, their
frequencies will increase as the luminosity
declines. Note, however, that during the decay
phase of such a burst, the luminosity changes
rapidly. If the frequencies of the QPOs track
this change, the QPO peaks in power spectra will
be smeared out, unless the spectra are
constructed from short ($\lta 0.1$~s) segments
of data.

So far, no kilohertz QPOs have been observed
near the maximum of type-I X-ray bursts, which is
consistent with what is expected in the
sonic-point model. More sensitive searches will
be required to determine if kilohertz QPOs are
present during the decay phase of bursts.

\subsection{Kilohertz QPO Amplitudes and Coherence}

{\em Amplitudes of oscillations at different
frequencies}.---A key question that any theory of
the kilohertz QPOs must answer is why at most two
strong QPOs have so far been seen simultaneously
in the kilohertz QPO sources. In addressing this
question, it is important to consider both the
{\em generation\/} of oscillations near the
neutron star and the effects of {\em
propagation\/} of the radiation through the gas
surrounding the neutron star. In order to be
detectable far away, an oscillation must be
strong at the source or have a small attenuation,
or both.

The QPOs at the sonic-point Keplerian and beat
frequencies are the only QPOs expected to have
moderately high amplitudes far from the neutron
star because the optical depth and luminosity
oscillations that produce them generate relatively
high amplitudes near the star, because these
frequencies (and their overtones) are the {\em
only\/} frequencies generated by the sonic-point
mechanism in lowest order, and because scattering
of photons by the gas surrounding the neutron
star selectively suppresses the already weak
higher-frequency oscillations generated at higher
orders.

As discussed in \S~3.3, to lowest (first) order,
the sonic-point mechanism generates oscillations
only at the sonic-point Keplerian and beat
frequencies (see Table~\ref{table:FreqGen}).
As explained there, only very weak overtones
of the sonic-point Keplerian frequency are likely
to be generated, because the angular distribution
of the radiation from the bright impact
footprints of the rotating density pattern
produced by clumps is expected to be very broad.

The radiation pattern that rotates with the star
is also expected to be broad, but at the sonic
radius, where it interacts with the orbiting gas
to generate the luminosity oscillation at the
sonic-point beat frequency, it is not likely to
be perfectly sinusoidal. Hence overtones of the
sonic-point beat frequency $\nu_{\rm Bs}$ may be
generated in the inward mass flux from the
clumps at the sonic radius, producing overtones
of $\nu_{\rm Bs}$ in the X-ray flux from the
star. As radiation from the stellar surface
propagates outward through the part of the
central corona that extends beyond the sonic
radius, the radiation pattern that rotates with
the star is likely to be broadened further (see
below), reducing the amplitudes of oscillations
at the stellar spin frequency and its overtones
seen by a distant observer.

In addition to the QPOs at the sonic-point
Keplerian and beat frequencies and their
overtones, in second order the sonic-point
mechanism also generates QPOs at $\nu_{\rm spin}$
and $2\nu_{\rm Ks}-\nu_{\rm spin}$ and in third order at
$\nu_{\rm Ks}-2\nu_{\rm spin}$, $\nu_{\rm
Ks}+\nu_{\rm spin}$, $3\nu_{\rm Ks}-2\nu_{\rm
spin}$, and $3\nu_{\rm
Ks}-\nu_{\rm spin}$ (again see
Table~\ref{table:FreqGen}). These high-order
oscillations are expected to be very weak, even
near the neutron star.

The attenuation of an oscillation depends
strongly on whether it is a {\it beaming}
oscillation or a {\it luminosity} oscillation
(see \S~3.6). A beaming oscillation is an
oscillation produced by rotation of an angular
radiation pattern, like the beam of a lighthouse;
in a pure beaming oscillation, the luminosity
remains constant as the radiation pattern
rotates. In contrast, a luminosity oscillation is
one produced by an oscillation of the luminosity
of the source; in a pure luminosity oscillation,
the radiation pattern remains static as the
luminosity oscillates. In the sonic-point model,
the sonic-point Keplerian frequency is the only
intrinsically strong beaming oscillation, and the
sonic-point beat frequency is the only luminosity
oscillation produced by the accretion flow.

In passing through a scattering region,
luminosity  oscillations are attenuated only by
time-of-flight smearing whereas beaming
oscillations are also attenuated by the
isotropization of the radiation pattern by the
scattering. Hence beaming oscillations are
weakened more by propagation through a scattering
corona. Figure~\ref{fig:AttenFactor} compares the
attenuation of luminosity and beaming
oscillations produced at the center of a uniform,
spherical, scattering cloud of radius $R_{\rm C}
= 3\ee6$~cm, as a function of the optical depth
of the cloud. The size of this cloud is
comparable to the dimensions of the small
scattering coronae with optical depths
$\sim\,$3--5 that are thought to surround the
neutron stars in the atoll and Z sources (Lamb
1989, 1991; see also \S~2.2). The frequencies of
the oscillations shown in
Figure~\ref{fig:AttenFactor} have been chosen to
represent a hypothetical accreting neutron star
with $\nu_{\rm Ks}=\,1100$~Hz and $\nu_{\rm
spin}=\,300$~Hz. The beat-frequency luminosity
oscillation is therefore at 800~Hz.  The
attenuation factors have been calculated using 
equations~(\ref{eq:LumAtten}) and
(\ref{eq:BeamAtten}).
Figure~\ref{fig:AttenFactor} shows that the
beaming oscillation at the sonic-point Keplerian
frequency must have a relatively high intrinsic
amplitude in order to produce a strong, 
observable QPO for a distant observer, whereas the
luminosity oscillation at the sonic-point beat
frequency needs only to have a moderate 
amplitude.

The amplitudes of the oscillations at the
overtones of either luminosity or beaming
oscillations seen by a distant observer are
expected to be far smaller than the amplitudes
seen at the fundamental frequencies, for several
reasons. For example, beaming patterns with
multiple lobes are much more  easily isotropized
by scattering than are beaming patterns  with
single lobes (see, e.g., Brainerd \& Lamb
[1987]). Moreover, for either beaming or
luminosity oscillations, overtones of an
oscillation are at higher frequencies than the
fundamental and are thus more susceptible to
time-of-flight smearing.

As a specific example, we consider
\fu{1728$-$34}. As interpreted within the
sonic-point model, the spin frequency of this
neutron star is 363~Hz (see Table~3). During one
observation, $\nu_{\rm Ks}$ was 1045~Hz. Assuming
that this neutron star is surrounded by a central
corona with a radius $R_{\rm C}=3\ee6$~cm and an
optical depth $\tau=5$, the amplitude of the
luminosity oscillation at the beat frequency
$\nu_{\rm Ks}-\nu_{\rm spin} = 682~\Hz$ seen by a
distant observer is about 85\% of the amplitude
at the neutron star. For comparison, the beaming
oscillation at $\nu_{\rm Ks}+\nu_{\rm spin} =
1408~\Hz$, which is generated only in third order
and is therefore likely to be very weak even near
the neutron star, has an amplitude at infinity
which is only about 20\% of its amplitude at the
star. The amplitudes of oscillations at the
overtones of the sonic-point Keplerian frequency
are all reduced by factors $\gta$20. For these
reasons, it is not surprising that the only
strong QPOs seen are at the sonic-point
Keplerian and beat frequencies.

Although oscillations at frequencies other than
the sonic-point Keplerian and beat frequencies
are likely to be very weak, their amplitudes are
surely not zero. We therefore expect that, if
neutron-star LMXBs are observed with sufficient
sensitivity, oscillations at other frequencies
will be detected. In particular, a careful search
near the spin frequency, at the first overtone of
the beat frequency, and near the sum of the
Keplerian and spin frequencies in sources with
pairs of high-frequency QPO peaks may reveal very
weak QPOs at these frequencies.

The oscillation at the beat frequency is a {\em
luminosity\/} oscillation, but in the sonic-point
model it is created by interaction of the weakly
{\em beamed\/} radiation pattern that is rotating
at the stellar spin frequency with the clumps of
gas that are orbiting at or near the sonic point.
It is therefore important to emphasize why the
beaming oscillation at the stellar spin frequency
is strong enough to modulate significantly the
mass inflow rate from clumps at the sonic point,
yet too weak to produce a significant peak in
power spectra constructed by a distant observer.

Previous modeling of the X-ray spectra (Lamb
1989, 1991; Psaltis et al.\ 1995; Psaltis \& Lamb
1998a, 1998b, 1998c) and the 1--100~Hz X-ray
variability of the atoll and Z sources (Lamb
1989, 1991; Miller \& Lamb 1992) strongly
indicates that a hot, central, Comptonizing
corona surrounds the neutron star, extending
several stellar radii from its surface (see
\S~2.2). The sonic radius is well inside this
corona, so the scattering optical depth from the
stellar surface to the sonic point is much less
than the scattering optical depth from the
stellar surface to infinity. The finite size of
the neutron star also diminishes the effect of
attenuation on the amplitude of X-ray brightness
oscillations for gas orbiting close to the star.
These factors make the brightness oscillation
produced by the radiation pattern that rotates
with the star much stronger at the sonic radius
than it is far away.

{\em Dependence of QPO amplitudes on accretion
rate}.---In the sonic-point model, the ratio of
the amplitude of the QPO at the sonic-point
Keplerian frequency to the amplitude of the QPO
at the sonic-point beat frequency generally
depends on the strength of the neutron star
magnetic field. If the magnetic field is too weak
to channel gas even near the stellar surface, the
amplitude of the oscillation at the beat
frequency will be too small to be detected (see
\S~3.1). However, the ratio of the amplitudes of
the QPOs at different frequencies is also
expected to depend on the mass accretion rate.

For example, as the mass accretion rate rises,
the electron density and optical depth of the gas
around the neutron star also rise. Hence, the
amplitude at infinity of the beaming oscillation
at the sonic-point Keplerian frequency will fall
faster relative to the amplitude of the
luminosity oscillation at the sonic-point beat
frequency, all else being equal; indeed, the QPO
at the Keplerian frequency may become undetectable
while the QPO at the beat frequency remains strong.
This expectation is consistent with the observed
behavior of the kilohertz QPO pair in
\fu{1728$-$34} (Strohmayer et al.\ 1996c),
assuming that the increasing mass accretion rate
causes an increase in the countrate (but see
\S~4.2). In this source, the amplitude of the
higher-frequency QPO decreases relative to the
amplitude of the lower-frequency QPO as the
countrate increases. 

Alternatively, if the sonic
point moves far enough away from the stellar
surface, as may happen at low accretion rates
(and hence countrates), the optical depth through
the flow from the stellar surface to the sonic
radius may become large enough to reduce greatly
the amplitude at the sonic point of the radiation
pattern that rotates with the star, thereby
suppressing the luminosity oscillation at
$\nu_{\rm Bs}$. This is consistent with an
observation of \fu{1728$-$34} in which the
lower-frequency QPO is not observed when the
frequency of the higher-frequency QPO is low
(Strohmayer et al.\ 1996c).

This analysis shows that if only a {\em single\/}
high-frequency QPO peak is observed, it could be
either the sonic-point Keplerian-frequency QPO or
the sonic-point beat-frequency QPO, depending on
the magnetic field of the neutron star and the
mass accretion rate at the time. In this case, a
secure identification can only be made by
considering other properties of the QPO and
comparing them with the properties of the QPOs
seen in other observations of the same source.

{\em Dependence of QPO amplitudes on photon
energy}.---In the sonic-point model, the relative
amplitudes of the Keplerian frequency and beat
frequency QPOs are expected to increase steeply
with photon energy over the $\sim\,$5--10~keV
energy range (see \S~3.5). This is because the
optical depth is expected to oscillate at both
frequencies: the optical depth along the line of
sight from the stellar surface oscillates at the
Keplerian frequency, while the total optical
depth of the scattering region oscillates at the
beat frequency as the density of the accreting
gas falling on the stellar surface oscillates at
this frequency. As shown in \S~3.5, a modest
oscillation in the optical depth produces a QPO
with a large relative amplitude at high photon
energies. This is consistent with the steep
increase of QPO amplitude with increasing photon
energy observed in the higher-frequency of the two
simultaneous QPOs in \fu{1636$-$536}
(Zhang et al.\ 1996), \fu{0614$+$091} (Ford  et
al.\ 1997b), and KS~1731$-$260 (Wijnands \& van
der Klis  1997), in the lower-frequency of the
two simultaneous QPOs in \fu{1608$-$52}
(Berger et al.\ 1996; M\'endez et al.\ 1998) and
\fu{1728$-$34} (Strohmayer et al.\ 1996c), and
in the kilohertz QPOs in the Z sources
\gx{5$-$1} (van der Klis et al.\ 1996e),
\gx{17$+$2} (Wijnands et al.\ 1997c), and
\cyg{2} (Wijnands et al.\ 1998).

In \S~3.5 we used a simple analytical model to
derive an expression for the photon energy
dependence of the rms amplitude of a QPO produced
by oscillations in the optical depth and in the
injection rate of soft photons.
Figure~\ref{fig:AmpVsEnergy} compares the results
of a more detailed numerical calculation
performed using the algorithm of Miller \& Lamb
(1992) with amplitude data for the
lower-frequency kilohertz QPO observed in
\fu{1608$-$52} and the higher-frequency kilohertz
QPO seen in \fu{1636$-$536}. In performing this
calculation we made the same assumptions as in
\S~3.5 and assumed further that the injection
rate of soft photons is constant in time, that
the spectrum of the injected photons is a
blackbody at temperature $kT=0.6$~keV, that the
electron temperature in the central corona is
$kT_e=9$~keV, and that the optical depth varies
from $\tau=3$ to $\tau=3.3$ during an
oscillation. The model matches the data well. The
geometry of the actual upscattering region is
undoubtedly much more complicated than assumed in
this calculation, but the excellent correspondence
with the data and the ubiquity of the steep
increase of amplitude with photon energy over the
5--10~keV energy range suggest that the model has
many elements in common with the true physical
situation.

As a source moves in the X-ray color-color
diagram and its spectrum changes (implying a
change in the average optical depth and electron
temperature of the Comptonizing region), we expect
that the photon energy dependence of the rms
amplitude of the oscillations will also change.
Moreover, as discussed in \S 3.5, we expect that,
in general, the dependence of the beat-frequency
QPO amplitude on photon energy will be different
from the dependence of the Keplerian-frequency
QPO amplitude. This is consistent with the
observations of \fu{0614$+$091} (Ford et al.\
1997b) in which amplitude versus photon energy
curve of the beat-frequency QPO has a minimum at
$\sim\,$10~keV, whereas the amplitude versus
photon energy curve of the Keplerian-frequency
QPO increases monotonically from 2~keV to 20~keV.
Because the QPOs amplitude depend strongly on
photon energy, to be meaningful a comparison of
QPO amplitudes measured for different sources or
at different times for the same source must
consider the same range of photon energies.

{\em Coherence of the kilohertz QPOs}.---In
\S~3.4 we showed that the high coherence of the
Keplerian and beat-frequency QPOs in the
sonic-point model is primarily a consequence of
the extremely sharp increase in the inward radial
velocity near the sonic point, which maps a small
range of orbital frequencies onto the stellar
surface. There we also considered the decoherence
caused by destruction of clumps by turbulent
dissipation within the disk flow, by advection to
the stellar surface, and by decay as gas is
stripped from them by the supersonic flow at the
sonic point. We concluded that the QPO peaks can
be as narrow as observed ($\nu/\delta\nu \sim
30$--200) for reasonable conditions in the
accretion flow.

The FWHM $\delta\nu_{\rm BF}$ of the sonic-point
beat frequency oscillation is expected to be
comparable to the FWHM $\delta\nu_{\rm KF}$ of
the sonic-point Keplerian frequency oscillation,
because the beat-frequency QPO is produced by the
beat of a nearly periodic signal at frequency
$\nu_{\rm spin}$ against the Keplerian frequency
(see  \S~3.4). We do not expect the FWHM of the
two kilohertz QPO peaks to be identical, however,
because there are processes that can affect the
FWHM of one but not the other. For example, the
FWHM of the beat-frequency peak depends in part
on the range of radii over which radiation forces
can affect the mass accretion rate significantly.
This range can be either less than or greater
than the range of radii over which the inward
radial velocity increases rapidly, and hence
$\delta\nu_{\rm BF}$ can be either less than or
greater than $\delta\nu_{\rm KF}$. This effect
can also displace the centroid of the
beat-frequency peak relative to the centroid of
the Keplerian frequency peak, so that the
observed frequency difference $\nu_{\rm
Ks}-\nu_{\rm Bs}$ is close to, but not exactly
equal to, the stellar spin frequency.

\subsection{Amplitudes of QPOs Produced by Stars
with Different Magnetic Field Strengths}

In the sonic-point model, the larger the
magnetosphere, the smaller the fraction of the
accreting gas that reaches the surface of the
star without coupling to the magnetic field, and
hence the smaller the amplitude of the kilohertz
QPOs. Therefore, we expect the rms amplitudes of
the kilohertz QPOs to be roughly anticorrelated
with the strength of the stellar magnetic field,
if all other physical quantities remain fixed. As
discussed in \S~2.2, the ``4U'' atoll sources are
thought to have the weakest magnetic fields, the
``GX'' atoll sources and the ``Sco-like'' Z
sources are thought to have somewhat stronger
fields, and the ``Cyg-like'' Z sources are thought
to have the strongest fields (Psaltis \& Lamb
1998a, 1998b, 1998c).  Hence, we expect kilohertz
QPOs to be common and strong in the ``4U'' atoll
sources, but weak or even undetectable with
current instruments in the ``GX'' atoll sources
and the Z sources.\footnote{After this specific
prediction was made in the originally submitted
and circulated version of the present paper,
kilohertz QPOs with very low amplitudes were
detected in all six of the originally identified
Z sources; see Table~\ref{table:ReportedQPOs}.
To date, kilohertz QPOs have not been detected
in any of the GX atoll sources.} This is indeed the
case, as is evident from 
Table~\ref{table:ReportedQPOs}. If, as indicated
by the spectral modeling discussed in \S~2, the
magnetic field of Cir~X-1 is weak, we expect it
to exhibit sonic-point QPOs when it is in its
low state. We also expect that kilohertz QPOs
will be undetectable by current instruments in
any sources that have strong magnetic fields and
therefore produce strong, periodic oscillations
at their spin frequencies.

The relation between kilohertz QPO amplitude and
magnetic field strength may be made
semi-quanti\-tative using the spectral
calculations (see \S~2.2) of Psaltis et al.\ (1995)
and Psaltis \& Lamb (1998a, 1998b, 1998c). These
calculations indicate that LMXBs containing 
neutron stars with weaker magnetic fields
generally have larger hard X-ray colors (e.g.,
the ratio of the 7--20~keV countrate to the
5--7~keV countrate). Based on this physical
picture, we expect the rms amplitudes of the
kilohertz QPOs produced by the sonic-point
mechanism to be higher in sources with larger
hard colors. This trend is evident in
Figure~\ref{fig:AmpVsHardColor}, which shows the
rms amplitudes versus the hard X-ray colors of
four sources observed with the PCA detector
onboard the \rxte satellite. If other sources
with kilohertz QPOs follow this same trend, this
will be strong support for the sonic-point model.

As discussed in \S~2.3, the observed
anticorrelation between kilohertz QPO amplitude
and magnetic field strength is also strong
evidence that the magnetospheric beat-frequency
mechanism {\em not\/} the correct explanation of
these oscillations.

\section{IMPLICATIONS FOR NEUTRON STARS AND DENSE
MATTER}

In the sonic-point model, the higher-frequency
QPO in a kilohertz QPO pair has a frequency
equal to the orbital frequency of gas near the
sonic point; the relatively high coherence of
this QPO is a consequence of the fact that the
gas clumps that generate it are in nearly
circular orbits. In this section we show that
the inferred existence of a nearly circular
orbit around a neutron star with a frequency in
the kilohertz range can be used to derive
interesting new upper bounds on the mass and
radius of the star and constraints on the
equation of state of the dense matter in all
neutron stars. For simplicity we discuss first
the case of a nonrotating star around which the
radial component of the radiation force is
negligible. We then consider the changes in the
mass and radius constraints caused by
frame-dragging when the star is spinning and by
the radial component of the radiation force. We
show that rotation at $\sim 300$~Hz typically
increases the bound on the mass by $\sim 20$\%
and the bound on the radius by a few percent. In
contrast, the radiation force {\em reduces\/}
the upper bounds on the masses and radii by a
few percent or less in the atoll sources but
perhaps by much larger percentages in the Z
sources.

Observations of kilohertz QPOs may be able to
establish the existence of an innermost stable
circular orbit around some neutron stars (see
\S~4.2). If this can be accomplished, it would be
the first evidence concerning a prediction of
general relativity in the strong-field regime.
If furthermore the frequency of a particular
kilohertz QPO can be securely established as the
orbital frequency at the radius of the innermost
stable circular orbit and if the spin frequency
of the neutron star can be determined, then the
frequency of the QPO can be used to fix the mass
of the neutron star for each assumed equation of
state, tightening the constraints on the
properties of dense matter and possibly ruling
out many currently viable equations of state. We
discuss how this can be done and the evidence
that would signal detection of a QPO with the
orbital frequency of the marginally stable orbit.

\subsection{Nonrotating Star}

Suppose that the frequency of the
higher-frequency QPO in a kilohertz pair is
$\nu_{\rm QPO2}$ and that, as in the sonic-point
model, $\nu_{\rm QPO2}$ is the orbital frequency
of gas in a nearly circular Keplerian orbit
around the neutron star. Assume for now that the
star is not rotating and is spherically
symmetric. Then the exterior spacetime is the
Schwarzschild spacetime. In this spacetime, the
orbital frequency (measured at infinity) of gas
in a circular orbit at Boyer-Lindquist radius
$r$ around a star of mass $M$ is (see
eq.~[\ref{eq:SonicKeplerFreq}])
 \begin{equation}
 \nu^\nonrot_{\rm K}(M,r)=(1/2\pi)(GM/r^3)^{1/2}\;.
 \label{eq:KeplerFreqZero}
 \end{equation}
 Here and below the superscript zero indicates
that the relation is that for a nonrotating
\hbox{($j=0$)} star. If the mass of the star is
known, equation~(\ref{eq:KeplerFreqZero}) may be
solved for the orbital radius $R_{\rm orb}$ where
the Keplerian frequency is $\nu_{\rm QPO2}$, with
the result
 \begin{equation}
 R^\nonrot_{\rm orb}(M,\nu_{\rm QPO2}) =
 (GM/4\pi^2\nu_{\rm QPO2}^2)^{1/3}\;.
 \label{eq:OrbitalRadiusZero}
 \end{equation}
 Conversely, if the orbital radius of the gas is
known, equation~(\ref{eq:KeplerFreqZero}) may be
solved for the mass of the star that gives an
orbital frequency equal to $\nu_{\rm QPO2}$, with
the result
 \begin{equation}
 M^\nonrot(R_{\rm orb},\nu_{\rm QPO2}) =
 (4\pi^2/G) R_{\rm orb}^3 \nu_{\rm QPO2}^2\;.
 \label{eq:OrbitalMassZero}
 \end{equation}
 If, as is so far the case for the kilohertz QPO
sources, neither $M$ nor $R_{\rm orb}$ are known,
equations~(\ref{eq:OrbitalRadiusZero})
and~(\ref{eq:OrbitalMassZero}) do not determine
$R_{\rm orb}$ or $M$ but do establish a relation
between them. In the radius-mass plane, this
relation is a curve that begins at the origin and
rises up and to the right.

As a specific example, the dashed curve marked
$M^\nonrot(R_{\rm orb})$ in
Figure~\ref{fig:PieSlice} shows the relation given by
equation~(\ref{eq:OrbitalMassZero}) for
$\nu_{\rm QPO2}=1220$~\Hz, a value of
$\nu_{\rm QPO2}$ observed in \fu{1636$-$536}
(W.~Zhang, personal communication). Hence, if
\fu{1636$-$536} were not rotating, the mass of
the neutron star in this source and the orbital
radius of the gas clumps producing the QPO
during this observation would have to correspond
to some point along this dashed curve.

Consider now the constraints on the neutron star
mass and radius that follow from the {\em
frequency\/} of the higher-frequency QPO.
Obviously, in order for gas to be in orbit, it
must be outside the star. This means that for an
orbit of given radius $R_{\rm orb}$, the mass $M$
of the star must be greater than
$M^\nonrot(R_{\rm orb},\nu_{\rm QPO2})$;
alternatively, for a star of given mass $M$, the
radius $R$ of the star must be less than
$R^\nonrot_{\rm orb}(M,\nu_{\rm QPO2})$. Thus,
the point that represents the mass and radius of
the neutron star must lie above the curve
$M^\nonrot(R_{\rm orb},\nu_{\rm QPO2})$ in the
radius-mass plane. The larger the value of
$\nu_{\rm QPO2}$, the higher the curve, so the
most stringent---and hence the
relevant---constraint on the mass and radius of
the neutron star in a particular source is given
by the {\em highest\/} value of $\nu_{\rm QPO2}$
ever observed in that source, which we denote
$\nu_{\rm QPO2}^\ast$. For example, 1220~Hz is
the highest value of $\nu_{\rm QPO2}$ seen so
far in \fu{1636$-$536}, so this is the current
value of $\nu_{\rm QPO2}^\ast$ for this source
(in fact, this is the highest value of $\nu_{\rm
QPO2}$ seen so far in {\em any\/} source).
Consequently, if the neutron star in
\fu{1636$-$536} were not rotating, the point in
the radius-mass plane that represents its mass
and radius would have to lie above the curve
\hbox{$M = M^\nonrot(R_{\rm orb}, 1220\,\Hz)$},
which is the dashed curve shown in
Figure~\ref{fig:PieSlice}.

So far, the only information we have used to
constrain the mass and radius of the star is the
frequency of the higher-frequency QPO in a
kilohertz pair. However, we have the additional
information that the {\em coherence\/} of the
higher-frequency QPO is high (\hbox{$\nu_{\rm
QPO}/ \delta\nu_{\rm QPO} \sim 100$}) in many
sources. Let us assume that the higher-frequency
QPO in a pair is generated by the same mechanism
in all sources that show such pairs. Then the
high coherence of these higher-frequency QPOs
imposes additional constraints on the neutron
star's mass and radius because, in order to
produce a QPO with such high coherence, the gas
that generates the QPO must be in a nearly
circular orbit. Hence $R_{\rm orb}(M,\nu_{\rm
QPO2}^\ast)$ must be greater than the radius
$R_{\rm ms}(M)$ of the innermost stable circular
orbit, because gas inside $R_{\rm ms}$ spirals
quickly inward to the stellar surface.

For a nonrotating star, the radius of the
innermost stable circular orbit is a function
only of the mass of the star and is given by
\hbox{$R^\nonrot_{\rm ms}(M) = 6\,GM/c^2$}.
Inverting this relation gives
 \begin{equation}
 M^\nonrot(R_{\rm ms}) = (c^2/6G)R_{\rm ms}\;.
 \label{eq:msMassZero}
 \end{equation}
 In the radius-mass plane,
relation~(\ref{eq:msMassZero}) is a straight line
of slope $+1$ through the origin and is the
dotted line marked $M^\nonrot(R_{\rm ms})$ in
Figure~\ref{fig:PieSlice}. This line intersects
$M^\nonrot(R_{\rm orb},\nu_{\rm QPO2}^\ast)$ at
the mass and radius values
 \begin{eqnarray}
  M^\nonrot_{\rm max} &\equiv&
  c^3(\sqrt{864}G\pi\nu_{\rm QPO2}^\ast)^{-1}
  \nonumber\\ \nonumber\\
  &=&2.2\,(1000~\Hz/\nu_{\rm QPO2}^\ast)\;\msun\;,\\
  \label{eq:MassBoundZero}
 \end{eqnarray}
 and
 \begin{eqnarray}
  R^\nonrot_{\rm max} &\equiv&
  c(\sqrt{24}\pi\nu_{\rm QPO2}^\ast)^{-1}
  \nonumber\\ \nonumber\\
  &=&19.5\,(1000~\Hz/\nu_{\rm QPO2}^\ast)\;{\rm km.}
  \label{eq:RadiusBoundZero}
 \end{eqnarray}
 If $M$ were larger than $M^\nonrot_{\rm max}$,
the orbital radius\break
 $R^\nonrot_{\rm
orb}(M,\nu_{\rm QPO2}^\ast)$ of the gas
generating the QPO would be less than
$R^\nonrot_{\rm ms}(M)$, so $M^\nonrot_{\rm
max}$ is an upper bound on the mass of the star;
$M^\nonrot_{\rm max}$ is inversely proportional
to $\nu_{\rm QPO2}$ but independent of the
star's radius and the orbital radius of the gas
clumps that are producing the higher-frequency
QPO.  Similarly,
$R^\nonrot_{\rm max}$ is an upper bound on the
radius of the star; $R^\nonrot_{\rm max}$ is
also inversely proportional to $\nu_{\rm QPO2}$
but independent of the star's mass and the
orbital radius of the gas clumps that are
producing the higher-frequency QPO. 

The heavy horizontal line plotted in
Figure~\ref{fig:PieSlice} shows the upper bound
on the mass of a nonrotating star for $\nu_{\rm
QPO2}^\ast=1220~\Hz$. This upper bound is
$1.8\,\msun$, so the mass of the neutron star in
\fu{1636$-$536} would have to be less than this
if it were not rotating. For
$\nu_{\rm QPO2}^\ast=1220~\Hz$, $R^\nonrot_{\rm
max}$ is 16.0~km, so the radius of the neutron
star in \fu{1636$-$536} would have to be be less
than this if the star were not rotating.

Suppose now that the frequency $\nu_{\rm
QPO2}^\ast$ of a particular QPO is securely
identified as the orbital frequency of gas in
the innermost stable circular orbit around a
particular neutron star. Then $R_{\rm orb} =
R_{\rm ms}$ for this QPO, so the representative
point of the orbit is at the intersection of the
diagonal $M^\nonrot(R_{\rm ms})$ line and the
$M^\nonrot(R_{\rm orb},\nu_{\rm QPO2}^\ast)$
curve. The mass of the star is therefore
$M^\nonrot_{\rm max}$. Hence {\em identification
of a QPO frequency with the frequency of the
innermost stable circular orbit immediately
determines the mass of the star}. The radius of
the star is not determined by such an
identification, but must still be less than
$R^\nonrot_{\rm max}$.

As a specific example, suppose that the 1220~\Hz
QPO observed in \fu{1636$-$536} is securely
identified as the orbital frequency of gas in
the innermost stable circular orbit around this
neutron star. Then the mass of this neutron star
would be determined as 1.8\,\msun, if it were not
rotating.

{\em These arguments apply to stars with
arbitrary spin rates as well as to nonrotating
stars}, although the expressions for $R_{\rm
orb}(M,\nu_{\rm QPO2}^\ast)$ and $R_{\rm ms}(M)$
are different for a rotating star and depend on
the star's spin rate as well as its mass. They
may be summarized as follows:

(1)~Stellar radii \hbox{$R>R_{\rm
orb}(M,\nu_{\rm QPO2}^\ast)$} are excluded,
because there is no Keplerian orbit with
frequency $\nu_{\rm QPO2}^\ast$ outside a star
with such a large radius; the value of $R_{\rm
ms}(M)$ is irrelevant.

(2)~If $R<R_{\rm orb}(M,\nu_{\rm QPO2}^\ast)$
but $R_{\rm ms}(M) >$\break$R_{\rm
orb}(M,\nu_{\rm QPO2}^\ast)$, there is a
Keplerian orbit with frequency $\nu_{\rm
QPO2}^\ast$ outside the star, but any
oscillation produced by gas in this orbit would
have a coherence much lower than that observed.

(3)~If $R<R_{\rm orb}(M,\nu_{\rm QPO2}^\ast)$
and $R_{\rm ms}(M) <$\break$R_{\rm
orb}(M,\nu_{\rm QPO2}^\ast)$, there is a
Keplerian orbit with frequency \hbox{$\nu_{\rm
QPO2}^\ast$} outside the star and the
oscillation produced by gas in this orbit can
have the required high coherence.

(4)~If the frequency $\nu_{\rm QPO2}^\ast$ of a
particular QPO is securely identified as the
orbital frequency of gas in the innermost stable
circular orbit around a particular neutron star,
the mass of the star is $M^\nonrot_{\rm max}$;
its radius must be less than $R^\nonrot_{\rm
max}$.

For a nonrotating star with \hbox{$\nu_{\rm
QPO2}^\ast = 1220~\Hz$}, the combinations of
stellar mass and radius allowed by condition~(3)
are the points in the radius-mass plane above the
curve \hbox{$M=M^\nonrot(R_{\rm orb},1220\,\Hz)$}
but below \hbox{$M=M^\nonrot_{\rm max}$}. This
region is outlined in Figure~\ref{fig:PieSlice}
by the heavy solid line. The point representing
the mass and radius of the neutron star in
\fu{1636$-$536} would have to lie in this region
if the star were not rotating. The allowed region
collapses to the heavy solid horizontal line if
the QPO frequency is identified as the orbital
frequency of gas in the innermost stable
circular orbit so that condition~(4) applies.

Figure~\ref{fig:EOSconstraints} compares the
mass-radius relations for nonrotating neutron
stars given by five equations of state for
neutron-star matter ranging from soft to hard
with the regions of the radius-mass plane
allowed if $\nu_{\rm QPO2}^{\ast}$ is a
Keplerian orbital frequency, for nonrotating
stars and three values of $\nu_{\rm
QPO2}^{\ast}$. In order to make possible
comparisons with previous studies of neutron
star properties (see, e.g, Pethick \& Ravenhall
1995), we show the mass-radius curve given by the
early Friedman-Pandharipande-Skyrme (FPS)
realistic equation of state (Friedman \&
Pandharipande 1981; Lorenz, Ravenhall, \&
Pethick 1993). The FPS equation of state uses a
different approach, but is similar to the
softest equations of state permitted by modern
realistic models of the nucleon-nucleon
interaction (Pandharipande, Akmal, \& Ravenhall
1998). These equations of state all give maximum
gravitational masses of about $1.8\,\msun$ for
nonrotating stars.

As an example of the mass-radius curves given by
later realistic equations of state, we show the
mass-radius curve predicted by the UU equation
of state (Wiringa, Fiks, \& Fabrocini 1988). Although
it is based on older scattering data, the UU
equation of state is similar to the recent ${\rm
A18} + {\rm UIX'} + \delta v_{\rm b}$ equation
of state (Akmal, Pandharipande, \& Ravenhall
1998), which is based on the most modern
scattering data. Like the ${\rm A18} + {\rm UIX'}
+ \delta v_{\rm b}$ equation of state, the UU
equation of state gives a maximum mass of about
$2.2\,\msun$ for a nonrotating neutron star.

As an example of the mass-radius curves predicted
by the relatively stiff equations of state
typically given by mean field theories, we
include the mass-radius curve for the mean-field
equation of state of Pandharipande \& Smith
(1975b; L in the Arnett \& Bowers [1977]
survey). The maximum mass of a nonrotating star
constructed using equation of state L is
$2.7\msun$. We also show the mass-radius curve
given by the very early tensor interaction (TI)
equation of state of Pandharipande \& Smith
(1975a; M in the Arnett \& Bowers [1977] survey)
and the Reid soft-core equation of state of
Pandharipande (1971; A in the Arnett \& Bowers
[1977] survey). The maximum masses of
nonrotating stars constructed using equations of
state M and A are $1.8\,\msun$ and
$1.65\,\msun$, respectively. Equations of state
A, L, and M are no longer of interest to nuclear
physicists and are included here primarily to
facilitate comparison with previous work on
mass-radius constraints.

For a particular source, QPO frequency, and
equation of state, the allowed portion of the
mass-radius relation is the segment within the
pie-slice shaped region analogous to the region
in Figure~\ref{fig:PieSlice} outlined by the
heavy line or, if the QPO frequency is
identified as the frequency of the innermost
stable circular orbit, simply the horizontal
heavy line. If the mass-radius relation given by
a particular equation of state intersects the
allowed region defined by the highest QPO
frequency seen in a given source, that
particular equation of state is viable. If that
equation of state is furthermore the correct
equation of state, the mass and radius of the
neutron star in the source must correspond to
one of the points along the segment of the
mass-radius relation that lies within the
pie-slice shaped region, so the mass and radius
of the star are bounded from above {\em and\/}
from below; for most equations of state, only a
narrow range of radii is allowed.

For example, the highest-frequency QPO so far
seen in any source is the 1220~Hz QPO observed in
\fu{1636$-$536} (W.~Zhang, personal
communication). Hence, if equation of state M is
the correct equation of state, then the mass and
radius of the neutron star in \fu{1636$-$536}
would have to satisfy $1.7\,\msun<M<1.8\,\msun$
and $11.6~\km<R<16.0~\km$ if it were
nonrotating. If equation of state M is the
correct equation of state and 1220~Hz is
identified as the frequency of the innermost
stable circular orbit, then the mass and radius
of the neutron star in \fu{1636$-$536} would be
determined as about $1.8\,\msun$ and between
11~\km and 12~\km, if the star were not rotating.

An equation of state that gives a mass-radius
relation that does {\em not\/} intersect the
region allowed for a given source is ruled out
for that source. We stress that {\em because the
equation of state of the matter in neutron stars
is expected to be essentially the same in all
such stars, an equation of state that is
inconsistent with the properties of any neutron
star is excluded for all neutron stars}. For
example, observation of a 1500~Hz Keplerian
orbital frequency in any source would rule out
equations of state L and M.

So far our discussion of constraints on the
neutron stars in the kilohertz QPO sources has
assumed that they are not rotating. As we
discuss in the next subsection, the changes in
these constraints caused by the spin of the star
are likely to be small in most sources.

\subsection{Effects of Stellar Rotation}

Rotation affects the structure of the star for a
given mass and equation of state and the
spacetime exterior to the star. Hence the
mass-radius relation, the orbital frequency at a
given radius, and the radius $R_{\rm ms}$ of the
innermost stable circular orbit are all
affected. As a result, the bounds on the mass
and radius of the star and the constraints on
the equation of state implied by observation of
a kilohertz QPO of a given frequency are
affected. Obviously, the size of these effects
depends on the spin rate of the star. Treating
these effects accurately will be particularly
important if the frequency of a kilohertz QPO is
ever securely identified as the Keplerian
frequency at the marginally stable orbit around
a neutron star (see \S~5.4), because the ability
to rule out some equations of state may depend
on it.

The parameter that characterizes the importance
of rotational effects is the dimensionless
quantity \hbox{$j \equiv cJ/GM^2$}, where $J$
and $M$ are the angular momentum and
gravitational mass of the star. The value of $j$
that corresponds to a given observed spin
frequency depends on the neutron star mass and
equation of state, and is typically higher for
lower masses and stiffer equations of state. For
example, at the 363~Hz spin frequency inferred
for \fu{1728$-$34} (Strohmayer et al.\ 1996c), a
$1.4\,\msun$ neutron star with the softer
equation of state A has \hbox{$j \approx 0.1$},
whereas a $1.4\,\msun$ neutron star with the
very stiff equation of state L has \hbox{$j
\approx 0.3$} (see Table~\ref{table:jValues}).

As discussed in \S~5.1, if the frequency of the
higher-frequency QPO in a kilohertz QPO pair is
an orbital frequency, then the region of the
radius-mass plane allowed for the neutron star in
that source is bounded below by the curve $R_{\rm
orb}(M)$, which gives the orbital radius at which
the Keplerian frequency is equal to the highest
observed frequency $\nu_{\rm QPO2}^\ast$ of the
higher-frequency QPO. Given $j$, it is
straightforward to compute the orbital radius for
any given orbital frequency as a function of $M$.
However, it is not $j$ but instead the star's
spin frequency $\nu_{\rm spin}$ that is
determined by the observations. Hence, {\em to
obtain the relevant $R_{\rm orb}(M)$ curve one
must vary $j$ in such a way that $\nu_{\rm
spin}$ is kept constant as $M$ is varied}.
Moreover, $j$ depends not only on $\nu_{\rm
spin}$ and $M$ but also on the equation of
state. Thus the relevant $R_{\rm orb}(M)$ curve
depends not only on the star's spin rate, but
also on the equation of state assumed.
Therefore, for rotating stars (unlike static
stars), one cannot present a single $R_{\rm
orb}(M)$ curve that constrains all equations of
state, even for a star with a given spin
frequency.

As also discussed in \S~5.1, the region of the
radius-mass plane allowed for the neutron star
in a given source is bounded above by the
horizontal line $M=M_{\rm max}$, which is the
mass at which the curve $R_{\rm orb}(M)$
intersects the curve $R_{\rm ms}(M)$; the radius
$R$ of the star is bounded above by $R_{\rm
orb}(M_{\rm max})$. The two curves $R_{\rm
ms}(M)$ and $R_{\rm orb}(M)$ are both affected
by rotation of the star, so $M_{\rm max}$
depends on the stellar spin rate.

Determining the mass at which \hbox{$R_{\rm
orb}(M)=R_{\rm ms}(M)$} is equivalent to
determining the mass at which $\nu_K(R_{\rm
ms})$ is equal to $\nu_{\rm QPO2}^\ast$. Here we
focus on the latter condition. The allowed mass
of a rotating star is bounded above if, as $M$
is increased at constant $\nu_{\rm spin}$, the
orbital frequency $\nu_K(R_{\rm ms})$ crosses
the QPO frequency $\nu_{\rm QPO2}^\ast$ from
above, once and only once. As we show below,
this is the case for slowly rotating stars.
However, this may not be the case for some
neutron star equations of state and spin rates.
(For a Kerr black hole with fixed spin
frequency, $\nu_{\rm K}(R_{\rm ms})$ first
decreases and then increases with increasing
mass.)

Fortunately, most of the neutron stars that
exhibit kilohertz QPOs appear to have spin
frequencies in the range 250--350~Hz (see
\S~4.2), which is low enough that a first-order
treatment of rotational effects is adequate. To
see this, note that neutron stars with masses
\hbox{$\sim 1.5 \dash 2\,\msun$} have moments of
inertia \hbox{$I \sim (1 \dash
3)\ee{45}~\gpsqcm$}, so stars with spin
frequencies \hbox{$\lta 350~\Hz$} have
$j$-values \hbox{$\lta 0.3$} (see
Table~\ref{table:jValues}). The lowest-order
changes in the structure of a rotating star are
${\cal O}(j^2)$ so, to first order in $j$, the
mass-radius relation and moment of inertia of a
rotating star are the same as for a nonrotating
star of the same mass (Hartle \& Thorne 1968).
Thus the error made by neglecting higher-order
terms is \hbox{$\lta 10$\%} for spin frequencies
\hbox{$\lta 350~\Hz$}. 

We now compute the mass and radius constraints
imposed on a slowly rotating neutron star by
observation of a kilohertz orbital frequency. The
calculation is simplified by the fact that to
first order in $j$, the spacetime outside a
uniformly and steadily rotating relativistic star
is the same as the Kerr spacetime for the same
$M$ and $j$ (Hartle \& Thorne 1968).

Consider first the effect of stellar rotation on
the motion of gas orbiting the star. To first
order in $j$, the orbital frequency (measured at
infinity) of gas in a prograde Keplerian orbit at
a given Boyer-Lindquist radius $r$ is
 \begin{equation}
   \nu_K(r,M,j) \approx
    [1-j(GM/rc^2)^{3/2}]\,\nu^\nonrot_K(r,M)
  \label{eq:KeplerFreqFirst}
 \end{equation}
 and the radius of the marginally stable orbit is
 \begin{equation}
  R_{\rm ms}(M,j) \approx
   [1-j(2/3)^{3/2}]\,R^\nonrot_{\rm ms}(M)\;,
  \label{eq:msRadiusFirst}
 \end{equation}
 where $\nu^\nonrot_K$ and $R^\nonrot_{\rm ms}$
are the Keplerian frequency and radius of the
marginally stable orbit for a nonrotating star.
Expressions~(\ref{eq:KeplerFreqFirst})
and~(\ref{eq:msRadiusFirst}) are first-order
expansions of the exact expressions for these
quantities given by Bardeen, Press, \& Teukolsky
(1972) for the Kerr spacetime. Hence, to first
order in $j$, the frequency of the prograde orbit
at $R_{\rm ms}$ around a star of given mass $M$
and dimensionless angular momentum $j$ is (cf.
Klu\'zniak, Michelson, \& Wagoner 1990)
 \begin{eqnarray}
 \nu_{\rm K, ms} &\approx&
   \left[
     1-j(1/6)^{3/2}
   \right]
   \left[
     1+j(2/3)^{1/2}
   \right]\nu^0_{\rm K, ms}
   \nonumber\\ \nonumber\\
 &\approx&
   2210\,(1+0.75j)(M_\odot/M)\,\Hz\;,
 \end{eqnarray}
 where $\nu^0_{\rm K, ms}$ is the Keplerian
orbital frequency at the radius of the innermost
stable circular orbit for a nonrotating star.
Therefore, {\em the net effect of the star's
rotation is to increase the frequency of the
prograde orbit at $R_{\rm ms}$}.

As explained above, the region of the radius-mass
plane allowed for a given star is not the region
allowed for constant $j$ but is instead the
region allowed for constant $\nu_{\rm spin}$. To
demonstrate that the mass and radius of the star
are bounded from above, it is sufficient to show
that $\nu_K(R_{\rm ms})$ decreases with
increasing $M$ at fixed $\nu_{\rm spin}$, or
equivalently, that $(d\nu_{\rm
K,ms}/dM)_{\nu_{\rm spin}}<0$ for all $M$ for
the equation of state under consideration. Now
to first order in $j$,
 \begin{eqnarray}
  \label{Slope}
   \left[
     \frac{d\nu_{\rm K,ms}(M,j)}{dM}
   \right]_{\nu_{\rm spin}}
    = \quad \left\{
    \left[\frac{\pd\nu_{\rm K,ms}}
       {\pd M}(M,0)\right]_{j}
   \right.
    \nonumber\\ \nonumber\\
   \left. 
     \mbox{}
     + \left[\frac{\pd \nu_{\rm K,ms}}
             {\pd j}(M,0)\right]_{M}
      \left(\frac{dj}{dM}\right)
  \right\}_{\nu_{\rm spin}}\\
    \nonumber\\
  \approx
 -\ \frac{\nu^0_{\rm K,ms}(M)}{M}
 \left[
  1 + 0.75j\left(2 - \frac{d\ln I}{d\ln M}\right)
 \right]_{\nu_{\rm spin}} .
 \end{eqnarray}
 Note that the derivatives on the right side of
equation~(\ref{Slope}) are to be evaluated at
$j=0$. In order to show that $(d\nu_{\rm
K,ms}/dM)_{\nu_{\rm spin}}$ is negative for
slowly rotating stars, it is sufficient (but not
necessary) to show that $({d\ln I}/{d\ln
M})_{\nu_{\rm spin}}$ is always less than 2. This
is the case for all the equations of state
tabulated by Cook et al.\ (1994) and is the case
even for incompressible matter, which is
unphysically stiff. (For a star made of
incompressible matter, $I \propto MR^2 \propto
M^{5/3}$, so \hbox{$(d\ln I/d\ln M) =
\frac{5}{3}<2$}.) Thus, {\em for slowly rotating
stars $\nu_K(R_{\rm ms})$ decreases with
increasing mass for constant $\nu_{\rm spin}$, so
the masses and radii of such stars are bounded
above}.

Computation of the mass and radius constraints is
straightforward but depends on the stellar spin
rate and equation of state. The upper bounds on
the mass and radius are given implicitly by
 \begin{equation}
   M_{\rm max} \approx
   [1+0.75j(\nu_{\rm spin})]M^\nonrot_{\rm max}
   \label{eq:FirstMaxMass}
 \end{equation}
 and
 \begin{equation}
   R_{\rm max} \approx
   [1+0.20j(\nu_{\rm spin})]R^\nonrot_{\rm
max}\;,
   \label{eq:FirstMaxRadius}
 \end{equation}
 where $j(\nu_{\rm spin})$ is the value of $j$
for the observed stellar spin rate at the maximum
allowed mass for the equation of state being
considered and $M^\nonrot_{\rm max}$ and
$R^\nonrot_{\rm max}$ are the maximum allowed
mass and radius for a nonrotating star (see
eqs.~[\ref{eq:MassBoundZero}]
and~[\ref{eq:RadiusBoundZero}]).
 Expressions~(\ref{eq:FirstMaxMass})
and~(\ref{eq:FirstMaxRadius}) show that {\em the
bounds are always greater for a slowly rotating
star than for a nonrotating star, regardless of
the equation of state assumed}.

Figure~\ref{fig:RotationEffects} illustrates the
effects of stellar rotation on the region of the
radius-mass plane allowed for a given star for
spin rates $\sim 300~\Hz$, like those inferred
for the kilohertz QPO sources, and
\hbox{$\nu_{\rm QPO2}^\ast=1220~\Hz$}, the
frequency of the highest-frequency QPO so far
observed in \fu{1636$-$536}, which is also the
highest-frequency QPO so far observed in any
source. Our calculations show that {\em the mass
of the neutron star in \fu{1636$-$536} must be
less than $\sim 2.2\,\msun$ and its radius must
be less than $\sim 17~\km$}. As explained above,
the precise upper bounds depend on the equation
of state assumed.

For rapidly rotating stars, $j$ is not small
compared to unity and the structure of the star
depends appreciably on its rotation rate.
Derivation of bounds on the mass and radius of a
given star for an assumed equation of state
therefore requires construction of a sequence of
stellar models and spacetimes for different
masses using the assumed equation of state, with
$\nu_{\rm spin}$ as measured at infinity held
fixed. The maximum and minimum possible masses
and radii allowed by the observed QPO frequency
can then be determined.

\subsection{Effects of the Radial Radiation
Force}

The luminosities of the Z sources are typically
$\sim 0.5 \dash 1\,L_E$, where $L_E$ is the
Eddington luminosity (see \S~2). Hence, in the Z
sources the outward acceleration caused by the
radial component of the radiation force can be a
substantial fraction of the inward acceleration
caused by gravity. Therefore, in these sources
the radial component of the radiation force must
be included in computing the Keplerian orbital
frequency near the neutron star and taken into
account when constraints on the mass and radius
of the star are derived using the procedures
discussed in \S~5.1 and \S~5.2.

The radially outward component of the radiation
force reduces the orbital frequency at a given
radius. For example, if the star is spherical and
nonrotating  and emits radiation uniformly and
isotropically from its entire surface, the
orbital frequency (measured at infinity) of a
test particle at Boyer-Lindquist radius $r$ is
 \begin{equation}
  \nu_{\rm K}(L)=
  \nu_{\rm K}(0)
  \left[1-{(1-3GM/rc^2)^{1/2}\over{(1-2GM/rc^2)}}
  {L\over{L_E}}\right]^{1/2}\;,
  \label{eq:RadForceFreq}
 \end{equation}
 where $L$ is the luminosity of the star measured
at infinity and $\nu_{\rm K}(0)$ is the Keplerian
frequency in the absence of radiation forces.
 Thus, the Boyer-Lindquist radius of a circular
orbit with a given frequency is smaller in the
presence of the radial radiation force and the
constraints on the mass and radius of the star
are therefore tightened.

For the atoll sources, which have
luminosities\break
 $L \lta 0.1\,L_E$, the change
in the Keplerian frequency is at most
$\sim\,$5\%. For the Z sources, on the other
hand, which have luminosities \hbox{$L \approx
L_E$}, the change may be much larger, although
the change in the sonic-point Keplerian
frequency may be smaller than would be suggested
by a naive application of
equation~(\ref{eq:RadForceFreq}), if a
substantial fraction of the radiation produced
near the star is scattered out of the disk plane
before it reaches the sonic point. A more
detailed analysis of the effect of the radial
radiation force on the constraints on the mass
and radius of the star will be reported
elsewhere.

\subsection{Signatures of the Innermost
Stable\break Circular Orbit}

As explained in \S~4.2, observations of
kilohertz QPO sources may be able to demonstrate
the existence of an innermost stable circular
orbit around some of the neutron stars in these
sources. If so, this would be the first
confirmation of a strong-field prediction of
general relativity and an important step forward
in our understanding of strong-field gravity.

If the frequency of a kilohertz QPO produced by
an LMXB can be securely established as the
orbital frequency at the radius of the innermost
stable circular orbit and the spin frequency of
the star can be determined, the mass of the
neutron star in that source can be determined
for each assumed equation of state. Depending on
the range of allowed equations of state and
masses, this could have profound consequences
for our understanding of the equation of state
of neutron stars.\footnote{Following submission
and circulation of the original version of the
present paper, Kaaret, Ford, \& Chen (1997) and
Zhang, Strohmayer, \& Swank (1997) considered the
implications if the frequencies of some of the
kilohertz QPOs that have already been observed
are the orbital frequencies of innermost stable
circular orbits.} For example, establishing a
mass of $2.0\,\msun$ for a slowly rotating
neutron star would rule out eight of the twelve
currently viable equations of state considered
by Cook et al.\ (1994).

Given the profound consequences that would
follow from identifying the frequency of a
kilohertz QPO with the orbital frequency of the
innermost stable orbit in any source, it is very
important to establish what would constitute
strong, rather than merely suggestive, evidence
of such a detection. Observations that would
signal detection of the innermost stable orbit 
include the following:

{\em QPO frequency signature}.---The strongest
evidence that a QPO with the orbital frequency
of the innermost stable orbit has been detected
would be reproducible observation of a fairly
coherent, kilohertz QPO with a frequency that
first increases steeply with accretion rate but
then, at a high frequency, becomes nearly
constant as the accretion rate continues to
increase (see \S~3.3, \S~4.2, and Fig.~9).

The QPO frequency will approach a constant
because the sonic point in the flow cannot
retreat closer to the star than the radius of the
marginally stable orbit and coherent QPOs are
not expected from the rapidly inspiraling gas
inside the sonic point. Although the count rate
above which the frequency becomes constant may
vary, the constant frequency itself should not
vary, because the orbital frequency of the
marginally stable orbit depends only on the mass
and rotation rate of the neutron star, which
remain almost constant over many years.

In the sonic-point model, the frequency of the
lower-frequency QPO in a kilohertz QPO pair is
the beat frequency and therefore should also
increase steeply with accretion rate at first
but then become approximately constant at the
same accretion rate at which the frequency of
the higher-frequency QPO becomes constant.

{\em QPO amplitude signature}.---A second
signature that the marginally stable orbit has
been detected would be reproducible observation
of a decrease in the amplitude of the
lower-frequency QPO in a kilohertz QPO pair or a
simultaneous decrease in the amplitudes of both
QPOs in a pair, at a QPO frequency that is
always the same in a given source.

The amplitude of the QPO at the beat frequency
is expected to decrease once the sonic point has
moved inward to the marginally stable orbit
because in the sonic-point model, the QPO at the
beat frequency is generated by the drag force
exerted by the radiation coming from the stellar
surface that reaches the sonic point. Therefore,
a strong QPO is expected at the sonic-point beat
frequency only if the drag force exerted by the
radiation is dynamically important at the sonic
point. If, however, the sonic point has moved
inward to the innermost stable orbit and remains
there as the accretion rate continues to
increase, the optical depth from the stellar
surface to the sonic point will continue to rise
and radiation drag will become less and less
important there, causing modulation of the
inflow from the sonic point by the radiation
force to weaken. Thus, a decrease in the
amplitude of the QPO at the sonic-point beat
frequency with increasing accretion rate would
signal the approach of the sonic-point to the
radius of the innermost stable orbit (see
\S~3.1).

If radiation forces play an important role in
creating or amplifying the clumps that produce
the QPOs (see \S~3.1), the amplitude of the
Keplerian-frequency QPO may also decrease with
increasing accretion rate, once the radius of
the sonic point has reached the radius of the
innermost stable orbit (see \S~3.2).

{\em Possible QPO coherence signature}.---A
possible signature that the orbit of the gas
that is generating the QPO has receded inside
the radius of the marginally stable orbit would
be a steep drop in the coherence of both QPOs in
a kilohertz QPO pair (or in the coherence of the
Keplerian-frequency QPO, if the beat-frequency
QPO is not visible), at a certain critical
frequency, as the frequencies of the QPOs
increase steadily with accretion rate. This
would occur if radiation forces are able to
generate clumping in the flow at a radius inside
the sonic point. If this occurs at all, clumps
are likely to be produced at a range of radii
and to last only a short time, so any
oscillations that may be generated are likely to
have low coherence (see \S~3.4). The critical
frequency would be the orbital frequency at the
marginally stable orbit and hence should always
be the same in a given source.

Approach of the sonic radius to the radius of
the innermost stable orbit can be distinguished
from approach of the sonic radius to the radius
of the star because coherent Keplerian- and
beat-frequency QPOs may continue in the fist
case but are very unlikely in the second.
Kilohertz QPOs are unlikely to be generated if
the sonic point moves close to the stellar
surface because of the disruptive effect of the
stellar magnetic field and the the viscous shear
layer that is expected to develop if the
Keplerian flow interacts directly with the
stellar surface (see \S~4.2).

\section{SUMMARY AND CONCLUSION}

The sonic-point model explains naturally the most
important features of the kilohertz QPOs observed
in the atoll and Z sources, including their
frequencies, large amplitudes, and high
coherence. It also explains the frequent
occurrence of two simultaneous QPOs, the
observed steep increase of kilohertz QPO
amplitudes with increasing photon energy in the
$\sim\,$5--10~keV energy range, and the
anticorrelation of kilohertz QPO amplitudes with
the strength of the stellar magnetic field 
inferred from spectral models. An attractive
feature of the sonic-point model is that the
magnetic fields, accretion rates, and scattering
optical depths that it requires are completely
consistent with those inferred previously from
observations and modeling of the X-ray spectra
and lower-frequency X-ray variability of the
kilohertz QPO sources.

The sonic-point model leads to several general
expectations about the kilohertz QPOs:

(1)~{\em Kilohertz QPOs with the properties seen
in the atoll and Z sources should not be
observed in black-hole LMXBs}, because collision
of the accretion flow with the stellar surface
plays an essential role in the sonic-point model.

(2)~{\em The rms amplitudes of the kilohertz QPOs
should be anticorrelated with the strength  of
the neutron star's magnetic field}. From this it
follows that kilohertz QPOs should be very weak
or undetectable with current instruments in the
GX group of atoll sources and in the Cyg-like Z
sources (which include \cyg2, \gx{5$-$1}, and
\gx{340$+$0}), and undetectable with current
instruments in any sources that have strong
magnetic fields and therefore produce strong,
periodic oscillations at their spin
frequencies.  (After this specific prediction
was made in the original version of this paper,
kilohertz QPOs with very low amplitudes were
detected in all six of the originally identified
Z sources; see \S~4.4.) Another consequence of
expectation~(2) is that the positive correlation
of rms amplitude with hard color shown by the
four sources plotted in
Figure~\ref{fig:AmpVsHardColor} should be found
to be general among sources showing kilohertz
QPOs.

(3)~{\em If kilohertz QPOs are detected in the
tails of X-ray bursts, their frequencies will
tend to be lower when the accretion rate is
higher.}

(4)~{\em Weak oscillations should eventually be
detected at overtones of the sonic-point beat and
Keplerian frequencies and perhaps at the stellar
spin frequency, but oscillations at other
frequencies, such as $\nu_{\rm Ks}+\nu_{\rm
spin}$, should be extremely weak.}

At a lower level of certainty, the sonic-point
model suggests that (1)~the dependence of QPO
amplitude on photon energy will change as the
X-ray spectrum of the source changes, (2)~the
amplitude of the higher-frequency QPO in a
kilohertz QPO pair will drop relative to the
amplitude of the lower-frequency QPO at high
luminosities, and (3)~either the lower-frequency
QPO or the higher-frequency QPO in a pair may be
undetectable even when the other QPO is
detectable.

In closing, we emphasize that measurement of
Keplerian  frequencies in the kilohertz range
provides interesting new upper bounds on the
masses and radii of the neutron stars in the
kilohertz QPO sources and important new
constraints on the equation of state of the
matter in all neutron stars, as we have
shown in \S~5. As we demonstrated there, if the
neutron star in the atoll source \fu{1636$-$536}
has a spin frequency of $\sim 290$~Hz, as
indicated by the difference between the spin
frequencies of its two high-frequency QPOs, the
1220~Hz QPO observed in this source constrains
its mass to be less than about $2.2\,\msun$ and
its radius to be less than about 17~km; the
precise bounds depend on the equation of state
assumed.

If at some future time we are able to
demonstrate the existence of an innermost stable
circular orbit around one or more neutron
stars using observations of kilohertz QPOs, this
would be the first confirmation of a prediction
of general relativity in the strong-field regime
and a major advance in our understanding of
strong-field gravity. If the frequency of a
kilohertz QPO is securely established as the
orbital frequency at the radius of the innermost
stable circular orbit in a particular system by,
for example, observing the signatures discussed
in \S~5, and if, in addition, the spin frequency
of the neutron star can be determined, then the
frequency of the QPO will fix the mass of the
neutron star for each assumed equation of state,
providing a better understanding of the
properties of dense matter and possibly ruling
out many currently viable equations of state.

\acknowledgements

It is a pleasure to thank Phil Kaaret, Vicky
Kalogera, Ed Morgan, Michiel van der Klis, Tod
Strohmayer, Tom Baumgarte, Stu Shapiro, and Tomek
Bulik for useful discussions. We especially thank
Rudy Wijnands for providing the PCA response
matrix and the X-ray spectral data used to
compute the X-ray colors plotted in Figures~11
and 12, Greg Cook for providing the mass-radius
relations for the equations of state plotted in
Figure~\ref{fig:EOSconstraints}, and Ron Taam for
discussions about black hole disk solutions.

This work was supported in part by NSF grants
AST~93-15133 and AST~96-18524, NASA grant
NAG~5-2925, and NASA RXTE grants at the
University of Illinois, NASA grant NAG~5-2868 at
the University of Chicago, and through the \gro
Fellowship Program, by NASA grant NAS~5-2687.

\newpage

\figcaption[]{\label{fig:SideViews}
 Schematic drawings of the accretion flows
expected around atoll and Z sources.
 {\it Top\/}:~Side view of an atoll source with
a dipole magnetic field $\sim 10^{8}\,$G and a
mass accretion rate $\sim 0.01\,\mdote$. The
star's spin axis and the direction of its
magnetic moment are indicated by the arrows
labeled $\Omega$ and $\mu$. Gas in the inner
part of the Keplerian disk is clumpy and
penetrates very close to the stellar surface
before some of it is channeled by the star's
magnetic field. Collision of the channeled gas
with the stellar surface produces slightly
brighter spots that rotate with the star. The
light shading indicates the hot gas in the
magnetosphere and the central corona that
surrounds the neutron star.  
 {\it Bottom\/}:~Side view of a Z source with a
magnetic field $\sim 10^{9} \dash 5\ee9\,$G and
a mass accretion rate $\sim\mdote$, showing how
some of the gas in the disk is channeled out of
the disk by the star's magnetic field at $\sim 2
\dash 3$ stellar radii but some continues to
flow inward in a Keplerian disk flow that
penetrates close to the  stellar surface.
Collision of the channeled gas with the stellar
surface produces brighter spots that rotate with
the star. The light shading again indicates the
hot gas in the magnetosphere and the central
corona that surrounds the neutron star. The
arrows indicate the cooler, approximately
radial inflow outside the central corona that
is present in the Z sources.}

\figcaption[]{\label{fig:ParameterSpace}
 Parameter space of dipole magnetic moment $\mu$
and mass accretion rate (in units of the
Eddington mass accretion rate $\mdote$ that
produces an Eddington luminosity) for neutron
star LMXBs. The axis on the right shows the
equivalent surface magnetic field at the pole for
a neutron star with a radius of 10~km. The thin
solid lines, plotted using equation~(1),
indicate the magnetic moment for which, at the
given accretion rate, the Keplerian frequency is
500~Hz (top line) and 1100~Hz (bottom line)
where a substantial fraction of the gas in the
disk first couples to the stellar field and is
channeled out of the disk. The break at
\hbox{$\mdot = 0.3\,\mdote$} illustrates the
effect of the expected transition from
gas-pressure-dominated flow at low accretion
rates to radiation-pressure-dominated flow at
high accretion rates. The lines in this \mdot\
range are dotted to indicate that they are not
accurate. The solid lines, plotted using
equation~(2) for electron temperatures
\hbox{$kT_e=5$~keV} (top line) and
\hbox{$kT_e=15$~keV} (bottom line), divide the
${\dot M}$--$B_d$ plane into the regions where
the dominant source of soft photons is cyclotron
emission in the neutron star magnetosphere
(above the line) or blackbody emission from the
neutron star surface (below the line). The
labels ``4U'', ``GX'',  and ``Z'' indicate the
approximate regions in the ${\dot M}$--$B_d$
plane occupied, respectively, by 4U~atoll
sources, GX~atoll sources, and Z sources. 
These curves were calculated assuming a
neutron star mass \hbox{$M=1.4\,M_\odot$}, an
effective cyclotron photosphere of radius
\hbox{$R_{\rm cyc} = 1.5\times 10^6$}~cm, a
neutron star radius of $10^6$~cm, and a
spectrum truncated at cyclotron harmonic number
$n=15$, appropriate for a scattering optical
depth \hbox{$\tau=6$}. For masses, radii, or
optical depths different from these values,
the curves can shift by as much as 50\% in
magnetic field at a given accretion rate.}

\figcaption[]{\label{fig:Spirals}
 A neutron star with a dynamically negligible
magnetic field, accreting via a Keplerian disk
that penetrates close to the star. The star and
disk are viewed along the rotation axis of the
disk, which is rotating counterclockwise in this
view. The panels show, in Boyer-Lindquist
$r,\phi$ coordinates,
  (a)~the spiral trajectory followed by a single
element of gas as it falls supersonically from
the sonic radius to the stellar surface and
  (b)~the spiral pattern of higher gas density   
formed by gas streaming inward along spiral
trajectories with the shape shown in panel~(a)
from a region of denser gas (a ``clump'')
orbiting near the sonic radius.
  The gas trajectories and resulting density
pattern were computed in full general relativity
assuming that the gas is exposed to radiation
from the star at a radius of $9M$ and that the
star has a radius of $4M$, is nonrotating,
radiates isotropically, and has a luminosity
measured at infinity of $0.005\,L_E$.
  The surface density of the disk flow is much
smaller inside the sonic radius (lighter shaded
region) than outside (darker shaded region),
because of the sharp increase in the inward
radial velocity at the sonic radius.
  Gas falling inward from the sonic radius along
spiral trajectories collides with the neutron
star around its equator, producing an X-ray
emitting equatorial ring, which is indicated by
the grey ring around the star.
  The white arc at the stellar surface indicates
the bright, arc-shaped ``footprint'' where the
denser gas from the clump collides with the
stellar surface and produces a beam of X-rays
(white dashed lines) that rotates around the
star at the sonic-point Keplerian frequency.
  The footprint generally moves with respect to
the stellar surface.}

\figcaption[]{\label{fig:KeplerQPO}
 Time sequence of four snapshots of a neutron
star with a dynamically negligible magnetic field
accreting via a Keplerian disk that penetrates
close to the star, showing schematically how the
QPO at the sonic-point Keplerian frequency is  
produced.
  The viewpoint, gas trajectories, and density
pattern are the same as in Fig.~3.
  The drop in the surface density of the disk
flow  at the sonic radius, the bright ring of
emission  around the stellar equator, the
pattern of denser infalling gas inside the sonic
radius, the bright footprint where the denser
gas collides with the stellar surface, and the
beam of radiation coming from the footprint are
all indicated in the same way as in Fig.~3.
  The clump is shown advancing in its orbit by
90\degree\ from one panel to the next. The spiral
pattern of higher density gas, its bright
footprint, and the resulting beam of X-rays all
rotate around the star with a frequency equal to
the orbital frequency of the clump, so they are
also shown advancing by 90\degree\ from one panel
to the next.
  The footprint generally moves with respect to
the stellar surface.}

\figcaption[]{\label{fig:BeatQPO}
 Time sequence of six snapshots of a spinning
neutron star with a weak magnetic field accreting
via a prograde Keplerian disk that penetrates
close to the star, showing schematically how the
QPO at the sonic-point beat frequency is
produced.
  The disk and the star are viewed along their
common rotation axes and are rotating
counterclockwise in this view.
  The beam of radiation produced by the collision
with the stellar surface of the gas channeled by 
the weak stellar magnetic field is indicated by
the faint white beam emerging from behind the
star; this beam rotates with the star.
  To make the sequence of events easier to
visualize, the snapshots show the sequence of
events as seen in a frame corotating with the
star, so the beam of radiation that rotates with
the star is always pointing in the same
direction (here, the vertical direction).
  The angular velocity of the clump, which is
orbiting near the sonic radius, is greater than
the angular velocity of the star, so with time
the clump advances counterclockwise relative to
the beam of radiation that rotates with the
star. In this sequence, the clump advances
relative to the star by 90\degree\ from one
snapshot to the next.   
  The drop in the surface density of the disk
flow at the sonic radius, the bright ring of
emission around the stellar equator, the
pattern of denser infalling gas inside the sonic
radius, the bright footprint where the denser
gas collides with the  stellar surface, and the
beam of radiation coming from the footprint are
all indicated in the same  way as in Fig.~3.
  The events occurring in each panel are
described in the text.}

\figcaption[]{\label{fig:AngMomVsRadius}
 {\em Solid curve\/}: Angular momentum per unit
mass $u_\phi$ (as measured at infinity in units
of the gravitational mass $M$ of the star) of an
element of gas in a circular Keplerian orbit
around a nonrotating neutron star, as a function
of the Boyer-Lindquist radial coordinate of the
orbit (in units of $M$).
 {\em Dashed curve\/}: $u_\phi$ for a rotating
star with dimensionless angular momentum
\hbox{$j=0.1$}, which is close to the $j$-value
given by the best modern neutron-star matter
equations of state for a neutron star with
\hbox{$M=1.4\,\msun$} and a spin frequency of
$300$~Hz as measured at infinity (see \S~5.2).
 The top axis shows the radius in kilometers for
\hbox{$M=1.4\,\msun$}.
 The radius $R_{\rm ms}$ of the innermost stable
circular orbit is the radius at which the
specific angular momentum is a minimum; for
$j=0$, $R_{\rm ms}=6M$, whereas for
\hbox{$j=0.1$}, \hbox{$R_{\rm ms}=5.7M$}. }

\figcaption[]{\label{fig:AngMomFractionVsRadius}
 {\em Solid curve\/}: Fraction $\eta_{\rm flow}$
of the specific angular momentum of an element
of gas in Keplerian circular orbit at
Boyer-Lindquist radial coordinate $r$ (measured
in units of the stellar mass $M$) that must be
removed in order for the gas to fall from $r$ to
the radius $R_{\rm ms}$  of the innermost stable
circular orbit. The curve shown is for a static
star \hbox{($j=0$)}; the curves for slowly
rotating stars would be little different.
 {\em Dashed curve\/}: Estimate of the largest 
fraction $\eta_{\rm rad}({\rm max})$ of the
specific angular momentum of an element of gas
that can be  removed by radiation coming from the
surface of a nonrotating, isotropically
radiating, spherical star of radius $5M$ (see
text).
 The top axis shows the radius in kilometers for
\hbox{$M=1.4\,\msun$}.}

 \figcaption[]{\label{fig:SimpleExamples}
 Numerical results for the structure of the
accretion flow near the neutron star in the
fully general relativistic model of gas dynamics
and radiation transport in the inner disk
described in the text.
 (a)~The inward radial velocity $v^{\hat r}$ of
the gas in the disk measured by a local static
observer.
 (b)~The angular velocity of the gas in the disk
measured by an observer at infinity.
 (c)~The  surface density $\Sigma_{\rm co}$ of
the disk flow measured by an observer comoving
with the flow.    
  (d)~The radial optical depth $\tau_r$ from the
stellar surface through the disk flow to the
radius shown on the horizontal axis.
 The radial coordinate is the Boyer-Lindquist
radius in units of the stellar mass $M$; the
radius of the star is $5M$.
 The four curves in each panel are labeled with
the assumed accretion rate $\mdoti$ through the
inner disk, measured in units of the accretion
rate \mdote\ that would produce an accretion
luminosity at infinity equal to the Eddington
critical luminosity.}

\figcaption[]{\label{fig:nuKvsL} Sonic-point
Keplerian and beat frequencies in Hertz as
functions of the accretion luminosity in units of
the Eddington critical luminosity, computed
using the accretion flow model of \S~3.2. The
neutron star spin rate is assumed to be 300~Hz.
The other assumptions are the same as in Fig.~8.
The sonic-point Keplerian frequency $\nu_{\rm
Ks}$ (solid line) increases steeply with
increasing accretion luminosity until it reaches
$\nu_{\rm K}(R_{\rm ms})$ (dashed horizontal
line), the orbital frequency at the innermost
stable circular orbit, at which point $\nu_{\rm
Ks}$ stops increasing. The sonic-point beat
frequency $\nu_{\rm Bs}$ (dotted line) increases
steeply with increasing accretion luminosity
until it reaches $\nu_{\rm K}(R_{\rm
ms})-\nu_{\rm spin}$, at which point it too
stops increasing.}

\figcaption[]
 {
 \label{fig:AttenFactor}
 Sample attenuation factors $A_\infty/A_0$ for
the relative amplitudes of the X-ray brightness
oscillations produced by a hypothetical accreting
neutron star with a spin rate $\nu_{\rm spin} =
300~\Hz$ and a sonic-point Keplerian frequency
$\nu_{\rm Ks} = 1100~\Hz$, at the center of a
uniform, spherical scattering cloud with a radius
of $3\ee6$~cm. The sonic-point beat frequency
$\nu_{\rm Bs}$ is assumed to be $\nu_{\rm Ks}
- \nu_{\rm spin}$ and is therefore 800~Hz. The
attenuation factors for the oscillations at
$\nu_{\rm Ks}$ (dashed line) and at the upper
sideband frequency $\nu_{\rm Ks} + \nu_{\rm spin}
= 1400~\Hz$ (dotted line), which are assumed for
the sake of illustration to be pure beaming
oscillations produced by a rotating pencil beam,
are $\gta5$ for scattering optical depths
$\tau\gta5$. In contrast, the attenuation factor
for the oscillation at $\nu_{\rm Bs}$ (solid
line), which is a pure luminosity oscillation, is
much smaller.}

\figcaption[]{
 \label{fig:AmpVsEnergy}
 Measured amplitudes of the high-frequency QPOs
seen in \fu{1608$-$52} (open circles) and
\fu{1636$-$536} (filled circles) as a function of
photon energy. Data for \fu{1636$-$536} were
kindly provided by W.\ Zhang (1997, personal
communication) and reflect corrections made
after the report by Zhang et al.\ (1996) was
published. The dotted curve shows the variation
of the rms amplitude with photon energy due to
the optical depth variations expected in the
sonic-point model (see text). This curve is not
a fit; instead, for illustrative purposes we
have assumed (consistent with the unified model
of neutron star LMXBs) that the input spectrum
is a blackbody of temperature $kT=0.6$~keV and
that the Comptonizing corona has a temperature
$kT=9$~keV and an optical depth that varies from
$\tau=3$ to $\tau=3.3$.}

\figcaption{
 \label{fig:AmpVsHardColor}
 Measured rms amplitudes of the high-frequency
QPOs seen in \sco1, \fu{1735$-$444},
\fu{1636$-$536}, and \fu{0614$+$091} plotted
against the PCA X-ray hard colors (defined as
the ratio of the counts in the 7--20 keV bin to
the counts in the 5--7 keV bin) of these
sources. The photon energy ranges used in
computing these amplitudes were 2--20~keV for
\sco1 (van der Klis et al. 1996d),
0.24--18.37~keV for  \fu{1735$-$444} (Wijnands
et al. 1996), 7--20~keV   for \fu{1636$-$536}
(van der Klis et al. 1996c), and  2--20~keV
for \fu{0614$+$091} (van der Klis et al.\ 1996c).
The observed steep increase of QPO amplitudes 
with photon energy (Fig.~11) means that a
direct, quantitative comparison of QPO
amplitudes in different sources can only be made
if they are all computed using the same photon
energy ranges. For example, the QPOs in
\gx{5$-$1} are not detected below 10~keV, but
between 10~keV and 50~keV the QPO amplitude is
as high as $\sim$7\% (van der Klis et al.
1996e). It is not straightforward to compare
this with the reported $\sim$1\% amplitude for
\sco1 in the 2--20~keV band. The range of hard
colors for \sco1 is that near the soft vertex.
The range of hard colors for \fu{1735$-$444}
has been artificially extended to reflect
possible systematic errors in the PCA response
caused by gain changes. In neutron star LMXBs,
the X-ray hard color is found to be greater for
sources with weaker magnetic fields (Psaltis et
al.\ 1995; Psaltis \& Lamb 1998), so the
correlation evident in this figure is striking
confirmation that the amplitudes of these QPOs
are lower for sources with stronger magnetic
fields. The PCA colors were kindly provided to
us by Rudy Wijnands.}

\figcaption[]{
 \label{fig:PieSlice}
 Radius-mass plane, showing how bounds on the
mass and radius of a nonrotating neutron star in
a QPO source with \hbox{$\nu_{\rm QPO2}^\ast =
1220~\Hz$} can be constructed. The dashed curve
that begins at the origin and rises up and to
the right is the relation
\hbox{$M=M^\nonrot(R_{\rm orb},\nu_{\rm
QPO2}^{\ast})$}, for \hbox{$\nu_{\rm
QPO2}^{\ast}=1220~\Hz$}; this curve gives the
mass for which the orbital frequency at the
radius shown on the horizontal axis is 1220~Hz.
The diagonal dotted line is the relation
\hbox{$M=M^\nonrot(R_{\rm ms})$}; this line
gives the mass for which the radius
$R^\nonrot_{\rm ms}$ of the innermost stable
circular orbit is equal to the radius shown on
the horizontal axis. The horizontal solid line
is the mass $M^\nonrot_{\rm max}$ at which
$M^\nonrot(R_{\rm ms})$ is equal to
$M^\nonrot(R_{\rm orb},\nu_{\rm QPO2}^{\ast})$
and hence $R^\nonrot_{\rm ms}(M)$ is equal to
$R^\nonrot_{\rm orb}(M,\nu_{\rm QPO2}^{\ast})$,
for \hbox{$\nu_{\rm QPO2}^{\ast}=1220~\Hz$}. 
The hatched region shows the combinations of
stellar mass and radius that are excluded.  The
region outlined by the heavy solid line shows
the combinations
allowed by the frequency and coherence of the
QPO (see text). Only masses less than
$M^\nonrot_{\rm max} = 1.8\,\msun$ and radii
less than \hbox{$R^\nonrot_{\rm max} =
16.0~\km$} are allowed. The allowed region
collapses to to the heavy solid horizontal line
if the QPO frequency is identified as the
orbital frequency of gas in the innermost stable
circular orbit (see \S~5.1). The steps in the
construction are the same for a rotating star
but the bounds are different (see \S~5.2).
 }

\figcaption[]{
 \label{fig:EOSconstraints}
 Comparison of the mass-radius relations for
nonrotating neutron stars given by five
representative equations of state for
neutron-star  matter with the regions of the
mass-radius plane allowed for nonrotating
stars (see Fig.~13 and text) in sources with
three values of $\nu_{\rm QPO2}^{\ast}$, the
highest observed frequency of the
higher-frequency QPO in a pair.
 The light solid curves show the mass-radius   
relations given by equations of state A
(Pandharipande 1971), FPS (Lorenz, Ravenhall, \&
Pethick 1993), UU (Wiringa, Fiks, \& Fabrocini
1988), L (Pandharipande \& Smith 1975b), and M
(Pandharipande \& Smith 1975a).
  Each allowed region is labeled by the value of
$\nu_{\rm QPO2}^{\ast}$ assumed in constructing
it. As in Figure~13, the hatched region shows
the combinations of stellar mass and radius that are 
excluded by $\nu^{\ast}_{\rm QPO2}$=1220~Hz.
The region bounded by the heavy line is the
region that would be allowed for the neutron
star in \fu{1636$-$536} if it were not rotating
(see text). The star's probable spin rate
affects the mass-radius relations hardly at all,
but enlarges the allowed region by $\sim 20$\%
(see \S~5.2 and Fig.~15).}

\figcaption[]{
 \label{fig:RotationEffects}
 Approximate constraints imposed on the mass,
radius, and equation of state of neutron stars
by the 1220~Hz QPO frequency observed in
\fu{1636$-$536}, when first-order effects of the
stellar spin are included. The QPO frequency is
assumed to be the Keplerian frequency of gas in a
prograde orbit. The light solid lines are the
mass-radius relations for the same equations of
state as in Figure~14. Allowed combinations of
mass and radius for the indicated values of the
dimensionless angular momentum $j$ are those
within the corresponding pie-shaped regions
bounded by the heavy dashed lines. For a
rotating star, accurate determination of the
allowed region requires computation of the   
bounding curves with $\nu_{\rm spin}$ held equal
to the observed value rather than with $j$ held
constant and therefore depends on the equation of
state assumed as well as the QPO frequency and
the mass of the star (see text).
 }


\begin{planotable}{lcccl}
\tablewidth{500pt}

\tablecaption{
 \label{table:ReportedQPOs}
 Known High-Frequency QPOs and Burst Oscillations$^{\rm a}$} 
\tablehead{Source&Type&Frequencies (Hz)&RMS
Amplitude (\%)&References}

\startdata

4U 0614$+$091&atoll&400\dash 600&$6\dash 15$&Ford et al.\ 1996, 1997a, 1997b\\
&&500\dash 1145&&van der Klis et al.\ 1996c\\
&&327&&M\'endez et al.\ 1997\\
&&630, 727&16.5, 15.8&\\
\hline
4U 1608$-$52&atoll&650\dash 890&5\dash 14&Berger et al.\ 1996\\
&&940\dash 1125&&M\'endez et al.\ 1998\\
&&570\dash 800&&Yu et al.\ 1997\\
\hline
4U 1636$-$536&atoll&840\dash 920&6.0& Zhang et al.\ 1996, 1997a\\
&&1150\dash 1220&6.6, 6.1&van der Klis et al.\ 1996c\\
&&580$^*$&&Wijnands et al.\ 1997b\\
\hline
4U 1728$-$34&atoll&637\dash 716&5.2\dash 6.9&Strohmayer et al.\ 1996a, 1996b,\\
&&500\dash 1100&5.5\dash 8.1&\quad 1996c, 1997a\\
&&363$^*$& 1.5\dash 5\\
\hline
KS 1731$-$260&atoll&524$^*$&12&Morgan \& Smith 1996\\
&&900, 1170\dash 1207&4\dash 5&Smith et al.\ 1997\\
&&&&Wijnands \& van der Klis 1997\\
\hline
4U 1735$-$444&atoll&1149&3.1&Wijnands et al.\ 1996\\
\hline
4U 1820$-$30&atoll&546\dash 796&3.2\dash 5.0&Smale et al.\ 1996, 1997\\
&&1066&&\\
\hline  
Aql~X$-$1&atoll&750\dash 830&\dash&Zhang et al.\ 1998a\\
&&549$^*$&&\\
\hline
Cyg~X-2&Z&730 \dash 1020& 3\dash 5& Wijnands et al.\ 1998\\
&&490\dash 530&&\\
\hline
GX 5$-$1&Z&567\dash 895&2.0\dash 6.7&van der Klis et al.\ 1996e\\
&&325\dash 448&&\\
\hline
GX~17$+$2&Z&470\dash 780&3\dash 5&van der Klis et al.\ 1997a\\
&&645\dash 1087&&Wijnands et al.\ 1997c\\
\hline
GX~340$+$02&Z&247 \dash 625&2.5&Jonker et al.\ 1998\\
&&625 \dash 820&2-5&\\
\hline
GX~349$+$2&Z&712&1.2&Zhang, Strohmayer, \& Swank 1998b\\
&&978&1.3&\\
\hline
Sco X-1&Z&570\dash 830&0.9\dash 1.2&van der Klis et al.\ 1996a,1996b\\ 
&&870\dash 1130&0.6\dash 0.9&\quad 1996d, 1997b\\
\hline
Unknown&unknown&589$^*$&2\dash 4&Strohmayer et al.\ 1996d\\
\enddata
$^{\rm a}$Complete as of 1998 May 31.
$^*$Burst oscillation.

\end{planotable}

\newpage  
\begin{planotable}{ccc}
\tablewidth{450pt}

\tablecaption{
 \label{table:FlowCases}
 Location of the Sonic Point}

\tablehead{Stellar radius&Strong radiation drag
force&Weak radiation drag force}

\startdata
\\
$R<R_{\rm ms}$&Sonic point at $R_{\rm aml}$ (Case 1a)&Sonic 
point at
$R_{\rm ms}$ (Case 1b){\rule[-2mm]{0mm}{6mm}}\\
\\
$R>R_{\rm ms}$&Sonic point at $R_{\rm aml}$ (Case 2a)&No sonic
point in Keplerian\\
&& disk flow (Case 2b)\\
{\rule[-2mm]{0mm}{6mm}}\\

\enddata
\end{planotable}

\begin{planotable}{lccc}
\tablewidth{500pt}

\tablecaption{
 \label{table:FreqGen}
 Amplitudes of frequencies generated by accretion flow}
\tablehead{Frequency&First Order&Second Order&Third Order}

\startdata

$\nu_{\rm Bs}$&$B_1$&---&---\\
$\nu_{\rm Ks}$&$K_1$&---&---\\
$2\nu_{\rm Bs}$&---&$B_2$&---\\
$2\nu_{\rm Ks}$&---&$K_2$&---\\
$\nu_{\rm spin}$&---&${1\over 2}K_1B_1$&---\\
$2\nu_{\rm Ks}-\nu_{\rm spin}$&---&${1\over 2}K_1B_1$&---\\
$\nu_{\rm Ks}-2\nu_{\rm spin}$&---&---&${1\over 2}K_1B_2$\\
$\nu_{\rm Ks}+\nu_{\rm spin}$&---&---&${1\over 2}K_2B_1$\\
$3\nu_{\rm Ks}-2\nu_{\rm spin}$&---&---&${1\over 2}K_1B_2$\\
$3\nu_{\rm Ks}-\nu_{\rm spin}$&---&---&${1\over 2}K_2B_1$\\

\enddata

\end{planotable}

\newpage
\begin{planotable}{lccl}
\tablewidth{500pt}
 
\tablecaption{
 \label{table:SpinFrequencies}
 Inferred Spin Frequencies of Neutron Stars with Kilohertz QPOs$^{\rm a}$}
\tablehead{Kilohertz QPO&$\nu_{\rm spin}$ (Hz)&$\nu_{\rm spin}$ (Hz)&References\\
 Source&from $\Delta\nu$&from $\nu_{\rm burst}$}

\startdata

4U 0614$+$091&$\sim$330&---&Ford et al.\ 1997a, 1997b\\
4U 1608$-$52&230\dash 290&---&M\'endez et al.\ 1998\\
4U 1636$-$536&$\sim$290&$\sim$580&Zhang et al.\ 1996, 1997\\
4U 1728$-$34&$\sim$360&$\sim$363&Strohmayer et al.\ 1996a, 1996b, 1996c\\
KS 1731$-$260&$\sim$260&$\sim$520&Smith et al.\ 1997, Wijnands \& van der Klis 1997\\
4U 1735$-$444&---&---&---\\
4U 1820$-$30&$\sim$275&---&Smale et al.\ 1997\\
Aql~X$-$1&---&$\sim$550&Zhang et al.\ 1998a\\
Cyg~X-2&$\sim$345&---&Wijnands et al.\ 1998\\
GX 5$-$1&$\sim$325&---&van der Klis et al.\ 1996e\\
GX~17$+$2&$\sim$295&---&Wijnands et al.\ 1997c\\
GX~340$+$0&$\sim$325&---&Jonker et al.\ 1998\\
GX~349$+$2&$\sim$266&---&Zhang et al.\ 1998\\
Sco X-1&$\sim$250\dash 300&---&van der Klis et al.\ 1996a, 1996b, 1996d, 1997b\\
Unknown&---&$\sim$590&Strohmayer et al.\ 1997a\\

\enddata
$^{\rm a}$Complete as of 1998 May 31. Here
$\Delta\nu$ is the difference between the
frequencies of the kilohertz QPOs seen
simultaneously in the persistent X-ray emission of
the source and $\nu_{\rm burst}$ is the frequency
of the brightness oscillation seen during type~I
(thermonuclear) X-ray bursts from the source.

\end{planotable}

\newpage
\begin{planotable}{ccc}
\tablewidth{450pt}

\tablecaption{
 \label{table:jValues}
 Dimensionless angular momenta for
 $\nu_{\rm spin}=363$~Hz}

\tablehead{EOS&Mass ($M_\odot$)&$j\equiv cJ/GM^2$}

\startdata

A&1.40&0.13\\
&1.66&0.10\\
FPS&1.40&0.16\\
&1.80&0.11\\
L&1.40&0.28\\
&2.70&0.17\\
M&1.40&0.31\\   
&1.80&0.13\\

\enddata

\end{planotable}

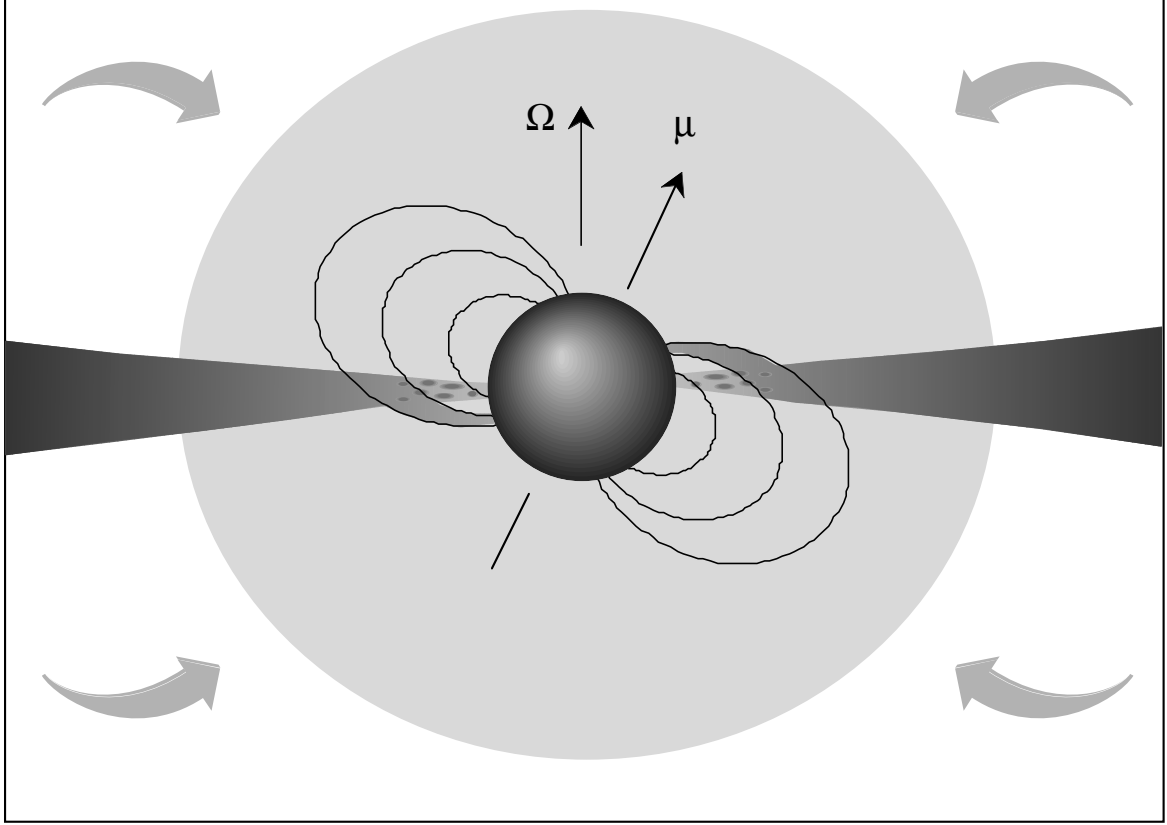
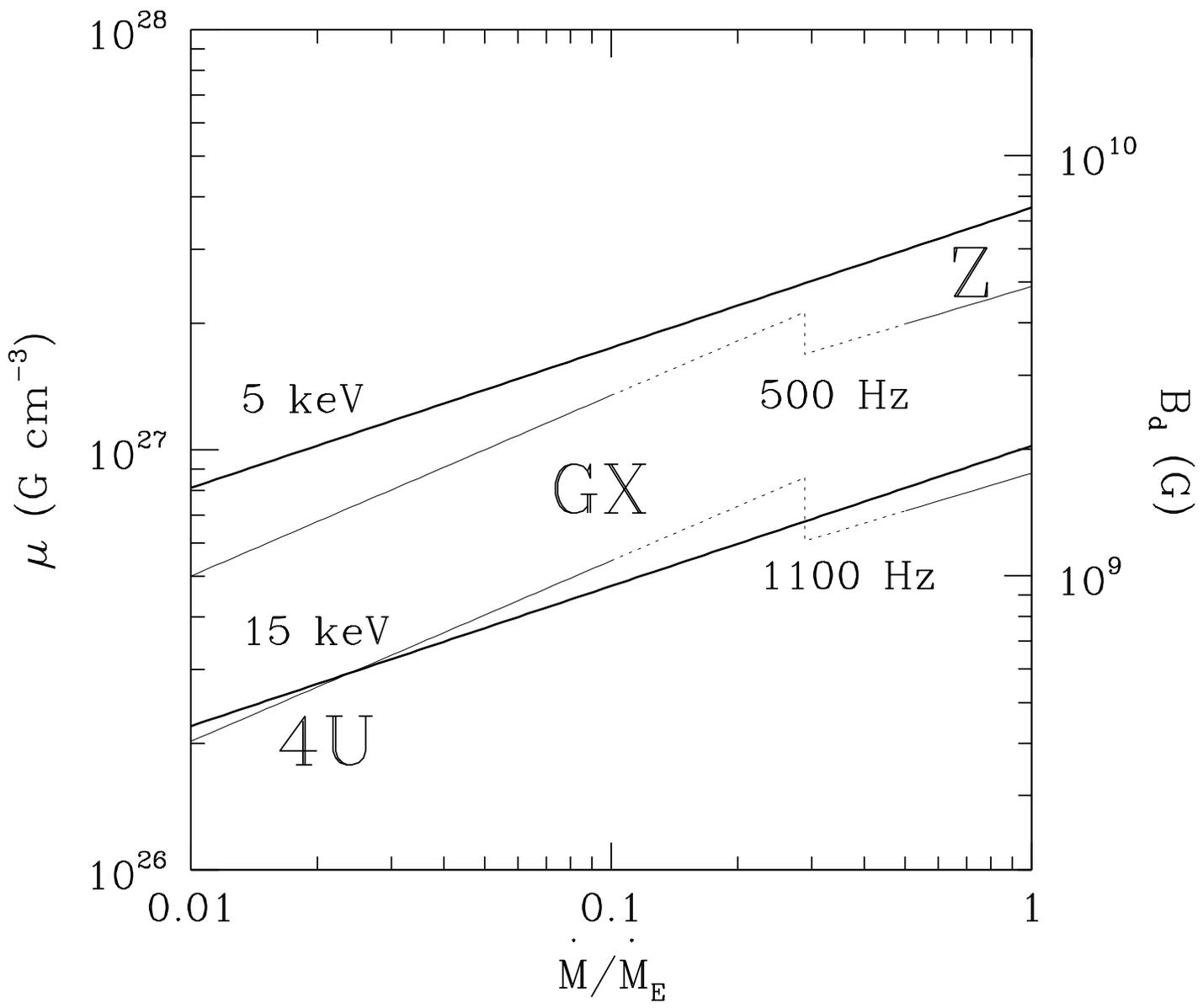
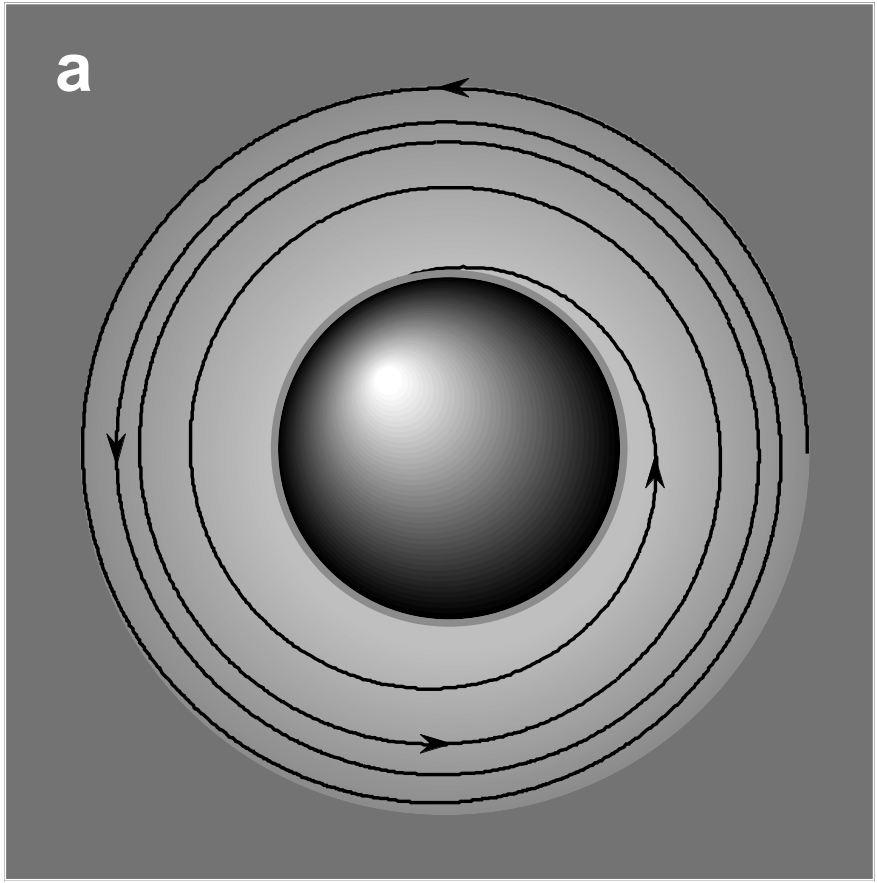
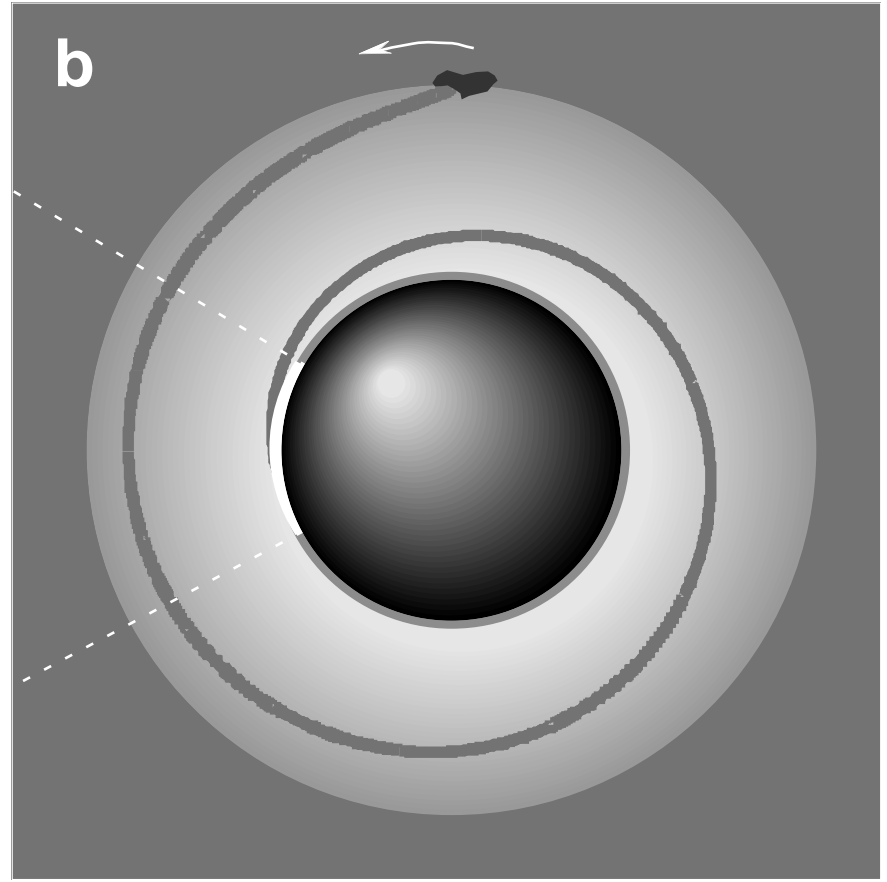
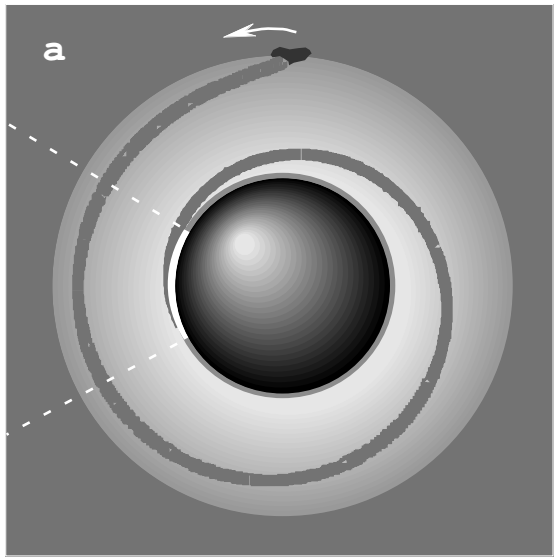
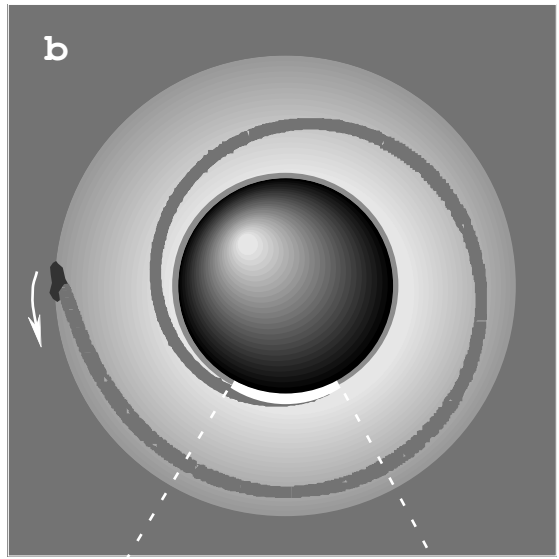
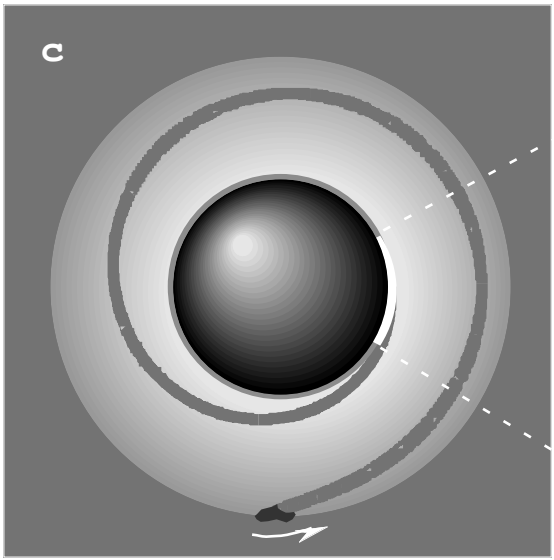
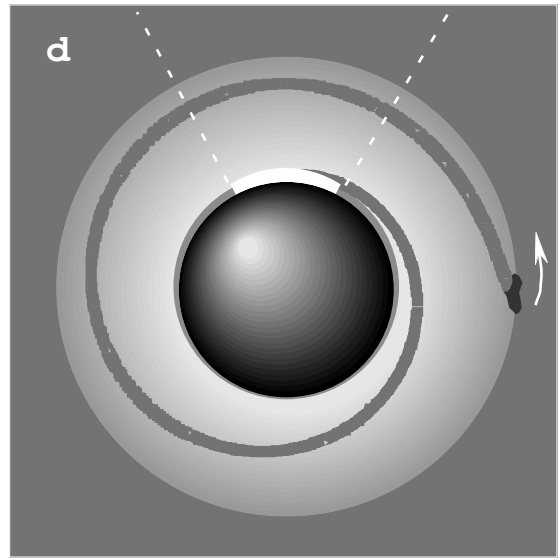
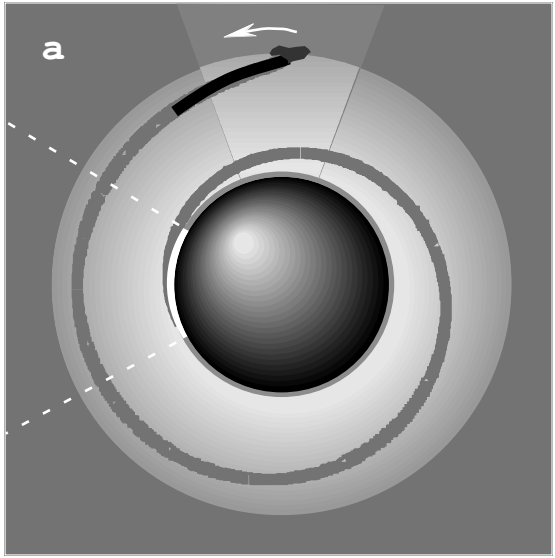
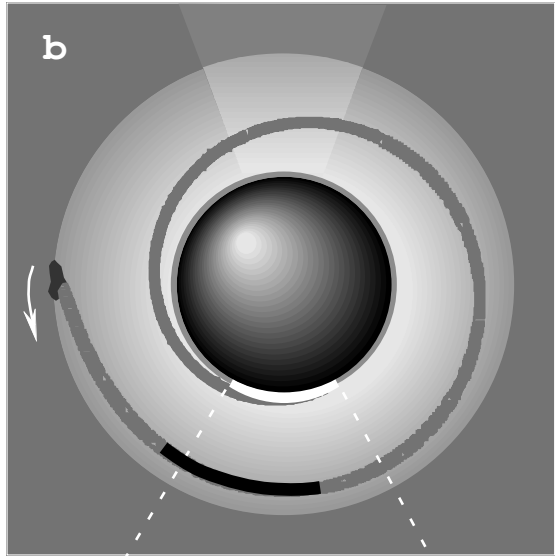
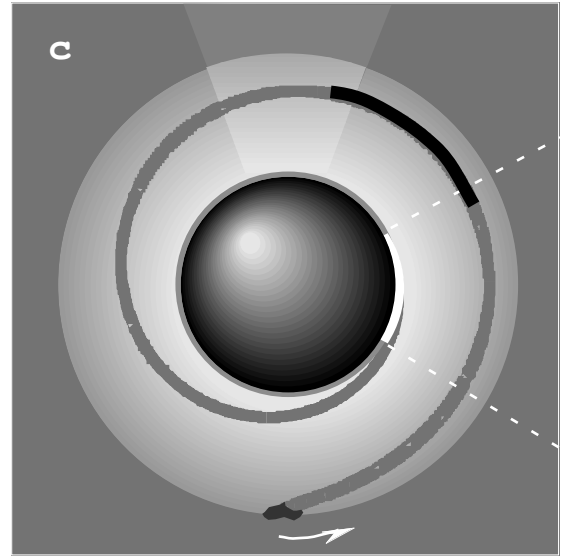
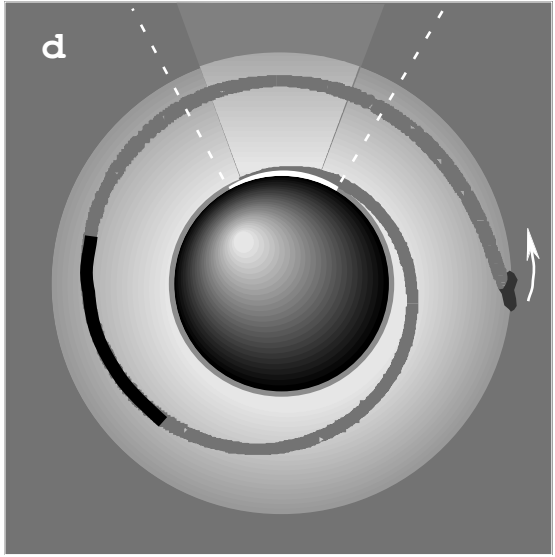
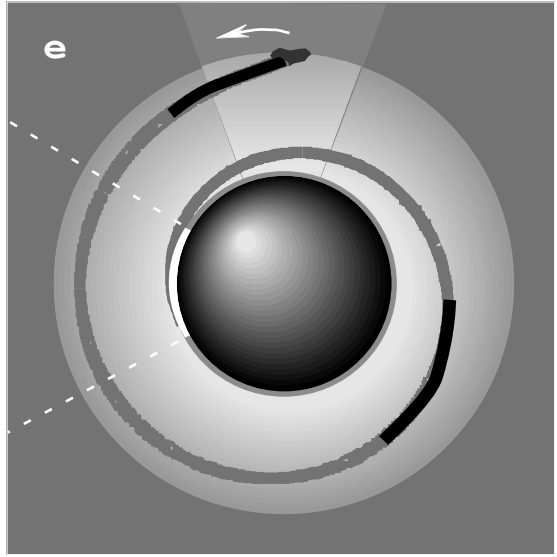
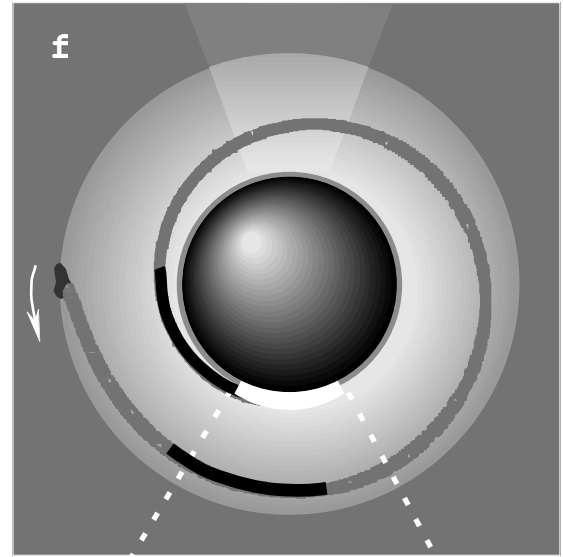
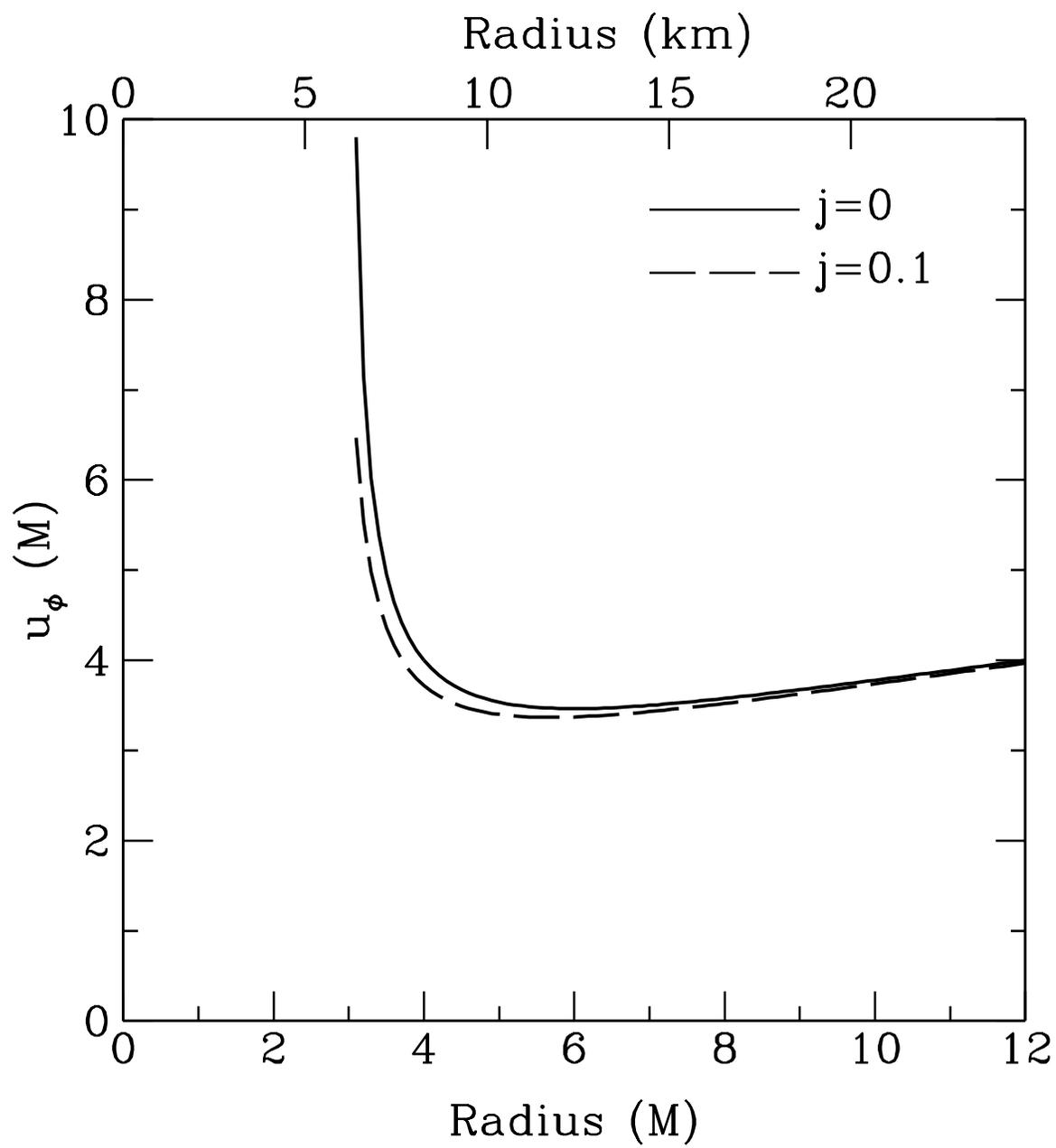
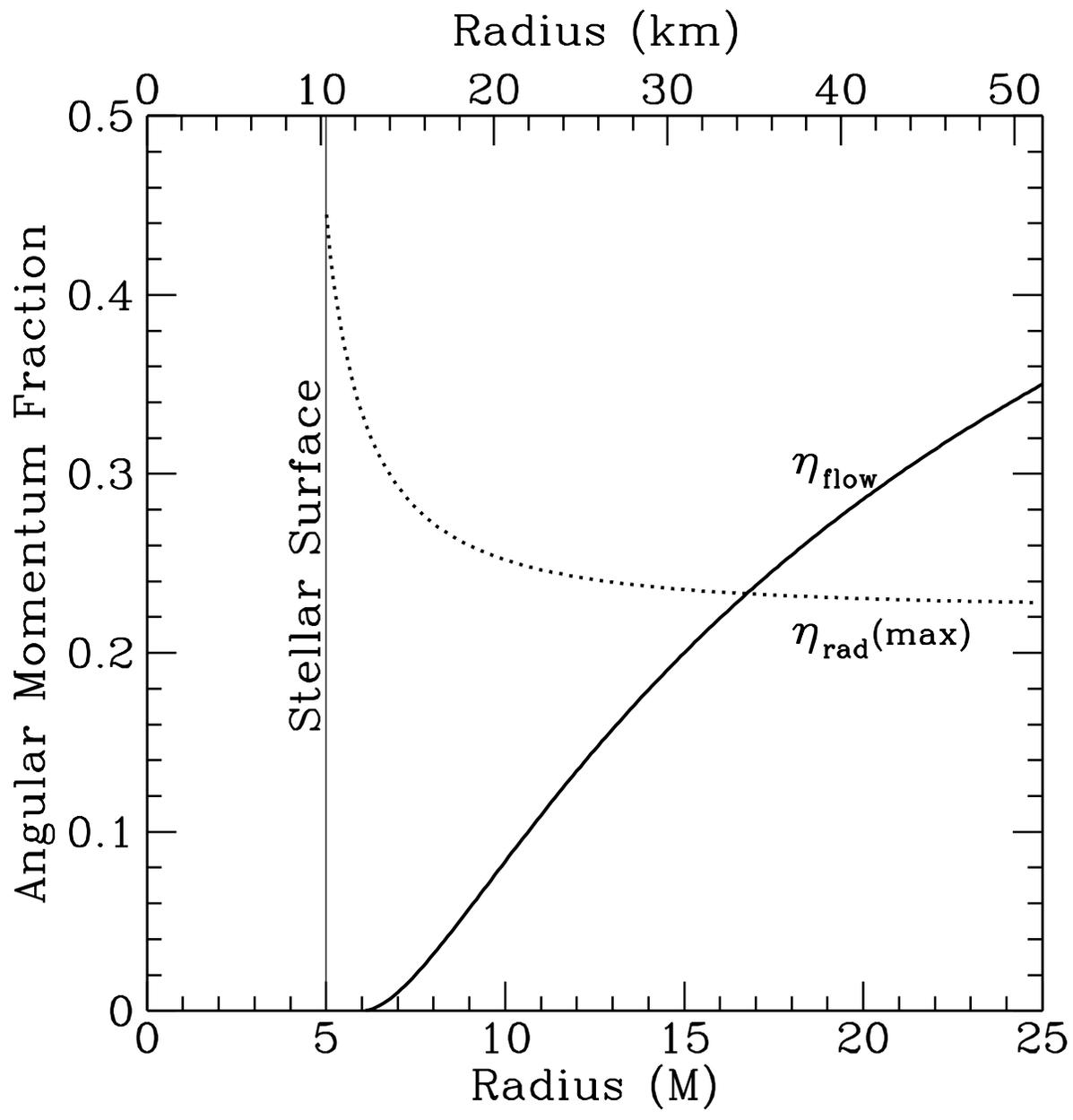
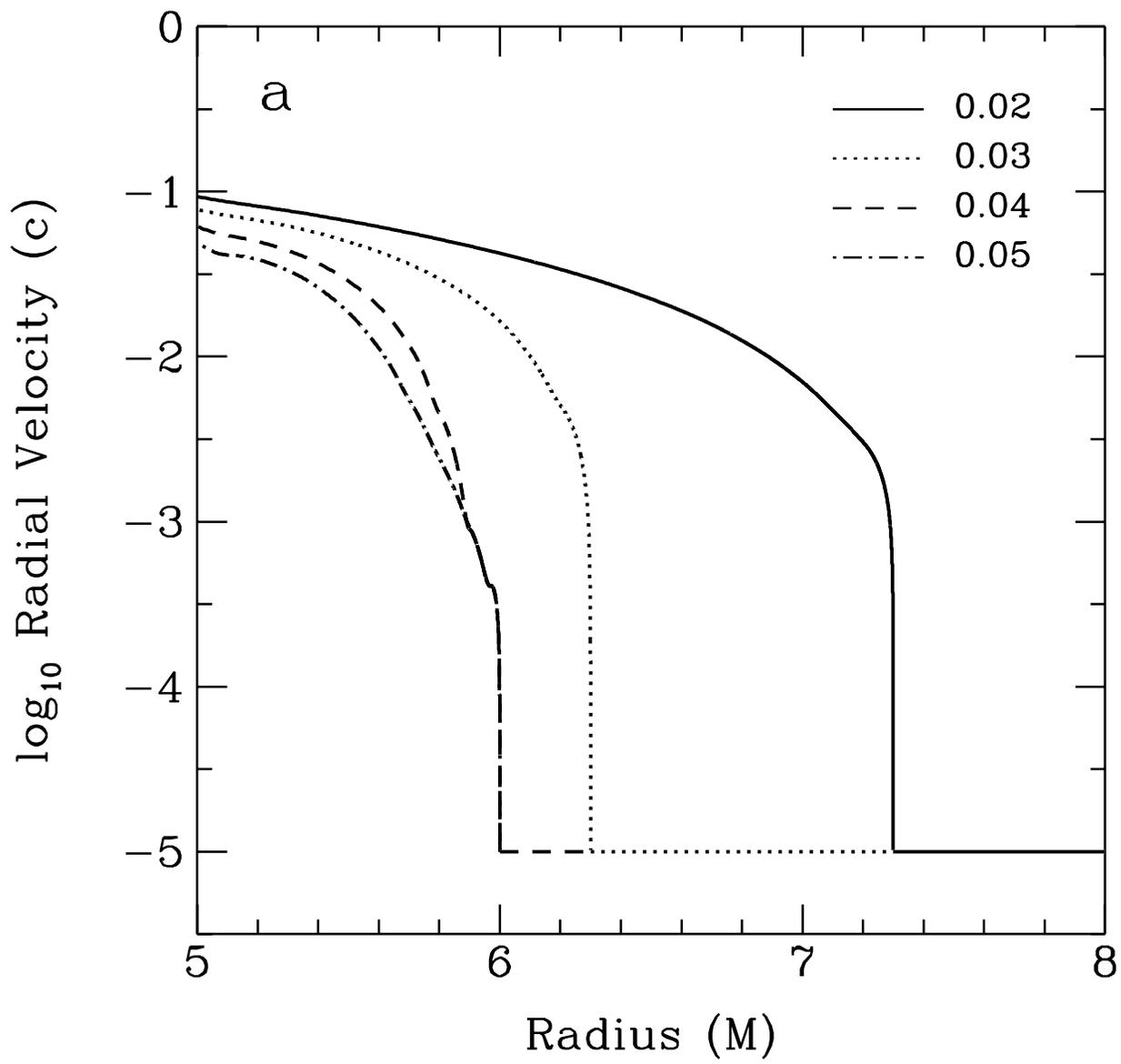
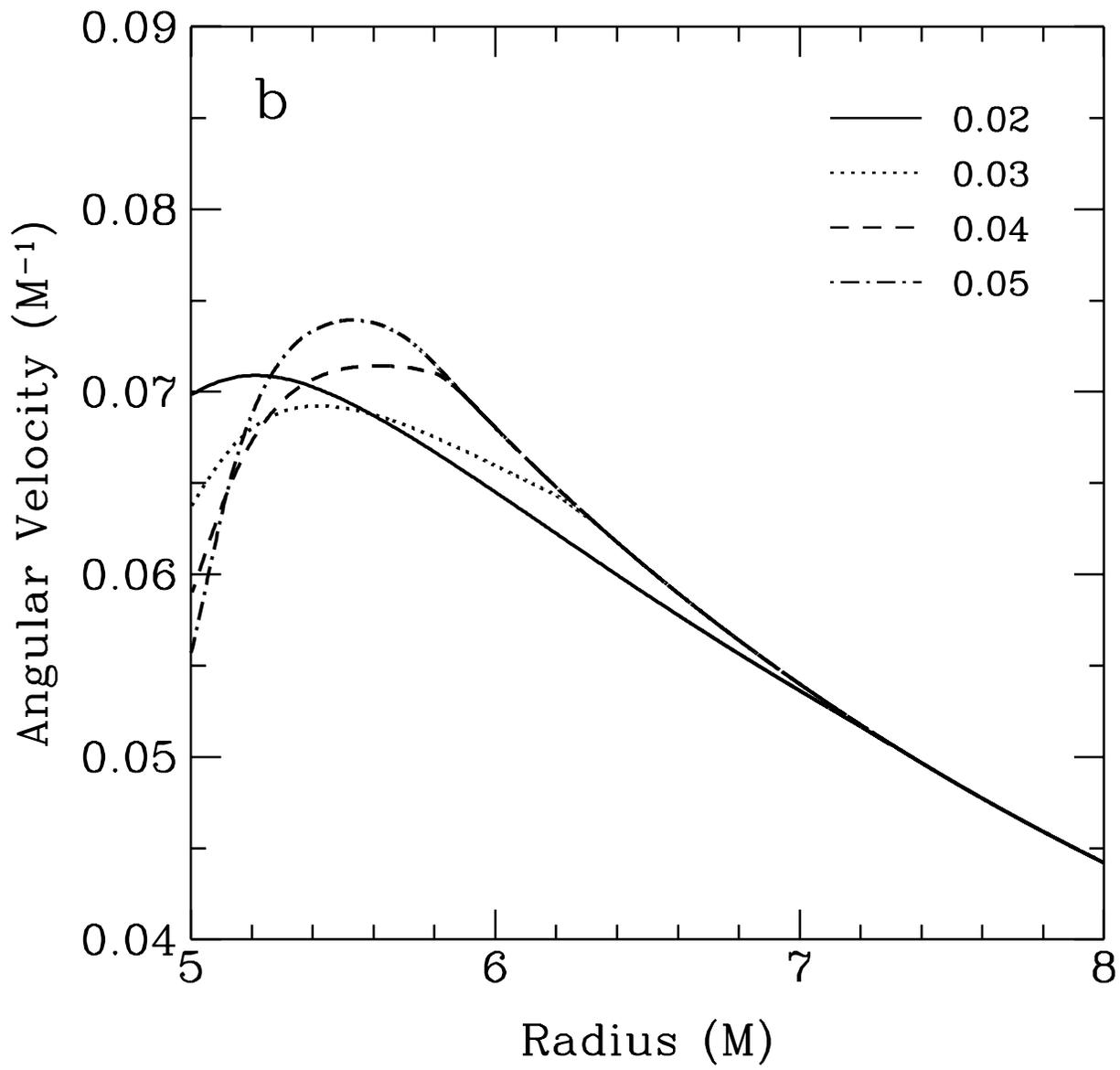
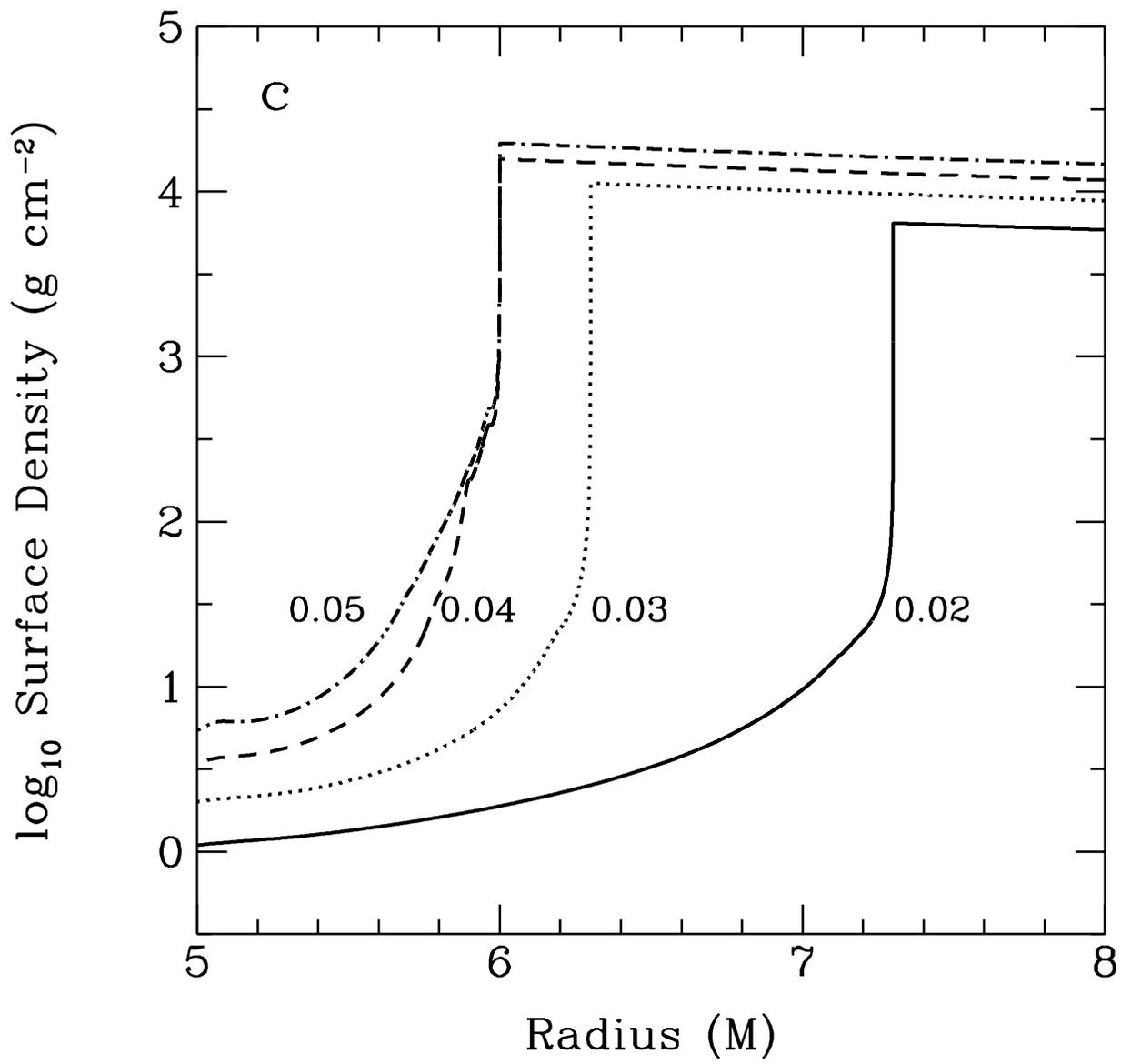
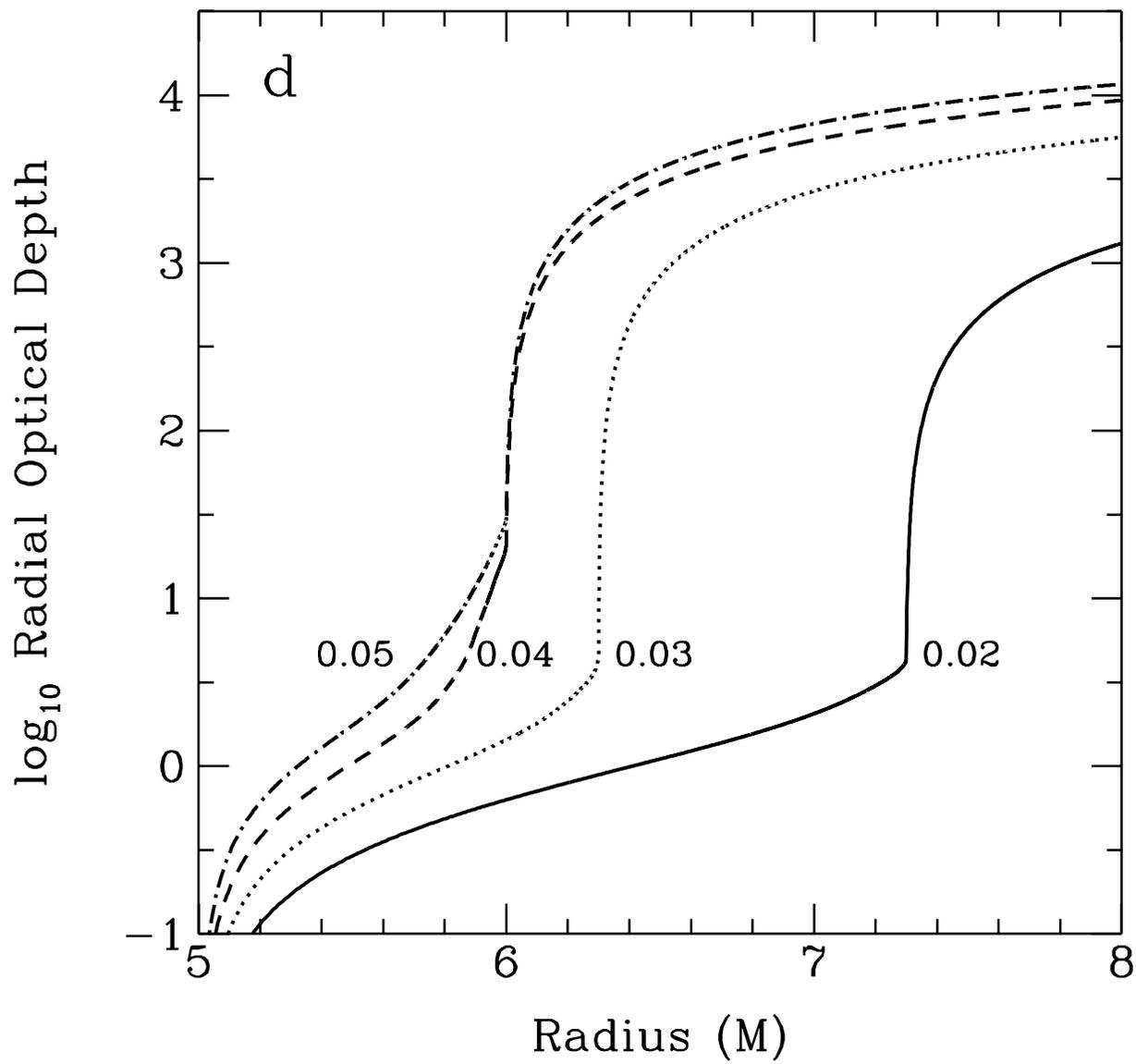
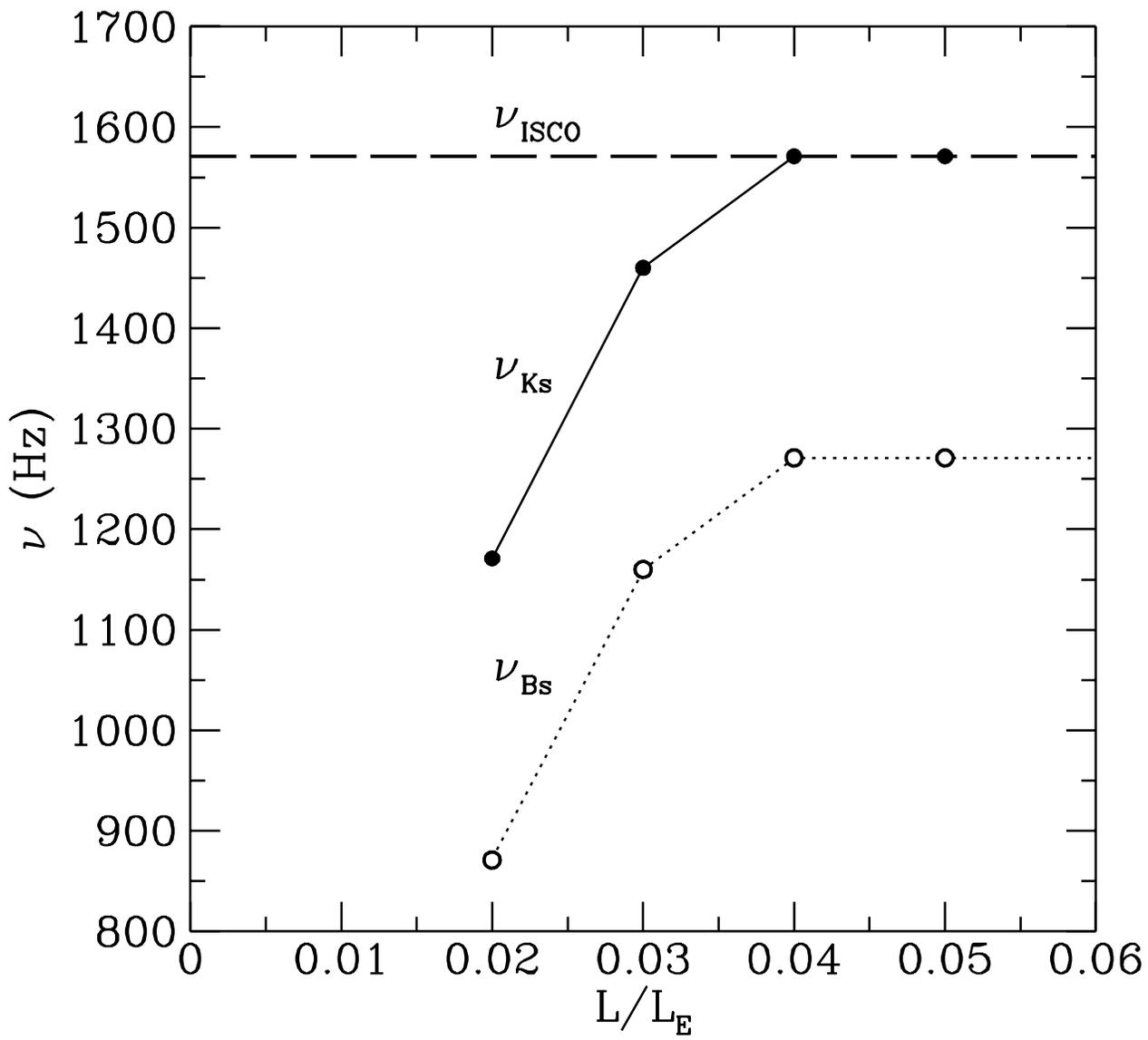
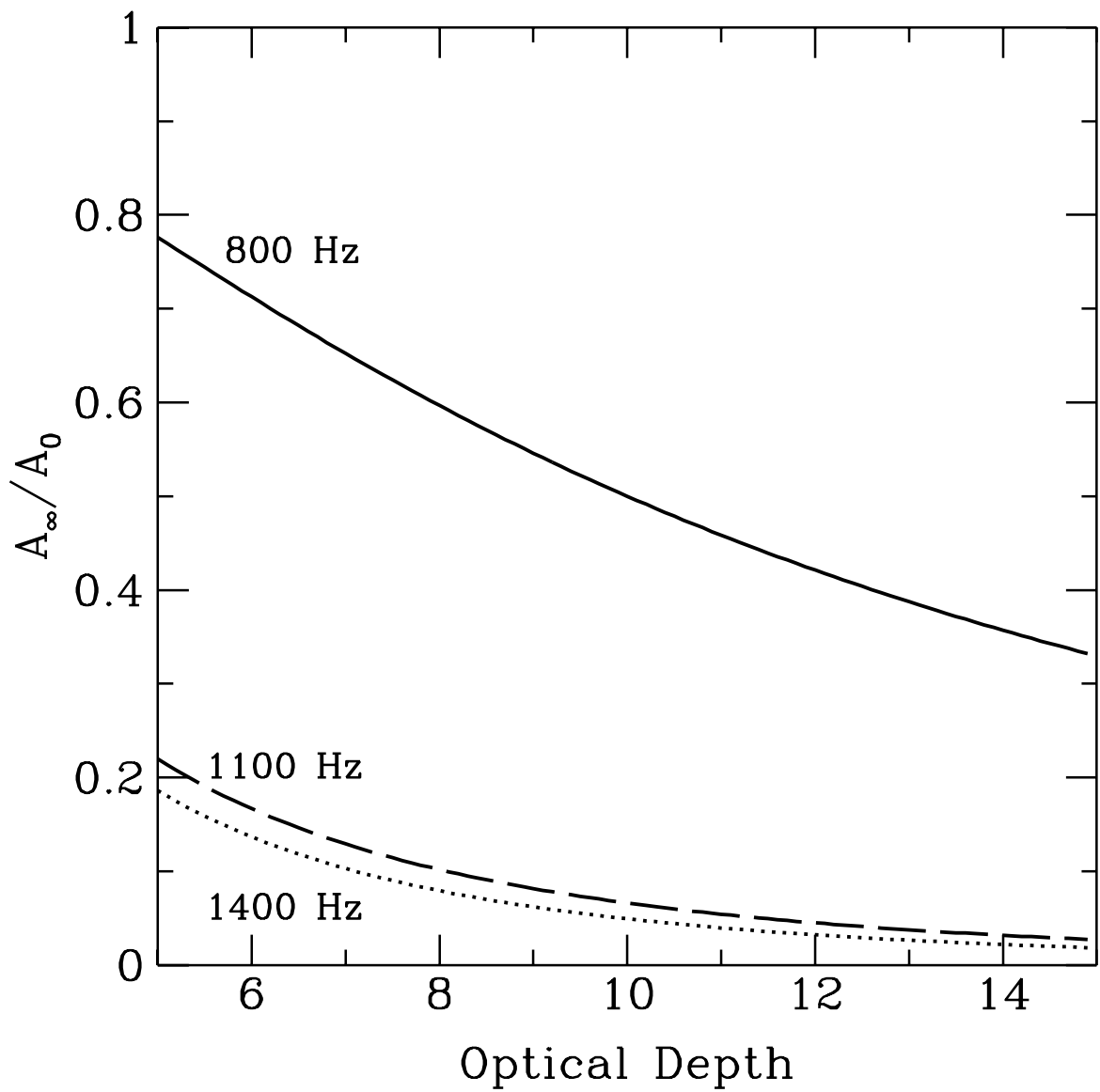
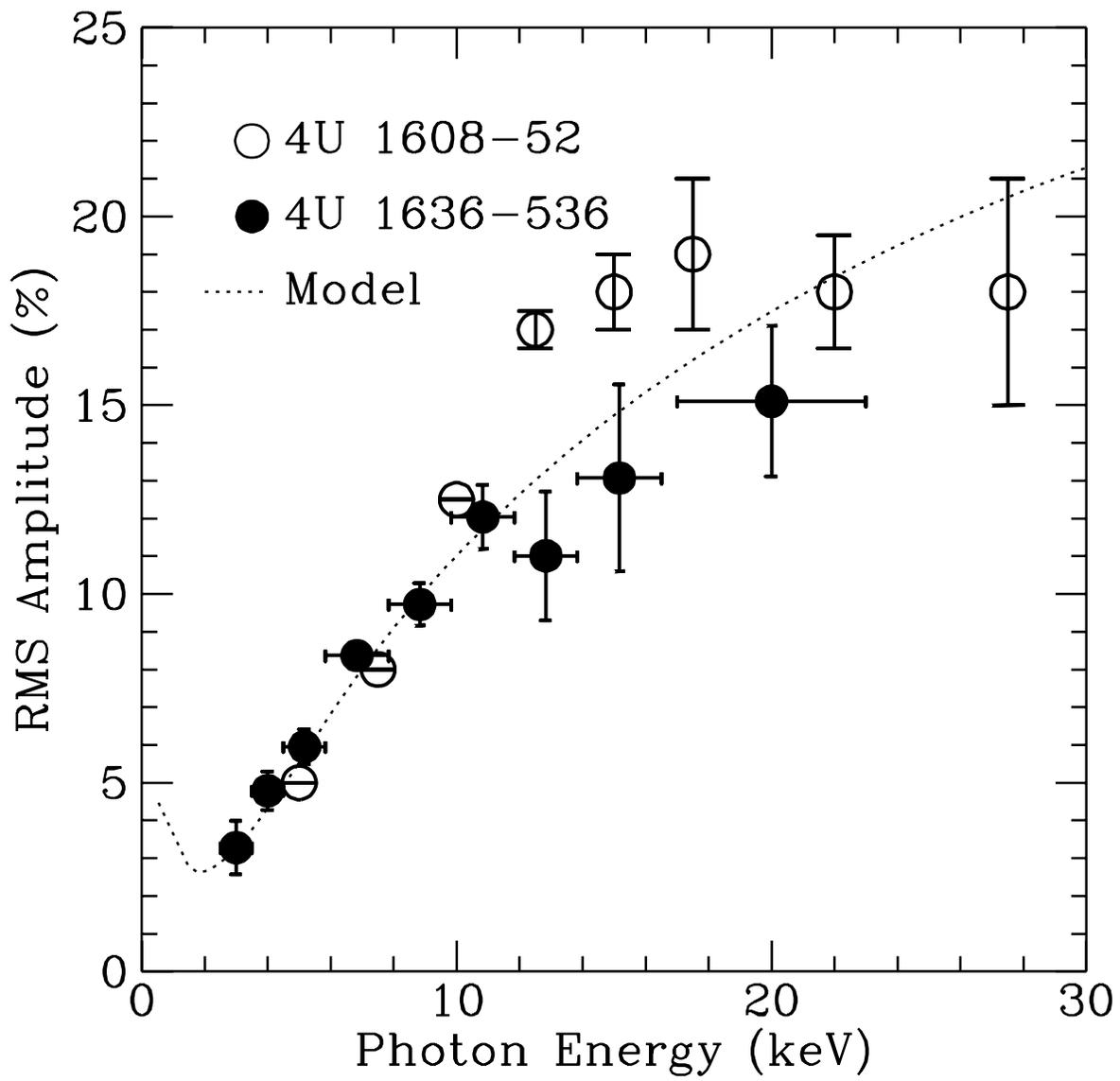
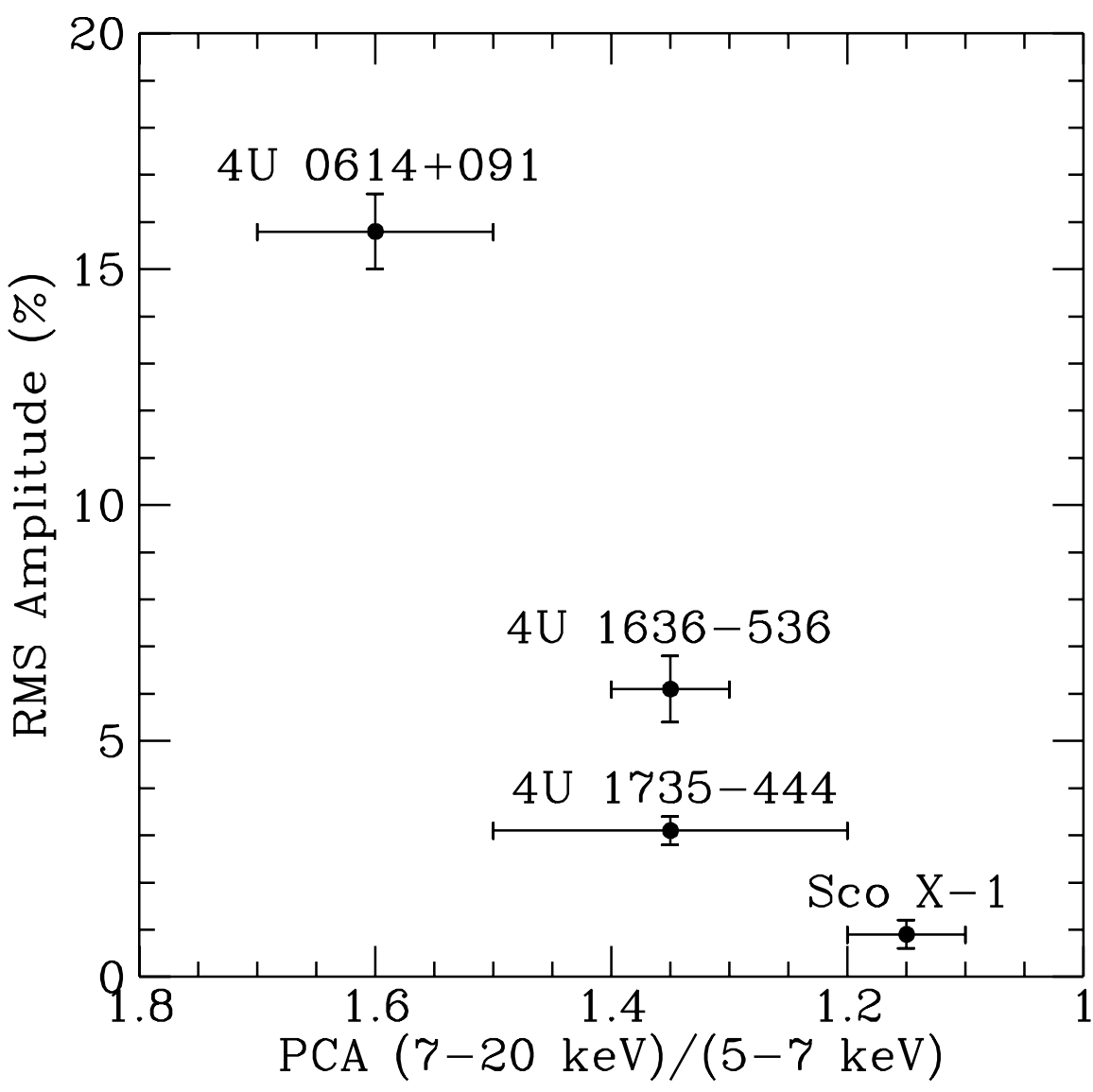
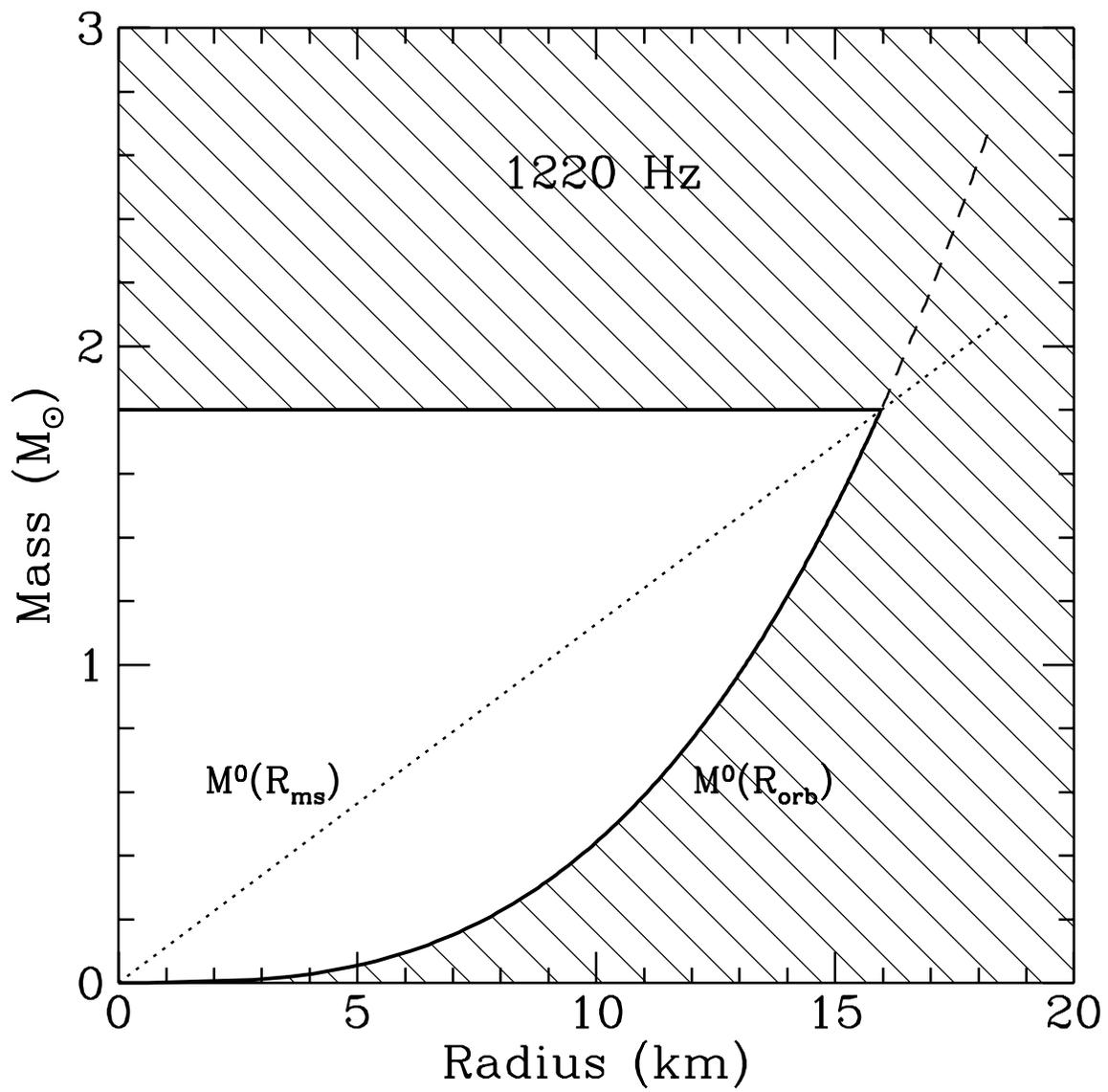
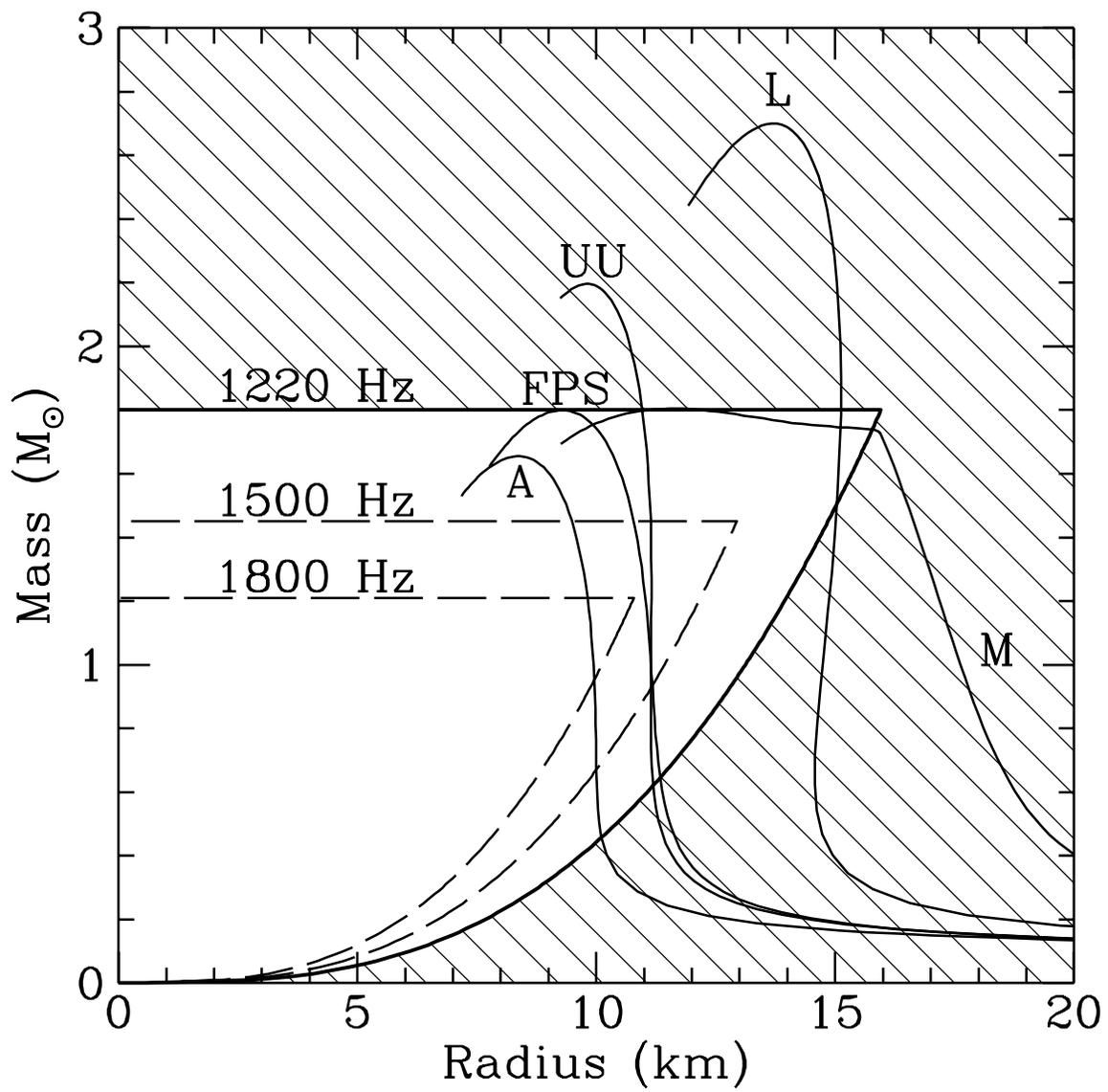
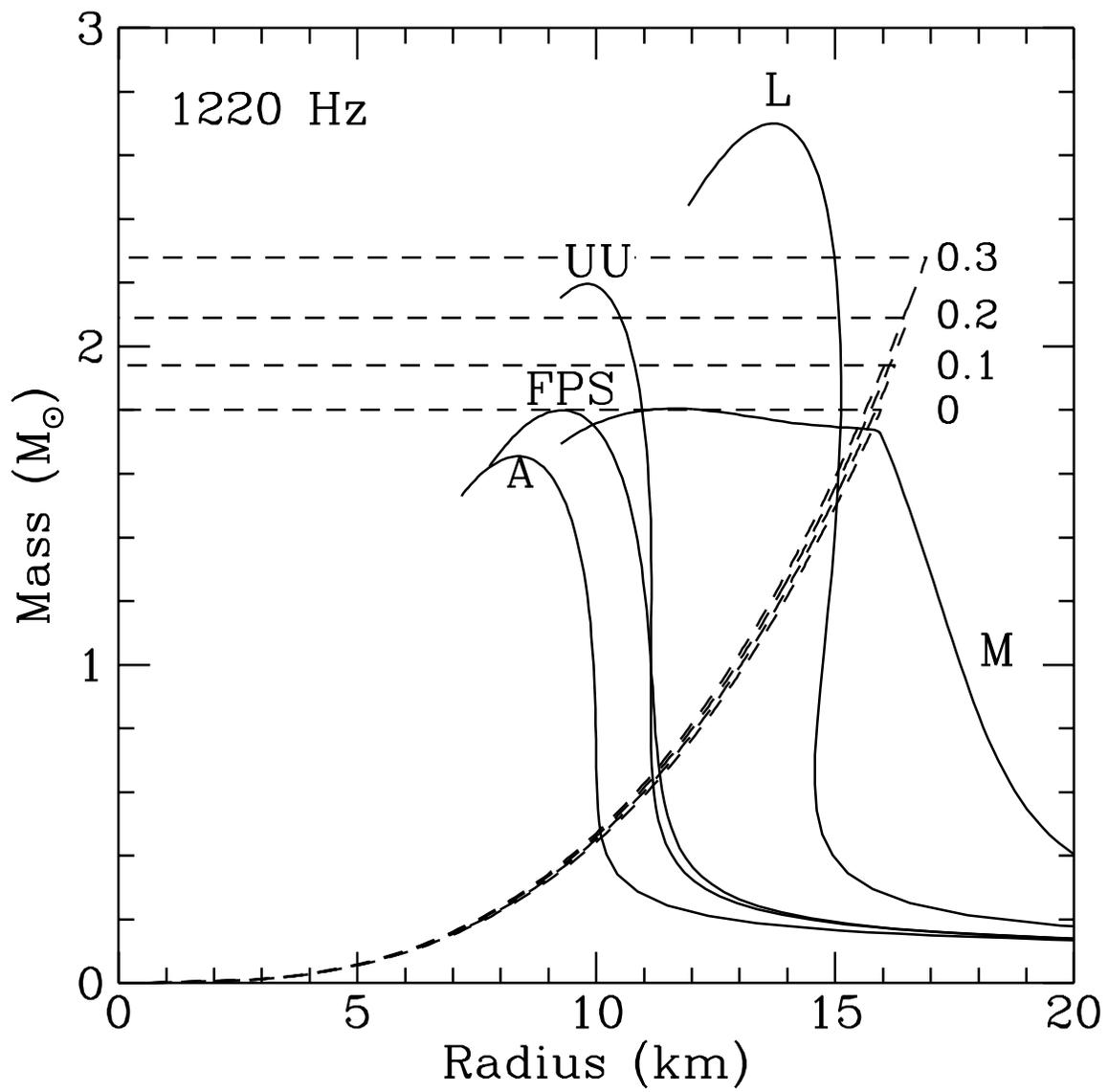
\end{document}